\documentclass[aps,prx,twocolumn,showpacs,superscriptaddress,amsfonts,amsmath,amssymb,assymb,eufrak]{revtex4-1}

\usepackage{epsfig}
\usepackage{color}
\usepackage{bm}
\usepackage{bbm}
\usepackage{dsfont}
\usepackage{braket}
\usepackage{sidecap}
\usepackage{floatrow}
\usepackage{braket}

\usepackage{graphicx}
\newcommand{\todayd}{\the\year/\the\month/\the\day}

\newcommand{\bib}{\bibitem}

\newcommand{\bel}{\begin{easylist}}
\newcommand{\eel}{\end{easylist}}
\def \({\left(}
\def \){\right)}

\newcommand{\sumtwo}[2]%
{\mathop{\sum_{#1}}_{#2}}
\newcommand{\sumthree}[3]%
{\mathop{\mathop{\sum_{#1}}_{#2}}_{#3}}
\newcommand{\sumfour}[4]%
{\mathop{\mathop{\mathop{\sum_{#1}}_{#2}}_{#3}}_{#4}} 
\newcommand{\prodtwo}[2]%
{\mathop{\prod_{#1}}_{#2}}
\newcommand{\mintwo}[2]%
{\mathop{\min_{#1}}_{#2}}
\newcommand{\maxtwo}[2]%
{\mathop{\max_{#1}}_{#2}}
\newcommand{\maxthree}[3]%
{\mathop{\mathop{\max_{#1}}_{#2}}_{#3}}
\newcommand{\limtwo}[2]%
{\mathop{\lim_{#1}}_{#2}}
\newcommand{\suptwo}[2]%
{\mathop{\sup_{#1}}_{#2}}
\newcommand{\supthree}[3]%
{\mathop{\mathop{\sup_{#1}}_{#2}}_{#3}}
\newcommand{\supfour}[4]%
{\mathop{\mathop{\mathop{\sup_{#1}}_{#2}}_{#3}}_{#4}} 
\newcommand{\inftwo}[2]%
{\mathop{\inf_{#1}}_{#2}}
\newcommand{\infthree}[3]%
{\mathop{\mathop{\inf_{#1}}_{#2}}_{#3}}
\newcommand{\inffour}[4]%
{\mathop{\mathop{\mathop{\inf_{#1}}_{#2}}_{#3}}_{#4}} 
\def\rnum#1{\resizebox{0.5em}{\height}{\expandafter{\romannumeral #1}}}
\def\Rnum#1{\resizebox{0.5em}{\height}{\uppercase\expandafter{\romannumeral #1}}}

\begin{document}

\title{Valence-bond solids, vestigial order, and emergent SO(5) symmetry \\ in a two-dimensional quantum magnet}

\author{Jun Takahashi}
\email{jt@iphy.ac.cn}
\affiliation{Beijing National Laboratory for Condensed Matter Physics and Institute of Physics, Chinese Academy of Sciences, Beijing 100190, China}
\affiliation{Department of Physics, Boston University, 590 Commonwealth Avenue, Boston, Massachusetts 02215, USA}

\author{Anders W. Sandvik}
\email{sandvik@bu.edu}
\affiliation{Department of Physics, Boston University, 590 Commonwealth Avenue, Boston, Massachusetts 02215, USA}
\affiliation{Beijing National Laboratory for Condensed Matter Physics and Institute of Physics, Chinese Academy of Sciences, Beijing 100190, China}

\date{\today}
             
\begin{abstract}
  We introduce a quantum spin-1/2 model with many-body correlated Heisenberg-type interactions on the two-dimensional
  square lattice, designed so that the system can host a four-fold degenerate plaquette valence-bond solid (PVBS) ground state
  that spontaneously breaks $\mathbb{Z}_4$ symmetry. The system is sign-problem free and amenable to large-scale quantum
  Monte Carlo simulations, thus allowing us to carry out a detailed study of the quantum phase transition between the
  standard N\'eel antiferromagnetic (AFM) and PVBS states. We find a first-order transition, in contrast to previously
  studied continuous transitions from the AFM phase into a columnar valence-bond solid
  (CVBS) phase. The theory of deconfined quantum criticality predicts generic continuous AFM--CVBS and AFM--PVBS transitions,
  and, in one version of the theory, the two critical order parameters transform under SO(5) symmetry.
  Emergent SO(5) symmetry has indeed been observed in studies of the AFM--CVBS transition, and here we show
  that the first-order AFM--PVBS transition also is associated with SO(5) symmetry. Such unexpected symmetry of the
  coexistence state, which implies a lack of energy barriers between the coexisting phases, has recently been observed
  at other first-order transitions, but the case presented here is the first example with SO(5) symmetry. The extended
  symmetry may indicate that the transition is connected to a deconfined critical point. We also discuss the
  first-order transition in the context of a recent proposal of spinons with fracton properties in the PVBS state, concluding
  that the fracton scenario is unlikely.
  Furthermore, we discover a novel type of eight-fold degenerate VBS phase, arising when the PVBS state
  breaks a remaining $\mathbb{Z}_2$ symmetry. This second phase transition, which is continuous, implies that the PVBS phase can be
  regarded as an intermediate ``vestigial'' phase, a concept recently introduced to describe multi-stage phase transitions
  involving a continuous symmetry. Here we construct a six-dimensional order parameter and also introduce a general graph-theoretic
  approach to describe the two-stage discrete symmetry breaking. We discuss different ways of breaking the symmetries in
  one or two stages at zero and finite temperatures. In the latter case we observe fluctuation-induced first-order
  transitions, which are hallmarks of vestigial phase transitions. We also mention possible connections of the AFM--PVBS transition
  to the SO(5) theory of high-$T_{\rm c}$ superconductivity.
\end{abstract}

\maketitle

\section{Introduction}

Competing interactions in a quantum antiferromagnet with spin-rotationally invariant interactions can lead to the destruction of the conventional N\'eel
antiferromagnetic (AFM) O(3) symmetry-broken ground state in two or higher dimensions. The most well studied transition into a gapped quantum paramagnet 
is in dimerized spin-1/2 systems (two spins per unit cell), where the ground state is non-degenerate and can be approximated as a product of singlets on
the dimers. This quantum phase transition has an experimental realization in TlCuCl$_3$ under pressure \cite{Merchant14}. For a uniform system with one spin
per unit cell, a uniform non-degenerate paramagnetic ground state is possible only under special conditions \cite{AKLT,Hastings04}. With a half-integer spin per unit
cell (e.g., a single spin-$1/2$) a more complex state must necessarily obtain which has either non-local entanglement and topological order
(a spin liquid \cite{Savary17}) or strong local entanglement leading to a degenerate symmetry-breaking pattern of singlets (a valence-bond solid,
VBS, sometimes also called valence-bond crystal) \cite{Majumdar69a,Majumdar69b,Misguich05,Read89,Read90,Sachdev96,Dagotto89,Misguich05}.

While experimental searches for spin liquids have been a dominant theme in quantum materials science for more than a decade \cite{Zhou17}, comparatively
less efforts have been devoted to quantum magnets with VBS states. A dimer VBS in the quasi-one-dimensional material GeCuO$_3$ was exhaustively studied in the
1990s \cite{Hase93}. More recently, signs of a VBS with singlets forming on four-spin plaquettes were detected in the quasi-two-dimensional system
SrCu$_2$(BO$_3$)$_2$ under high pressure \cite{Zayed17}. This material also hosts an adjacent AFM phase at still higher pressures \cite{Guo19}, thus for
the first time opening prospects for detailed experimental studies of a direct transition between AFM and VBS states in two dimensions.

\begin{figure}[t]
\includegraphics[width=8.2cm]{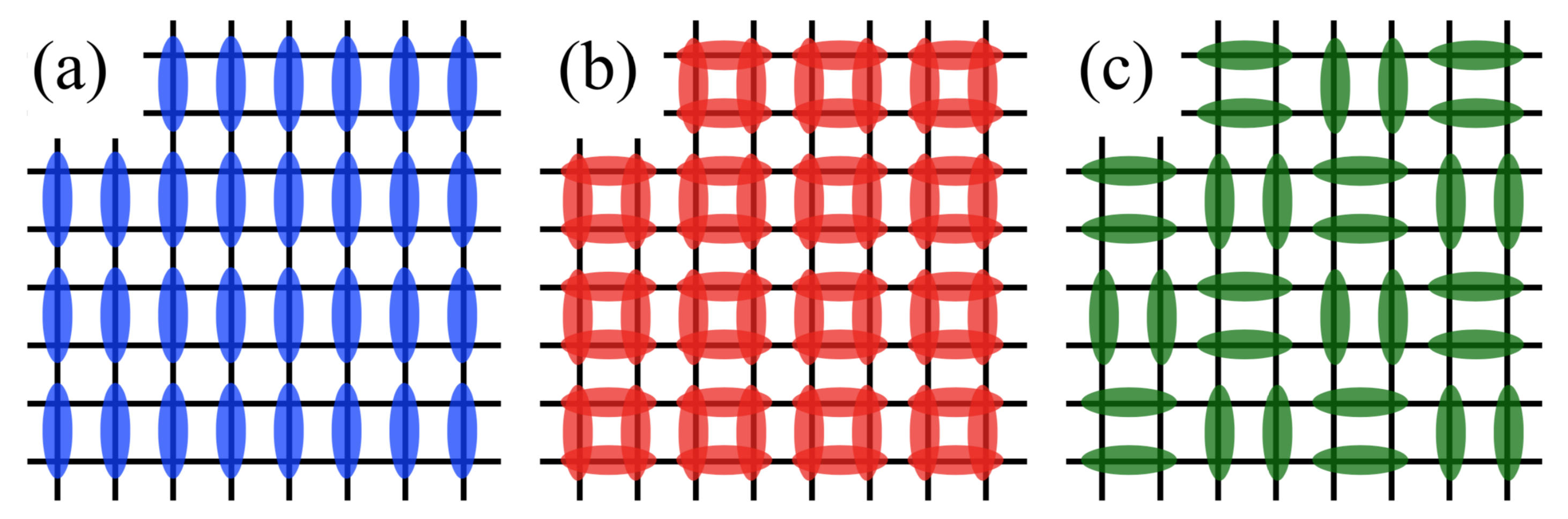}
\caption{\label{fig:VBS}
  Depiction of the different types of VBS states discussed in this work; (a) four-fold degenerate CVBS, (b) four-fold degenerate PVBS, and (c)
  eight-fold degenerate AVBS. Long ovals in (a) and (c) represent spontaneously ``frozen'' singlets and the ovals forming squares in (b)
  represent plaquette singlets, which can be expressed as  resonating pairs of horizontal and vertical singlet pairs. }
 \end{figure}

The AFM state has an obvious classical counterpart, and its low-energy
properties, including its quantum phase transition into the dimer paramagnet, can be understood within essentially classical field theories in one higher
dimension (imaginary time, to account for quantum fluctuations) \cite{CHN89,Chubukov94}. In contrast, spin liquids and VBSs are exotic from the classical
standpoint, and low-energy descriptions of them and their quantum phase transitions require more drastic deviations from the conventional field theories
of statistical physics \cite{Sachdev08,Wen19}. The mathematical complexity of the quantum-field theories used to described exotic states of quantum magnets
is formidable, and direct computational studies of the low-energy properties of lattice models are indispensable for testing and guiding analytical approaches,
and can also open new research directions \cite{Kaul13}.

In this article we introduce a two-dimensional (2D) quantum spin model in which the AFM ground state successively transitions into two different types of
VBS states; a four-fold degenerate plaquette VBS (PVBS) followed by an eight-fold degenerate state we name the alternating VBS (AVBS). These states
are schematically depicted in Fig.~\ref{fig:VBS}, along with the more commonly studied columnar VBS (CVBS) state. In the idealized PVBS state illustrated in
Fig.~\ref{fig:VBS}(b), a plaquette is a resonating pair of valence bonds, with equal amplitude for horizontal and vertical orientation. Such a singlet state
also goes under other names, e.g., plaquette-singlet solid or valence-plaquette solid. Here we adopt the name VBS as a generic term for non-magnetic states
breaking lattice symmetries, forming ordered unit cells with either static or resonating valence bond descriptions. 
In general, quantum and/or thermal fluctuations perturb the completely ideal singlet product states in Fig.~\ref{fig:VBS}, 
but the depicted patterns survive in actual ground states by having finite valued order parameters. 
The pattern can also be explicitly visualized by observing the singlet-density maps, as we will do later. 

\begin{figure}[t]
\includegraphics[width=8.17cm]{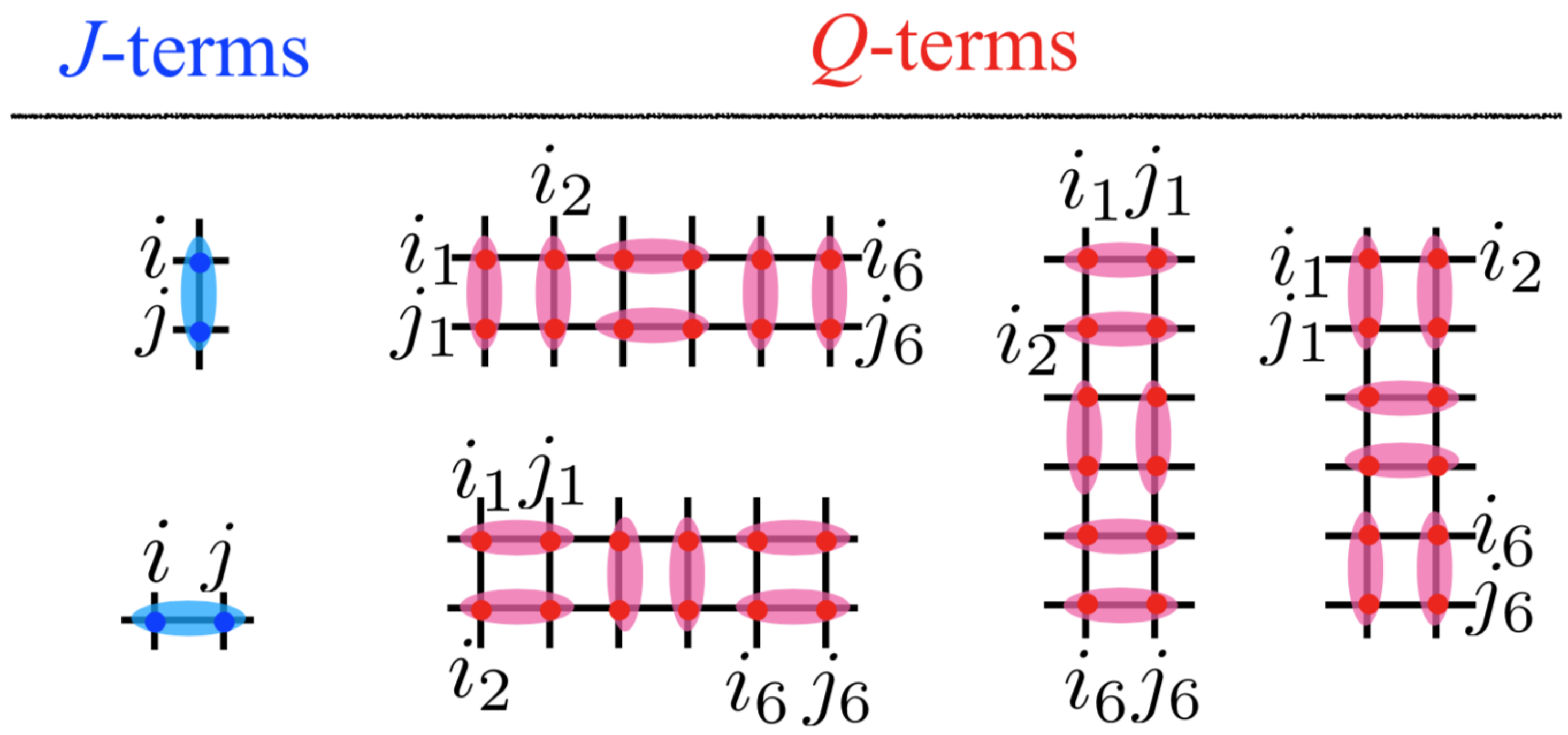}
\caption{\label{fig:JQ6}
  The $J$-$Q_6$ model in which PVBS and AVBS ground states (Fig.~\ref{fig:VBS}) are realized. The two types of interactions between the spins on the
  square lattice are defined using two-spin singlet projectors, here depicted with ovals. The standard Heisenberg exchange of strength $J$ between nearest
  neighbor spins is equivalent to projectors on the horizontal and vertical links,  while the interaction of strength $Q$ introduced here is expressed with
  products of six projectors, arranged in four different ways. The site labels are used in the formal definition of the Hamiltonian.}
\end{figure}

Using quantum Monte Carlo (QMC) simulations of the  $J$-$Q_6$ model illustrated in Fig.~\ref{fig:JQ6}, we find an unusual first-order AFM--PVBS quantum
phase transition with emergent SO(5) symmetry of the five-component combined AFM (three components) and PVBS (two components) order parameters. To analyze
the second transition, a continuous PVBS--AVBS transition, we introduce a new unified generic graph-theoretic description of multi-stage discrete symmetry
breaking transitions, which we use here to analyze the PVBS and AVBS order parameters and their possible symmetry breaking paths.

The AFM--PVBS transition is an interesting analogy to the
AFM--superconducting transition within the SO(5) theory of the cuprates \cite{Zhang97,Demler04}. Overall our results connect to several of the major concepts
currently under debate from the field-theory perspective; deconfined quantum critical (DQC) points \cite{Wang17}, spinons with fracton properties \cite{You19},
and vestigial phase transitions \cite{Fernandes19}. In the remainder of this introductory section we provide further background on these notions and
summarize our aims and findings.

\subsection{Valence-bond solid states and deconfined quantum criticality}

VBSs are often regarded as less exotic and interesting than spin liquids, because they have local order
parameters. However, the objects that form the long-range order in a VBS---the singlets---are not the elementary microscopic degrees of freedom of the model but
emergent entangled composites with no direct counterparts in conventional classical spin models. It was realized in 1980s and 1990s that continuum field-theory
descriptions of one-dimensional (1D) \cite{Haldane83,Affleck85,Affleck87} and 2D \cite{Haldane88,Murthy90,Read90} VBS states require careful consideration of
topological defects and their quantum interference, which is still an active field of research \cite{You19,Vishwanath04,Banerjee11,Xu11}. While the early
field-theory studies already indicated that 2D AFM--VBS transitions may be unusual, it was only after intriguing numerical results were found \cite{Sandvik02,Motrunich04}
that it was recognized that direct quantum phase transitions between AFM and VBS ground states may fall outside the standard Landau-Ginzburg-Wilson
(LGW) paradigm \cite{Senthil04a,Senthil04b} (as was already known to be the case with many transitions in 1D systems). Concurrently, some other exotic
non-LGW 2D transitions related to VBSs were also proposed in quantum-dimer systems \cite{Fradkin04}.

Within the LGW framework, direct transitions between two
ordered phases with unrelated symmetries are generically first-order. However, the two order parameters at the AFM--VBS transition are not independently
fluctuating but represent different manifestations of the confinement of more fundamental objects---spinons---which deconfine at the phase transition (the DQC
point). The expected first-order transition may then be supplanted by a generically continuous transition.

Although the DQC scenario predicts fascinating connections between the AFM and VBS states, it relies on theoretical assumptions such as the conservation
of skyrmion numbers at the transition point  \cite{Motrunich04,Senthil04a,Senthil04b}. Some of the assumptions are hard to justify rigorously, and,
thus, unbiased computational research plays a crucial role in this developing field. Numerical confirmations of the DQC phenomenon face several challenges, however.
To begin with, it is difficult to distinguish a continuous transition from a weak first-order transition in general. In order to deal with this issue in a reliable
manner, one needs to utilize a method which can reach large lattice sizes while still being free from systematic errors. QMC methods serve
such purposes, but in many cases they suffer from the sign problem \cite{Loh90,Chandrasekharan99} and are not useful.

The conditions for DQC transitions are at
first sight most naturally realized in geometrically frustrated SU(2) symmetric quantum spin systems, which is one of the model classes for which the QMC sign-problem
is most severe \cite{Nakamura98,Henelius00}. The lattices accessible with exact numerical diagonalization methods are too small to even allow definite characterization
of the non-magnetic states in these systems \cite{Dagotto89,Misguich05,Schulz96}. While in principle the density-matrix renormalization (DMRG) method or variational
states based on tensor networks can be used, and some results for potential DQC systems have been reported \cite{Gong14,Yu12,Zhao12,Haghshenas18}, these
techniques have not yet reached the length scales and reliability of modern QMC techniques. 
  
The invention of the $J$-$Q$ class of Hamiltonians \cite{Sandvik07,Lou09} has been a solution to the challenge of QMC simulations of the DQC transition in quantum
spin systems. In these models the Heisenberg interaction $J$ is supplemented by multi-spin correlated singlet projectors $Q$, which open possibilities of designing
sign-problem free Hamiltonians with the desired phase diagrams containing AFM and VBS phases. A large number of QMC studies with different variants of 2D $J$-$Q$
models have been reported \cite{Sandvik07,Melko08,Jiang08,Lou09,Sen10,Sandvik10a,Kaul11,Kaul12,Harada13,Chen13,Block13, Pujari15,Shao16,Suwa16,Qin17,Ma18,Zhao19,Ma19a}.
Concurrently, various discrete versions of the proposed DQC action have also been studied numerically \cite{Kragset06,Motrunich08,Kuklov08}, and three-dimensional
classical loop \cite{Nahum15a,Nahum15b} and dimer \cite{Sreejith14,Sreejith19} models with arguably the same low-energy fixed points have also been studied
extensively. Although the observed quantum phase transitions in many of these models appear to be continuous, there are puzzling scaling anomalies that have
been interpreted either as signs of an eventual weakly first-order transition (with discontinuities presumably developing on larger lattices than currently reachable)
\cite{Kuklov08,Wang17,Ma19,Nahum19}, or as manifestations of new physics associated with the DQC scenario---novel finite-size and finite-temperature scaling behaviors
related to the presence of two divergent length scales \cite{Shao16}.

While the ultimate continuous or first-order nature of the AFM--VBS transition is a question of fundamental interest,
it should be noted that the DQC scenario does not stand or fall with it---what matters is whether spinon deconfinement takes place on large length
scales, and this is now beyond doubt. The critical or near-critical state is at the very least a good approximation to a particular case of a gapless Dirac spin
liquid \cite{Suwa16,Ma18,Ma19a}, as expected at the DQC transition. Experimental realizations of spinon deconfinement \cite{Piazza15,Shao17} and DQC transitions
are plausible \cite{Zayed17,Guo19,Lee19}, though no conclusive observations have been reported as of yet.

The broader interest in the DQC point and related
phenomena, such as emergent symmetries and anomalous scaling, lies in the many unresolved puzzles in quantum matter where ``beyond-LGW'' physics may be at
play. For instance, some of the unusual metallic properties of the high-$T_c$ cuprates may arise from doping the DQC point \cite{Kaul07,Kaul08}. Being
accessible to large-scale QMC simulations, the DQC phenomenon and other exotic aspects of VBS states and transitions offer unique opportunities to
study important cases of beyond-LGW physics in detail. With a suitable sign-free ``designer Hamiltonian'' \cite{Kaul13}, definite numerical
results can be obtained and compared with approximate and often speculative predictions of field theories, and unique insights can be gleaned from
the simulations in their own right.

\subsection{$J$-$Q_6$ model and main findings}

Stimulated by the success of designer Hamiltonians suitable for QMC studies of the DQC phase transitions and VBS physics,
many other sign-free quantum spin models have recently been constructed for studies of a wide range of exotic quantum phases and quantum phase
transitions, e.g., symmetry-enhanced first-order transitions \cite{Zhao19}, $\mathbb{Z}_2$ spin liquids \cite{Block19}, and Haldane nematics \cite{Desai19}. In this
article we introduce a new type of $J$-$Q$ Hamiltonian that exhibits a host of fascinating phenomena within a single phase diagram. Our variant of the $J$-$Q_6$
model (where the subscript denotes the number of singlet projectors) contains 12-spin interactions (see Fig.~\ref{fig:JQ6}) and may appear contrived from the
standpoint of experimental realizations. However, in the spirit of designer Hamiltonians \cite{Kaul13}, it opens access to studies of quantum states and quantum phase
transitions which can likely be realizable also with other microscopic interactions (e.g., frustrated exchange interactions)
and whose quantum field-theory descriptions are of great current interest.

Being sign free, the $J$-$Q_6$ model illustrated in Fig.~\ref{fig:JQ6} (and define in detail in Sec.~\ref{sec:ModMethod}) is amenable to QMC simulations on
lattices with thousands of spins. We will construct the ground-state phase diagram as a function of the ratio $Q/J$ of the twelve-spin and Heisenberg couplings.
On increasing $Q/J$, we first find a direct quantum phase transition between the standard Heisenberg AFM ground state and a four-fold degenerate PVBS state
[Fig.~\ref{fig:VBS}(b)]. When further increasing the control parameter, the $\mathbb{Z}_4$ symmetry breaking PVBS state undergoes a $\mathbb{Z}_2$-breaking
continuous quantum phase transition into another state, the eight-fold degenerate AVBS state [Fig.~\ref{fig:VBS}(c)]. The initial aim of our study was to design a model
hosting an AFM--PVBS transition as a possible realization of a DQC transition. The $J$-$Q_6$ model was the first and so far only successful sign-free Hamiltonian
realizing a four-fold degenerate PVBS state, and in the course of studying it we also discovered the unexpected AVBS state.

Previous QMC studies of DQC transitions have focused on the CVBS state [\ref{fig:VBS}(a)], but the likewise four-fold degenerate PVBS is also a possible DQC
candidate on equal footing with the CVBS state according to theory \cite{Senthil04a,Sachdev08}. However, the AFM--PVBS transition found here is clearly
first-order though it exhibits the emergent SO(5) symmetry which was proposed in one variant of the DQC theory \cite{Senthil06}. Emergent symmetry would
not in general be expected at a first-order transition \cite{Nahum15b,Gazit18}, but recently other examples have been found \cite{Zhao19,Serna19,Yu19} where
the coexistence state appears to be described by a vector or pseudo-vector combining all the components of the two different order parameters and transforming
under a spherical symmetry. In the case at hand here the combined order parameter comprises three AFM components and two VBS components.

The emergent SO(5) symmetry may indicate the proximity of a continuous DQC transition with this symmetry. We will also discuss possible relevance of the first-order
transition to a recently proposed alternative fracton theory \cite{You19} of the PVBS state and AFM--PVBS transition. Furthermore, we demonstrate that the
two-stage breaking of the symmetries of the PVBS and AVBS phases fits into the framework of ``vestigial'' phase transitions \cite{Fernandes12,Fernandes19}, where our
case is the first example where both phases break discrete symmetries---the previous cases involved the breaking of a discrete symmetry followed by the
breaking of a continuous symmetry. We describe the double-discrete ground-state vestigial transition using a six-dimensional order parameter and also by a novel
graph-theoretic approach. The first-order nature of the transitions at finite (non-zero) temperature from a paramagnet into either the PVBS or AVBS state lends
further support to the vestigial phase scenario.

\subsection{Article outline}

The remainder of the article is structured as follows: In Sec.~\ref{sec:VBS} we provide further background on VBS states and numerical
studies of the AFM--VBS transition, setting the stage for our new developments related to PVBS ordering, emergent symmetries, and vestigial
phase transitions. In Sec.~\ref{sec:ModMethod} we define our $J$-$Q_6$ model in detail and briefly describe the QMC method we use to study
it. In Sec.~\ref{sec:DQCresult} we discuss results for the AFM--PVBS ground-state transition and contrast it with the often studied AFM--CVBS
transition. In Sec.~\ref{sec:Altresult} we present results for the second phase transition into the AVBS state. In Sec.~\ref{sec:OrderGraph} we
construct an order parameter that captures both the PVBS and AVBS phases, and also present our graph-theoretic approach (the ``order graph'')
for classifying the two-stage symmetry breaking. We present results for finite temperature in Sec.~\ref{sec:finT} and explain them based on the graph approach
and the scenario of vestigial phase transitions. Finally, in Sec.~\ref{sec:Discussion} we discuss implications of our results to existing theories, 
including the fracton scenario and the analogy of the AFM--PVBS transition to the cuprate SO(5) theory \cite{Zhang97}. 
In Appendix \ref{app:d4} we provide a detailed discussion on the subtle
issue of exactly what symmetries are broken in the CVBS and PVBS phases. In Appendix \ref{app:OrderGraph} we further discuss the 
motivations behind the concept of the order graph, and also describe its symmetry properties in more detail.

\section{Valence-bond solids and emergent symmetries}
\label{sec:VBS}

The square-lattice CVBS depicted in Fig.\ref{fig:VBS}(a) is the most well-studied non-magnetic state in the context of the DQC transition. In addition
to pointing to a continuous quantum phase transition between the CVBS and AFM states, numerical studies of $J$-$Q$ and related models have revealed that the
fluctuations among the four degenerate dimer patterns develop emergent U(1) symmetry as the critical point is approached \cite{Sandvik07,Jiang08,Lou09,Sandvik12,Nahum15b}.
This emergent symmetry confirms an important aspect of the DQC scenario, where the ordered CVBS patterns can be assigned angles $\phi = n\pi/2$, $n=0,1,2,3$,
and tunneling between these discrete angles corresponds to $n$ traversing a range of continuous values (alternatively, the order parameter is a complex scalar)
in an effective potential $\propto \cos(4\phi)$ \cite{Levin04}. The emergent U(1) symmetry corresponds to a flat distribution of the coarse-grained
CVBS angle $\phi$ at the critical point.

The proposal of emergent U(1) symmetry, which is also directly related to the conjectured absence of topological defects at the DQC transition
\cite{Senthil04a,Senthil04b}, was partially motivated by an analogy between the VBS order parameter and the magnetization vector of a classical
3D XY model with a four-fold symmetric ($q=4$) potential $\propto \cos(q\theta_i)$ for the microscopic spin angles $\theta_i$. In these clock models,
when $q\ge 4$ the cosine perturbation of the U(1) symmetric XY model is ``dangerously irrelevant'', i.e., it reduces the symmetry in the ordered state but
not at the critical point (in the thermodynamic limit when the order parameter is coarse-grained over large regions), and the universality class of the phase transition
remains to be that of the 3D XY model \cite{Jose77,Oshikawa00,Hove03,Lou07,Okubo15,Leonard15,Shao19}. In analogy, at the AFM--VBS transition the effective
low-energy interactions responsible for locking the dimerization to one of the four static patterns may be dangerously irrelevant, so that the
coarse-grained VBS order parameter can take any angle between $0$ and $2\pi$ when the critical point is approached from the VBS side.

If there indeed is emergent U(1) symmetry, then the field theory describing the AFM--VBS critical point does not need to include the ingredients causing
the four-fold degeneracy---quadrupled monopole instanton events in the path integral \cite{Read90,Murthy90}. In the originally proposed DQC field theory, the
CP$^{1}$ model \cite{Senthil04a,Senthil04b}, this aspect can be taken into account by not including any topological defects in the U(1) gauge field corresponding
to the continuously fluctuating VBS order parameter; hence the proposal that the transition is described by the non-compact CP$^{1}$ model. Here it should be
noted that the analogy with the clock model is not precise, because the critical points are different, and it has not been demonstrated within the field
theory that the quadrupled monopoles really are dangerously irrelevant.

Numerically, emergent U(1) symmetry has been established up to the largest length scales reachable in simulations of the $J$-$Q$ model
\cite{Sandvik07,Jiang08,Lou09,Sandvik12} and in related classical 3D loop models \cite{Nahum15a}. Moreover, there are also indications of an SO(5) symmetry
of the combined O(3) AFM and emergent U(1) VBS order parameters \cite{Nahum15a,Suwa16}. While the original DQC scenario did not involve any such symmetry,
only O(3)$\times$U(1), and allowed for different exponents $\eta_{\rm A}$ and $\eta_{\rm V}$ of the critical AFM and VBS correlation functions, a later
proposal was explicitly formulated with $\eta_{\rm A}=\eta_{\rm V}$. In this alternative (perhaps dual) theory the three components of the AFM order parameter
and the two VBS components are treated on equal footing as a five-component vector transforming as SO(5) \cite{Senthil06}. This treatment is possible only with
SU(2) spins, and the SU(N) generalizations of the CP$^{N-1}$ theory do not appear to allow such a higher symmetry. Numerically, $\eta_{\rm A} \approx \eta_{\rm V}$ has been
observed for SU(2) spins \cite{Lou09,Kaul12,Sandvik12} and in the loop model \cite{Nahum15a}, while generalizations of the $J$-$Q$ model and other related
models to SU(N) spins show clearly $\eta_{\rm A} \not= \eta_{\rm V}$, with the values of both exponents in remarkably good agreement with
$1/N$ expansions of the SU($N$) theory for large $N$ \cite{Lou09,Kaul12,Dyer15}.
Whether or not the observed SO(5) symmetry for SU(2) spins is exact or only approximate (i.e., breaking down on some large length scale) is still an open
question, especially in light of the fact that the numerical values of the exponents $\eta_{\rm A}$ and $\eta_{\rm V}$ of the lattice models do not satisfy
a bound obtained from conformal bootstrap calculations with SO(5) symmetry \cite{Nakayama16}.

%%%%%% DQC with PVBS %%%%%%
Although the DQC scenario has been probed in detail with models hosting CVBS ground states, the PVBS state depicted in Fig.~\ref{fig:VBS}(b) still
remains to be studied thoroughly, mainly due to the lack of microscopic models realizing it without QMC sign problem. Models in which a PVBS has been proposed
include the Heisenberg model with first- and third-neighbor AFM interactions \cite{Capriotti00}, which has a severe QMC sign problem that has prohibited
detailed studies of the putative AFM--PVBS transition. Resonating VBS states which spontaneously breaks the lattice symmetry similarly to the
square-lattice PVBS state have also been studied on the honeycomb lattice, with frustrated Heisenberg Hamiltonians with sign problems \cite{Li12,Ganesh13}
and $J$-$Q$ models without \cite{Banerjee11,Harada13} sign-problem. While it is apparently easier to construct sign-free $J$-$Q$ type Hamiltonians with PVBS ground
states on the honeycomb lattice, the symmetry broken in that case is $\mathbb{Z}_3$ instead of $\mathbb{Z}_4$ on the square lattice. The smaller number of
degenerate states has an additional complication in whether it allows for emergent U(1) symmetry or not \cite{Pujari15}. We will only
discuss the square lattice here.

The DQC scenario does not distinguish in any crucial manner between the CVBS and PVBS states, as they both are four-fold degenerate and
subject to the same mechanism of emergent U(1) symmetry--essentially the difference is in the sign of the effective cosine potential experienced
by the coarse-grained VBS order parameter. Contrary to the expectation of a DQC transition, there is a recent suggestion that spinons in
the PVBS states will be immobile fractons \cite{You19}, thus prohibiting the deconfinement mechanism underlying the emergent U(1) symmetry
and the continuous quantum phase transition into the AFM state. In the simulations of the $J$-$Q_6$ model presented here, we indeed find a first-order
AFM--PVBS transition. Nevertheless, we will argue that the absence of spinon deconfinement is not necessarily confirming the fracton picture, and that the
fracton mechanism is not universal for PVBS states and may only be realized under very particular conditions.

%%%% Our work : PVBS and alternating VBS %%%%%
Moving deeper into the PVBS phase, the $J$-$Q_6$ model exhibits another phase transition out of the PVBS phase into the eight-fold degenerate AVBS state
illustrated in Fig.~\ref{fig:VBS}(c). This previously not anticipated transition breaks an additional $\mathbb{Z}_2$ symmetry, and, thus the four-fold
degenerate PVBS phase can be regarded as an intermediate phase that only partially breaks the $D_4$ symmetry of the square lattice. This two-stage symmetry
breaking fits into the scheme of ``vestigial" transitions \cite{Fernandes12,Fernandes19}, although previous examples of
such multi-stage transitions have involved the breaking of a discrete (normally $\mathbb{Z}_2$) symmetry followed by the breaking of a continuous symmetry, in
contrast to both transitions breaking discrete symmetries in our case. We will introduce a general systematic way of describing such multi-stage discrete symmetry
breaking using a graph-theoretic approach.

In addition to serving as important testing grounds for the theory of quantum magnetism and quantum phase transitions, PVBS states are important also
considering potential realization of DQC transitions in real materials. So far, the most promising candidate is the quasi-2D quantum magnet
SrCu$_2$(BO$_3$)$_2$ under high pressure \cite{Zayed17,Guo19}, which appears to realize a certain type of PVBS state (though there are also
views opposing this notion \cite{Boos19}). The 2D magnetic interactions in SrCu$_2$(BO$_3$)$_2$ realize the Shastry-Sutherland model \cite{Shastry81},
and the ratio of the inter- and intra-dimer coupling constants change with pressure in such a way that the three phases of the model are
realized within the accessible pressure range; first a non-degenerate dimer singlet state, then a PVBS followed by an AFM state, with the latter further
stabilized by inter-layer couplings \cite{Guo19}. However, the Shastry-Sutherland PVBS state is only two-fold degenerate owing to the structure of the
lattice, while the original DQC scenario applies to a four-fold degenerate PVBS such as the one illustrated in  Fig.~\ref{fig:VBS}(b).
Nevertheless, there are theoretical proposals suggesting that also a two-fold degenerate VBS state may give rise to a DQC point or DQC-like physics
\cite{Lee19,Metlitski18}. 

Beyond the sign-problematic (in the most interesting parameter regimes) Shastry-Sutherland model and extensions of it \cite{Boos19a},
sign-free 2D $J$-$Q$ models with two-fold degenerate PVBS ground states can also be designed. In a ``checker-board'' $J$-$Q$ (CBJQ) model, an unusual
first-order transition between the AFM state and a $\mathbb{Z}_2$ breaking PVBS ground state was found \cite{Zhao19}. Surprisingly, an emergent O(4) symmetry of
the combined O(3) AFM and scalar PVBS order parameters in the coexistence state of the system was observed, despite the clear presence of
discontinuities at the transition.
While until recently emergent symmetries had only been expected at continuous transitions \cite{Nahum15b,Gazit18}, subsequently a first-order transition
with enhanced symmetry was also found in a $\mathbb{Z}_2$ deformed classical 3D loop model \cite{Serna19}. It is currently not clear whether the enhanced
symmetry is only manifested up to some large length scale or truly asymptotically exact, but a theory in a different context of boson-fermion supersymmetry
does suggest that first-order transitions can be associated with exact emergent symmetry \cite{Yu19}. Here we will demonstrate SO(5) symmetry at the
first-order AFM--PVBS transition of the $J$-$Q_6$ model, thus providing yet another example of the emergence of enhanced symmetries under unexpected conditions.

\section{Model and method}
\label{sec:ModMethod}
  
%%%% The JQ6 Model %%%%
The $J$-$Q_6$ model we consider here is defined with $S=1/2$ spins on a 2D square lattice. As illustrated in Fig.~\ref{fig:JQ6}, the model combines
the standard Heisenberg exchange of strength $J$ with a 12-body interaction of strength $Q$ in the Hamiltonian
\begin{equation}\label{eq:Ham}
H=-J\sum_{\langle i,j\rangle} P(i,j) 
-Q\sum_{\mathbb{H}} ~\prod_{k=1}^6 P(i_k,j_k),
\end{equation}
were, $P(i,j)$ denotes a bond operator; a singlet projector on the pair of sites $i$ and $j$, 
\begin{equation}\label{eq:pij}
P(i,j) = 1/4 - \mathbf{S}_i \cdot \mathbf{S}_j. 
\end{equation}
Thus, a single $J$-term $-J P(i,j)$ is equivalent to an AFM Heisenberg coupling, up to an additive constant. The sum over $\langle i,j\rangle$ runs
over all nearest-neighbor bonds. The $Q_6$ terms act on twelve spins specified by the set $\mathbb H$ of groups of twelve sites
$i_1,\ldots,i_6,j_1,\ldots,j_{6}$, with their relative positions and pairings into bonds with singlet projectors illustrated in Fig.~\ref{fig:JQ6}. In order to
preserve all symmetries of the square lattice, $\mathbb{H}$ includes all possible translations of the four bond arrangements shown. 
Note that two of the four patterns correspond to a $\pi/2$ rotation of the other two, preserving the four-fold rotational symmetry of the lattice. 
We use $L\times L (=N)$ square lattices with periodic boundary condition, yielding $2N$ $J$-terms and $4N$ $Q$-terms. In
some cases we will also consider open boundary conditions, in which case we remove all operators acting on sites beyond the $L \times L$ edge.

Previously studied $J$-$Q$ models have $Q$-terms with two or three bond operators and were occasionally referred to as $J$-$Q_2$ or $J$-$Q_3$ models, respectively
\cite{Lou09}. Different $J$-$Q$ models are further distinguished by the relative arrangement of the bond operators of the $Q$ terms, with columnar and
stair-case arrangement considered in the past---the former leading to DQC transitions, as discussed in the previous section, and the latter type of
$Q_3$ interaction causing a strongly first-order transition between the AFM state and a staggered VBS \cite{Sen10} (and a stair-case $Q_2$ interaction
does not destroy the AFM order). The combination of vertical and horizontal singlet projectors in the new $Q_6$ terms introduced here is the feature that promotes
the formation of a PVBS instead of the CVBS obtaining if all projectors are stacked in columns. For comparison, we will also present some results for
the case of a columnar $Q_6$ interaction, in which case we find a strongly first-order transition into a CVBS phase. We also note that we have
not been able to realize a PVBS state with less than six bond operators.

For all calculations reported here we used the stochastic series expansion (SSE) method \cite{Sandvik10b}, a QMC method without discretization errors that
incorporates efficient loop updates \cite{Sandvik99,Evertz03}. In order to search the entire sign-free ground state phase diagram (positive $J$ and $Q$),
we vary $Q \in [0, 1]$ in Eq.~(\ref{eq:Ham}) while keeping $J+Q=1$ fixed to serve as the unit of energy in the simulations. Unless otherwise noted, the
inverse temperature is set to $\beta =L$, which is sufficient for studying ground state ordering and scaling properties of quantum phase transitions
with dynamic critical exponent $z = 1$, which is the expected $z$ at both the DQC point and a quantum Ising critical point (which corresponds to the
symmetry-breaking at the PVBS--AVBS transition). The choice of $\beta(L)$ is also suitable for detecting a first-order phase transition.

\section{Direct AFM-to-PVBS transition}
\label{sec:DQCresult}

\begin{figure*}[t]
\includegraphics[width=18.17cm]{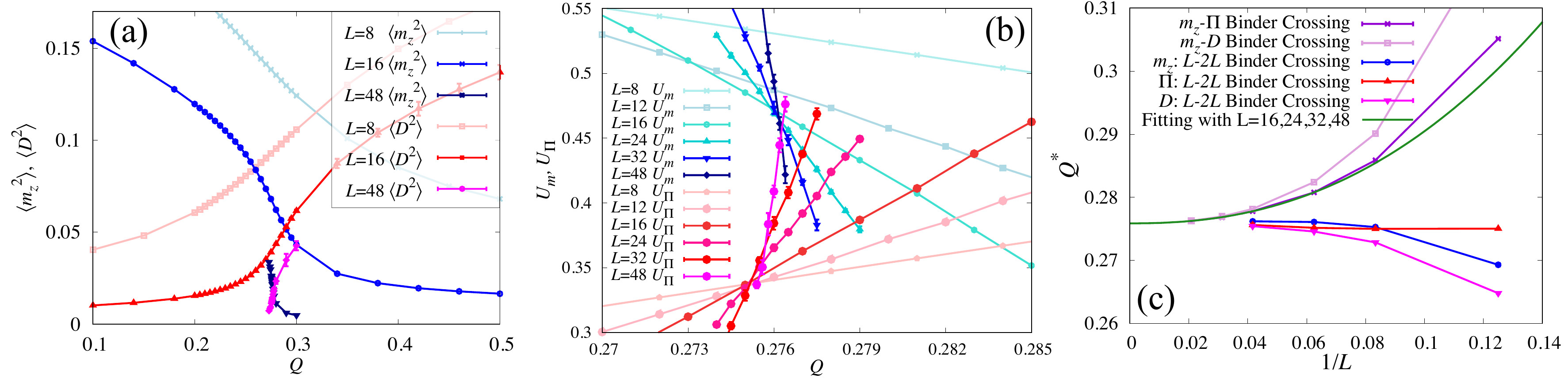}
\caption{\label{fig:HowToDQC} 
  (a) Squared order parameters in the neighborhood of the AFM--PVBS transition point. Results for three different system sizes, $L=8,16$, and $48$, are
  shown in each case (note that, for $L=48$ we have results only very close to the transition point). (b) Binder cumulants of the order parameters
  for the two phases, shown for system sizes $L=8,12,16,24,32,48$. (c) Finite-size transition points extracted using crossing points defined with several
  different quantities. The green curve is a fit to the estimates obtained from the same-size crossings between the AFM and PVBS cumulants and gives
  $Q_c=0.2758(5)$ for the infinite-size transition point, where the number in parenthesis here and elsewhere indicates the statistical error (one standard
  deviation of the mean) of the preceding digit. The other lines connecting points are only guides to the eye. When not visible, error bars are
  smaller than the graph symbols.}
\end{figure*}

The $J$-$Q_6$ model exhibits a direct quantum phase transition from the standard AFM phase to the PVBS phase at a point $Q_c \approx 0.2758$,
making this transition a candidate for 
the DQC mechanism. However, we here instead find a first-order transition with clearly finite coexisting order parameters at the transition point.
This is in sharp contrast to previously studied variants of the $J$-$Q$ model, which host apparently continuous AFM--CVBS transitions, or possibly
very weak first-order transitions with coexisting order parameters too small to be detected currently. Here we will first present our numerical
simulation results in detail and leave discussion of implication of our result to the DQC and fracton theories mostly to Sec.~\ref{sec:Discussion}.

\subsection{Order parameter definitions}

Let us first define the two components $\Pi_0$ and $\Pi_1$ of the PVBS order parameter $\Pi = (\Pi_0, \Pi_1)$ as 
\begin{eqnarray}
\Pi_{a} =\frac{2}{N}\sum_{x+y \equiv a } (-1)^{x}P^{z}_{x,y},
\label{Piadefs}
\end{eqnarray}
where $a=0,1$ is a sublattice label corresponding to a checker-board pattern of the lattice coordinates $(x,y)$, i.e., $a \equiv x+y ~{\rm mod}~ 2$.
$P^z_{x,y}$ is a projection operator to zero $z$-magnetization of the four spins on the plaquette with its low-left corner at $(x,y)$, i.e.,
\begin{equation}
  P^z_{x,y}= \left \{
  \begin{array}{ll}
    1, & {\rm if~} S^z_{x,y} + S^z_{x+1,y} + S^z_{x,y+1} + S^z_{x+1,y+1} = 0 \\
    0, & {\rm otherwise}.
  \end{array}\right.
\end{equation}
Eq.~(\ref{Piadefs})
is just one out of many possible order parameters capable of detecting the PVBS state. Ideally, one might prefer a spin-rotation invariant definition,
e.g., some direct measure of the singlet density. Evaluating such off-diagonal operations involving more than two spins is very time consuming, however
\cite{Beach07}. Though the condition for $P^z_{x,y}=1$ is a necessary but not sufficient condition for the four sites to form a singlet, the order parameter
$\Pi$ still detects a modulation of the mean singlet density.

It is useful to compare the plaquette order parameter with the frequently used dimer order parameter $D=(D_x,D_y)$, where, to conform with the plaquette order
parameter in Eq.~(\ref{Piadefs}), we also use a diagonal definition (instead of the rotationally invariant definition that can also be evaluated
efficiently \cite{Beach07}):
\begin{eqnarray}
D_x & = & \sum(-1)^{x} {S}^z_{x,y} {S}^z_{x+1,y}, \nonumber \\
D_y & = & \sum(-1)^{y} {S}^z_{x,y} {S}^z_{x,y+1}.
\label{dxdydefs}
\end{eqnarray}
For a CVBS or PVBS ordered state, the plaquette order parameter $(\Pi_0,\Pi_1)$ essentially behaves as a $\pi/2$ rotated columnar order parameter $(D_x,D_y)$,
and both squared order parameters are non-vanishing. This is a consequence of the fact that the CVBS and PVBS states both break the $D_4$ lattice
symmetry into $\mathbb{Z}_2$. To be more precise, when we regard symmetry transformations of the model that do not change the order parameter, they will naturally
form the eight-component dihedral group $D_4$, which breaks into $\mathbb{Z}_2$ in both the CVBS phase and the PVBS phase. The two remaining $\mathbb{Z}_2$
symmetries are isomorphic via an automorphism of the $D_4$ group. This point is explained in detail in Appendix \ref{app:d4}. For the discussion in this
section, it suffices to recognize that $\Pi$ and $D$ are both valid order parameters for detecting CVBS and PVBS order, and these orders can be distinguished by
examining the 2D distribution of either of the order parameters; here we will analyze both of them.

\subsection{First-order transition}

Examples of SSE results used in order to locate the AFM--PVBS transition point are shown in Fig.~\ref{fig:HowToDQC}.
Fig.~\ref{fig:HowToDQC}(a) shows the squared order parameters for three system sizes versus $Q$. It is apparent that the point where the AFM order dies out is
also where the PVBS order emerges, implying a direct transition. To precisely analyze the transition, we show in Fig.~\ref{fig:HowToDQC}(b) the Binder cumulants
$U_m$ and $U_\Pi$ of the order parameters, defined such that $U_X \to 1$  with increasing system size if there is long-range order of the $X$ kind
($X=m$ or $X=\Pi$) and  $U_X \to 0$ otherwise. The Binder cumulant of the AFM order parameter is given by
\begin{equation}
U_{m}=\frac{5}{2}\left (1-\frac{1}{3}\frac{\langle m_z^{4}\rangle}{\langle m_z^{2} \rangle^2} \right ),
\end{equation}
where the factors are chosen for a single component of the three-component AFM order parameter,
\begin{equation}
m_z = \frac{1}{N}\sum_{(x, y)} (-1)^{x+y} S^z_{x, y}.
\label{eq:mzdef}
\end{equation}
For the two-component plaquette order we have
\begin{equation}
 U_{\Pi}=2-\frac{\langle \Pi^4 \rangle}{\langle \Pi^2\rangle ^2},
\end{equation}
where $\Pi=(\Pi_0,\Pi_1)$, with the components defined in Eq.~(\ref{Piadefs}).
The crossing value $Q=Q^*$ between $U_X(Q,L)$ and $U_X(Q,2L)$ as the control parameter $Q$ is varied (for any order parameter $X$), is known to have much smaller finite-size
drifts compared to other quantities defining finite-size transition points. $L$-$2L$ crossing  values $Q^*(L)$
obtained from data such as those in Fig.~\ref{fig:HowToDQC}(b)
are graphed versus the inverse system size in Fig.~\ref{fig:HowToDQC}(c). The estimated transition point $Q_c=Q^*(L\to \infty)$
in the thermodynamic limit agrees well between different order
parameters. Another quantity, which serves even better as the $Q_c$ estimator for this system, is the crossing between the two different Binder cumulants computed
with the same lattice size, thus providing a larger number of $L$ points with the available data sets. We also show this estimate in Fig.~\ref{fig:HowToDQC}(c).
It is well described by a single power law correction for $L \ge 16$, and we use it for the most precise extrapolation of the transition point in the thermodynamic
limit; $Q_c=0.2758(5)$. 

\begin{figure}[t]
\includegraphics[width=8.5cm]{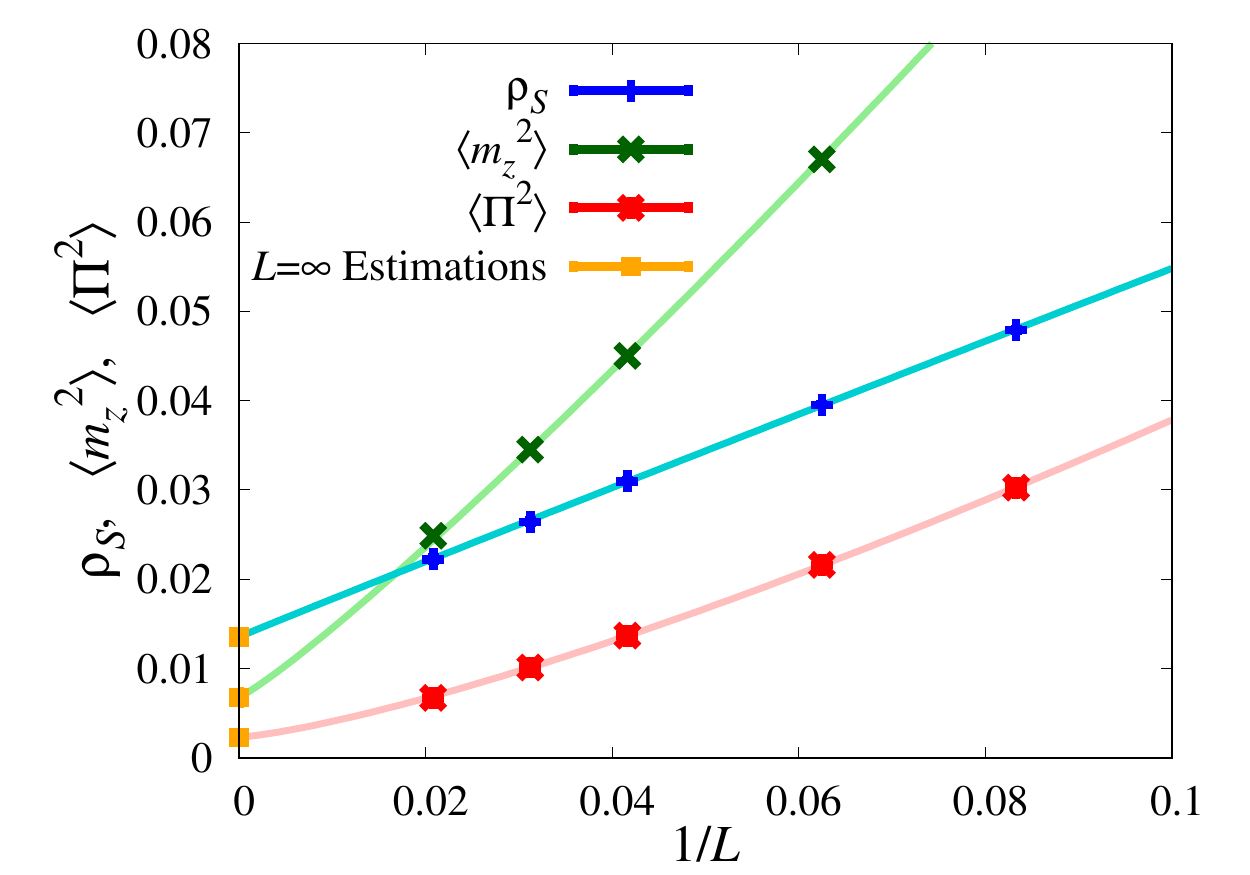}
\caption{\label{fig:OrderParametersAtDQC}
Order parameters characterizing the AFM and PVBS states graphed vs the inverse system size at the estimated transition point,
$Q_c=0.27581$. The curves show polynomial fits based on which infinite-size extrapolated values are obtained (the yellow squares shown at $1/L=0$).
When not visible, error bars are smaller than the graph symbols.}
\end{figure}

With the estimated transition point at hand, the nature of the AFM--PVBS transition can be further studied. We find that both order parameters approach finite
values with increasing system size at the estimated $Q_c$ value, as shown in Fig.~\ref{fig:OrderParametersAtDQC}. Here we also show another quantity
characterizing the AFM state; the spin stiffness $\rho_S$. In SSE calculations it is obtained using the simple formula \cite{Sandvik10b}
\begin{equation}
\rho_S = \frac{1}{\beta N}\left \langle (n_x^+ - n_x^-)^2 \right \rangle, 
\end{equation}
where $n_x^\pm $ denotes the number of $S^\pm_{x,y}S^\mp_{x+1,y}$ operators in the SSE operator string. The stiffness also clearly extrapolates to a finite value.

Finite values of both the AFM and PVBS order parameters at a single point implies phase coexistence and a first-order transition. However, even though the
extrapolated order parameters are substantial, the Binder cumulants of these quantities [Fig.~\ref{fig:HowToDQC}(b)] did not show the negative peaks that are
typically present at first-order transitions. A negative peak in a cumulant originates from the double-peaked distribution of the squared order parameter; one
peak at $0$ reflecting the disordered phase and a second peak at a non-zero value reflecting the ordered phase. In the conventional classical case \cite{Binder81,Vollmayr93},
the volume increase in free-energy barriers between the two degenerate phases implies that the negative peak grows linearly with the system volume, and in the
quantum case a corresponding behavior is expected on account of the increasing tunneling barrier with the system size (and this has been observed at the
phase transition in a staggered $J$-$Q_3$ model \cite{Sen10}). The tunneling barrier is absent if the order parameters form an enlarged spherical symmetry at
the transition point, so that moving from one phase to the other corresponds to rotating the order parameter without energy cost. Such an unexpected mechanism
at play at a first-order quantum phase transition was recently proposed to explain results for the transition between the AFM state and a two-fold degenerate
PVBS in the CBJQ model \cite{Zhao19}, and subsequently other potential cases were also identified \cite{Serna19,Yu19}.

\subsection{Emergent symmetry}

\begin{figure}[t]
\includegraphics[width=7cm]{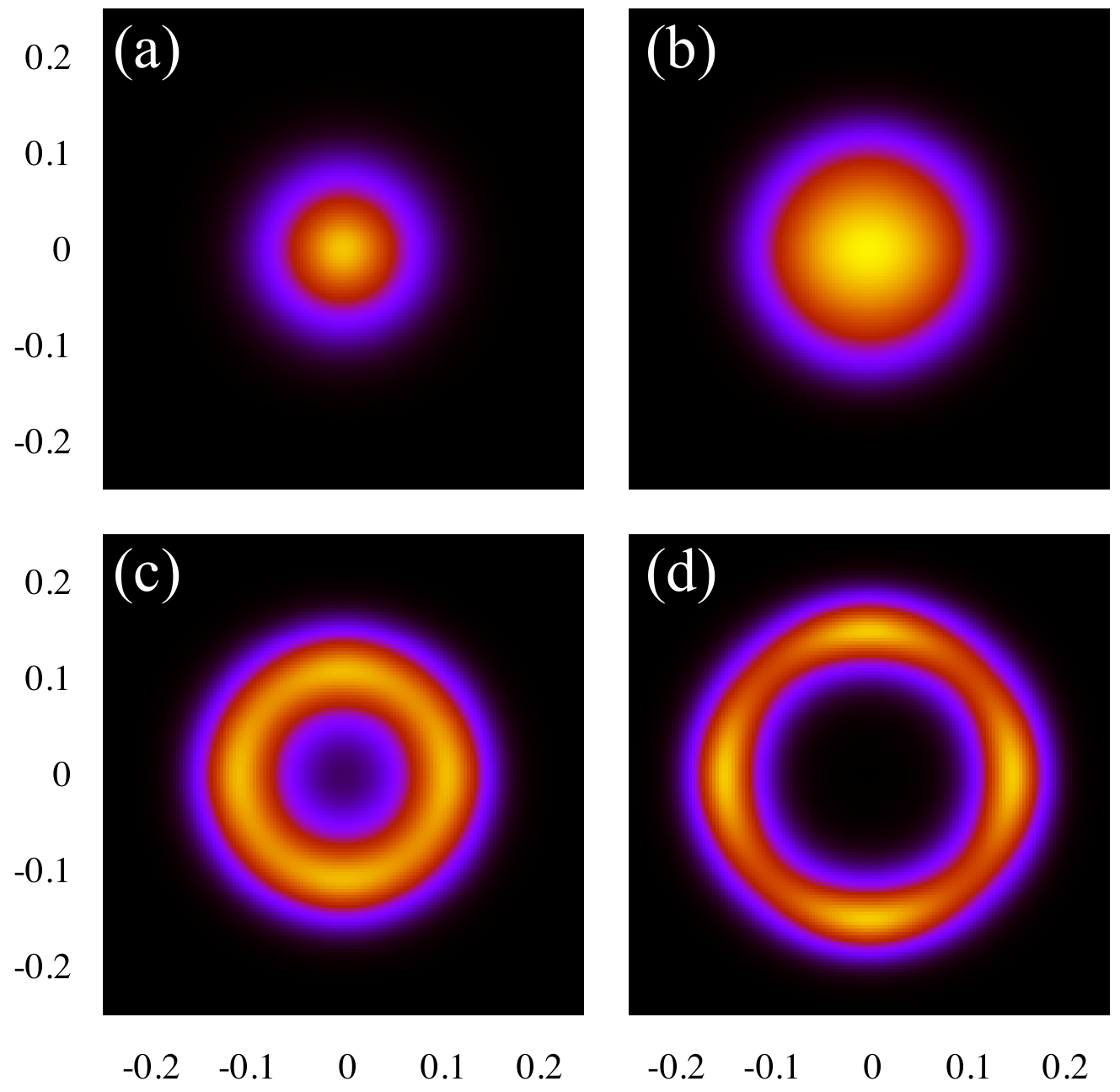}
\caption{\label{fig:DQCHistogram}
PVBS order-parameter distribution $P(\Pi_0,\Pi_1)$, with the components defined according to Eq.~(\ref{Piadefs}), accumulated in SSE
simulations of $L=48$ lattices. (a) At $Q=0.273$, close to the AFM--PVBS transition, slightly inside the AFM phase. (b) At the AFM--PVBS transition point,
$Q_c=0.2758$, (c) At $Q=0.28$, slightly inside the PVBS phase. (d) Further inside the PVBS phase at $Q=0.3$.}
\end{figure}

Anticipating an emergent symmetry also in the present case, we proceed to examine the symmetry properties of the order parameters at the AFM--VBS transition.
Fig.~\ref{fig:DQCHistogram} shows the probability distribution of the PVBS order parameter $P(\Pi_0,\Pi_1)$ near the transition point and deeper inside
the PVBS phase. Since we here focus on qualitative aspects of the angular dependence of the distributions,
the (linear) color scales in the histograms are unimportant and not shown
for simplicity. The fact that the histogram has four peaks on the horizontal and vertical axes once the system is sufficiently far inside the VBS phase,
seen clearly in Fig.~\ref{fig:DQCHistogram}(d), implies that this phase is indeed a PVBS and not a CVBS. Furthermore, an emergent U(1) symmetry in the histogram
at the critical point is clearly visible in Fig.~\ref{fig:DQCHistogram}(b). In Fig.~\ref{fig:DQCHistogram}(c) the distribution takes an approximate ring shape,
with maximum weight away from the center, indicating a large magnitude of the order parameter. The distribution is still nearly U(1) symmetric.

\begin{figure}[t]
\includegraphics[width=7cm]{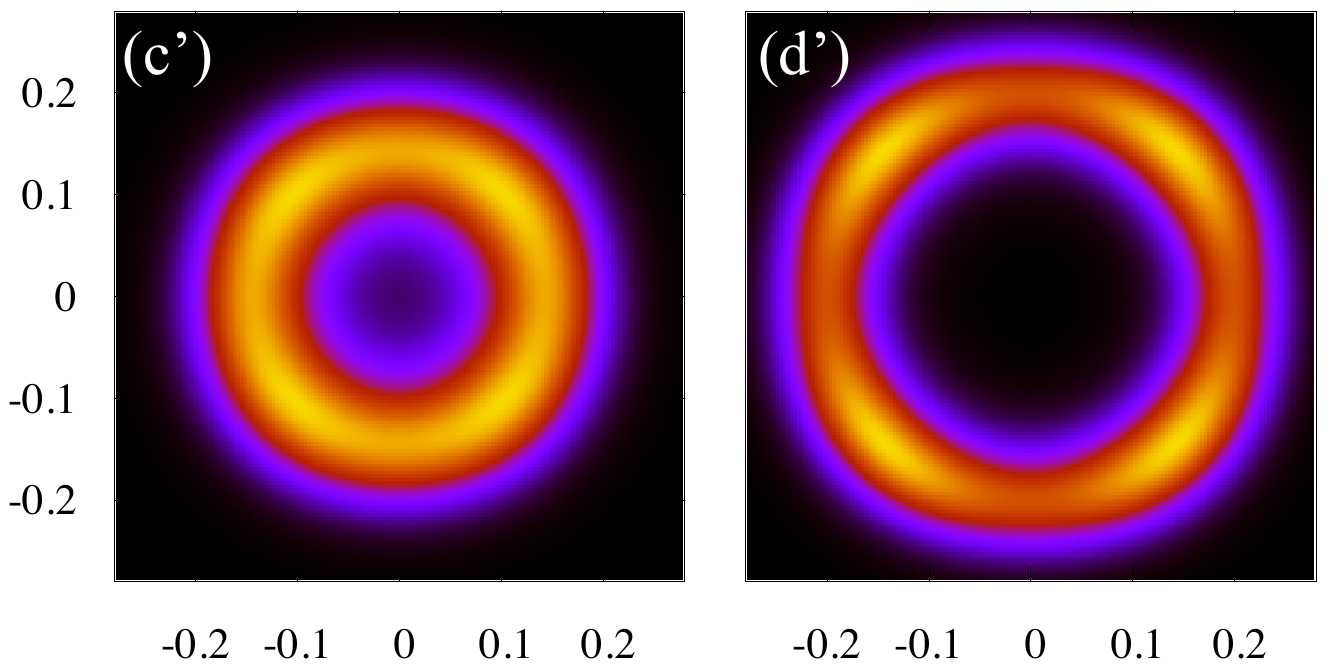}
\caption{\label{fig:DQCHistogram_DxDy} 
  Distribution of the dimer order parameter $P(D_x,D_y)$, with the components defined according to Eq.~(\ref{dxdydefs}), accumulated in the same simulations as
  the plaquette order distributions in Figs.~\ref{fig:DQCHistogram}(c,d); (c') and (d') are for $Q=0.28$ and $Q=0.3$, respectively.} 
\end{figure}

Figure \ref{fig:DQCHistogram_DxDy} shows the probability distribution of the dimer order parameter $P(D_x,D_y)$. In the VBS phase the four peaks of $D$ are 
located at the diagonal angles, also indicating PVBS order (which implies coexisting $D_x$ and $D_y$ dimer order). The near-U(1) symmetry of the distribution
$P(D_x,D_y)$ is similar to that seen in $P(\Pi_0,\Pi_1)$.

Emergent U(1) symmetry is a characteristic feature of the DQC phenomenon \cite{Levin04,Sandvik07}, and there is a divergent (on approach to the critical point)
length-scale associated with the break-down of the symmetry inside the VBS phase. This length scale is reflected in an increasing sharpness of the four peaks of
the order-parameter distribution when the system size is increased inside the VBS phase, which can be used to extract the exponent governing the relevant
length scale \cite{Lou09} (in a way which has recently been further refined in the context of classical clock models \cite{Shao19}). However, as we showed in
Fig.~\ref{fig:OrderParametersAtDQC}, in the present case the transition is clearly first-order, and the associated $L$ dependence of the histogram features
inside the PVBS phase should then be dictated by an exponent analogous to the dimensionality of a classical system \cite{Zhao19}. We have not studied the $L$
dependence inside the PVBS systematically in the present case, but qualitatively the four peaks become sharper as $L$ increases (approaching $\delta$-functions
as $L \to \infty$). We will next show that the emergent U(1) symmetry at the transition point actually reflects an even higher SO(5) symmetry, thus establishing
the AFM--PVBS transition in the $J$-$Q_6$ model as another example of a symmetry-enhanced first-order transition. The spherical symmetry allows the AFM and VBS
order parameters to continuously rotate into each other at fixed energy, which explains the absence of negative peaks in the Binder cumulants.

In the case of the CBJQ model, the O(3) symmetric AFM order and scalar (two-fold degenerate) PVBS order combine into an O(4) vector or SO(4) pseudo-vector.
The methods used could not distinguish between the two possible cases \cite{Zhao19} as physical reflections are not accessible in the simulations (while the
meandering of the order parameter vector on the sphere is easily accessible). In the $J$-$Q_6$ model, we similarly expect O(5) or SO(5) symmetry of the
components $(\Pi_0,\Pi_1,m_x,m_y,m_z)$. This kind of emergent SO(5) symmetry was previously proposed within the the DQC scenario \cite{Senthil06,Nahum15b}.

\begin{figure}[t]
\includegraphics[width=7cm]{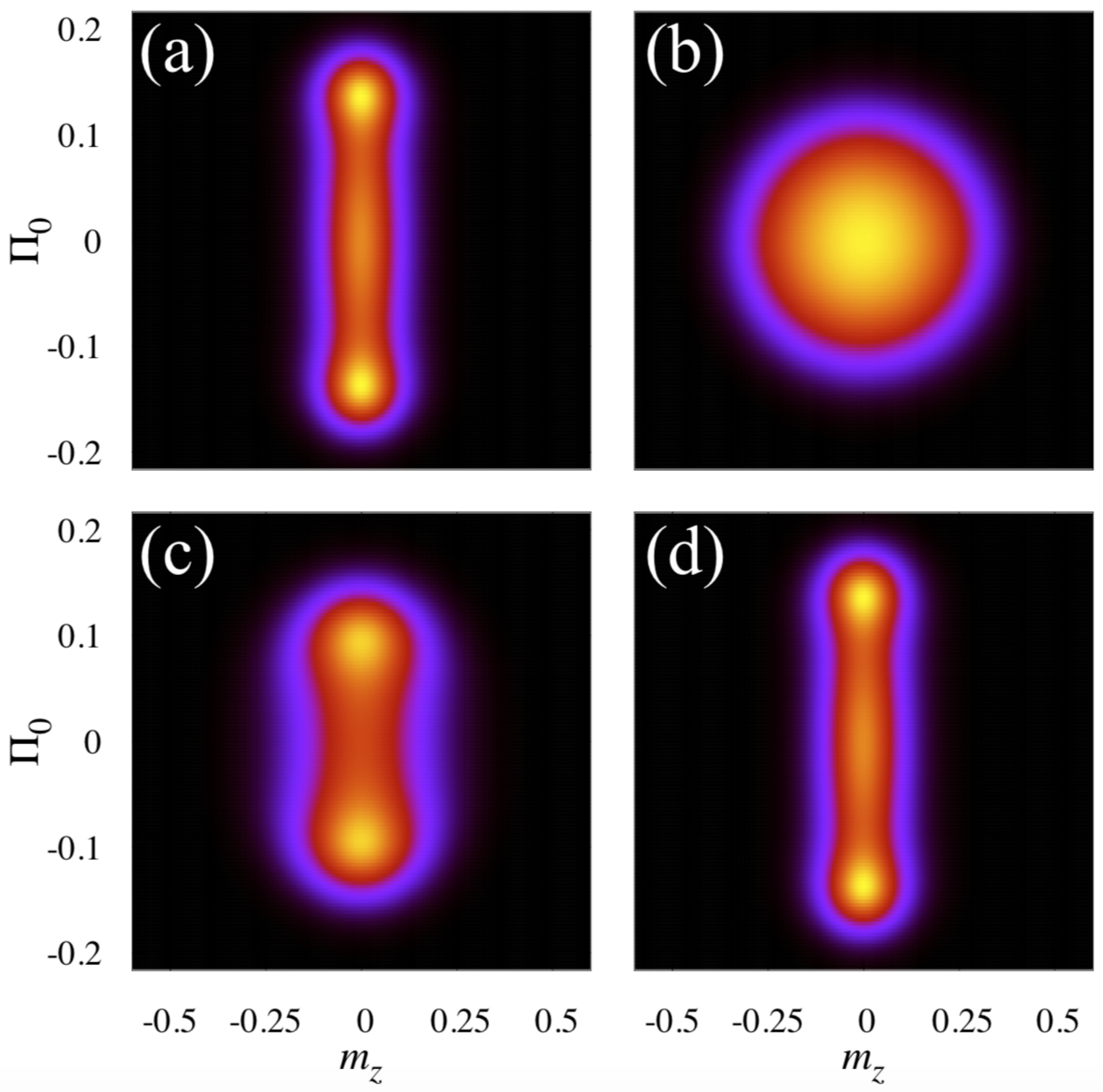}
\caption{\label{fig:ESO5}
Joint probability distribution $P(m_z,\Pi_0)$ of the AFM and VBS order parameters for system size $L=48$. The $Q$ values are the same
as in Fig.~\ref{fig:DQCHistogram}; (a) slightly inside the AFM phase at $Q=0.273$, (b) at the transition point $Q_c=0.2758$, (c) slightly inside
the PVBS phase at $Q=0.28$, and (d) well inside the PVBS phase at $Q=0.3$.} 
\end{figure}

To demonstrate
the unexpected enlarged symmetry at the first-order transition, we show the joint probability distribution of $m_z$ and $\Pi_0$ at and near the AFM--PVBS
transition point in Fig.~\ref{fig:ESO5}. Inside the AFM phase, Fig.~\ref{fig:ESO5}(a), long-range order with O(3) symmetry implies a line segment when the
five-dimensional distribution is projected down to the plane $(m_z,\Pi_0)$. Since we are near the transition, where the finite-size fluctuations of the amplitude
of the order parameter are significant, the line is broadened. In the VBS phase, Figs.~\ref{fig:ESO5}(c,d), the fluctuation of $\Pi_0$ among two non-zero
values (one positive and one negative) results in two blobs on the vertical axis. The central probability maximum visible in Fig.~\ref{fig:ESO5}(d)
corresponds to $\Pi_0$ being small when $|\Pi_1|$ is large, and the significant probability density remaining between the maximas show that tunneling between the
four PVBS patterns is still rather prominent at this $Q$ value for this system size. 
At the transition point, Fig.~\ref{fig:ESO5}(b), we observe an U(1) symmetric histogram between $m_z$ and $\Pi_0$ after simple rescaling. 
Along with the exact SO(3) symmetry of the AFM order parameter and the numerically observed O(2) symmetry of the PVBS order parameter,
this implies O(5) or SO(5) symmetry of the combined order parameter.

Note that, in constructing Fig.~\ref{fig:ESO5}, we had to account for the fact that the definitions of the two different order parameters are associated with
essentially arbitrary factors. The spherical symmetry at the transition point is apparent only if one of the order parameters is suitably rescaled by the
ratio of the standard deviations of the two order parameters. Away from the transition point, no such rescaling results in a rotationally symmetric histogram. The
features away from the transition point are most clearly visible if the scale factor is held fixed at its value calculated at the transition point, which is
what we have done in Fig.~\ref{fig:ESO5}.

We have carried out numerous tests of the angular uniformity of the histograms near the transition point,
including those discussed in Ref.~\cite{Zhao19}, for different system sizes.
Here we present a different method, analyzing conditional probabilities to test the expected form of the radial distribution and to quantify some aspects
of the angular distribution. We use the accumulated data points $(m_z,\Pi_0)$ for which $m_z=0$ to collect the conditional probability $P(\Pi_0|m_z=0)$, and
similarly accumulate $P(m_z | \Pi_0=0)$ and $P(\sqrt{\Pi_0^2+m_z^2}| m_z=\Pi_0)$. Here we point out that the computed equal-time values of $m_z$ and $\Pi_0$
in a given SSE sampled configuration are integer-based, and the above conditional probabilities are unambiguously defined. Related to this issue, note that the
$m_z=\Pi_0$ case does not correspond exactly to a diagonal cut in histograms such as Fig.~\ref{fig:ESO5}(b), because of the different factors needed to scale
from an elliptical to a circular-symmetric histogram. The required scale factor is not too far from unity, and $m_z=\Pi_0$ without rescaling (which we
consider for this analysis) corresponds to a radial line cut at an angle of about $\approx 70^{\circ}$ in Fig.~\ref{fig:ESO5}(b).

The first-order transition with higher symmetry corresponds to the order parameter vector living on the surface on an O(5) sphere, in contrast to the
case of a spherical symmetry at a critical point, where the radial distribution reflects critical fluctuations and are centrally peaked. The finite radius
of the sphere reflects the non-zero magnitude of the coexisting order parameters in the thermodynamic limit. 
The radius of the sphere observed in simulations (defined using the sum of squared order parameters, including the scale factor discussed above) will not be
constant because of finite-size effects, but will exhibit relative fluctuations that vanish when $L\to\infty$. 

\begin{figure}[t]
\includegraphics[width=8.8cm]{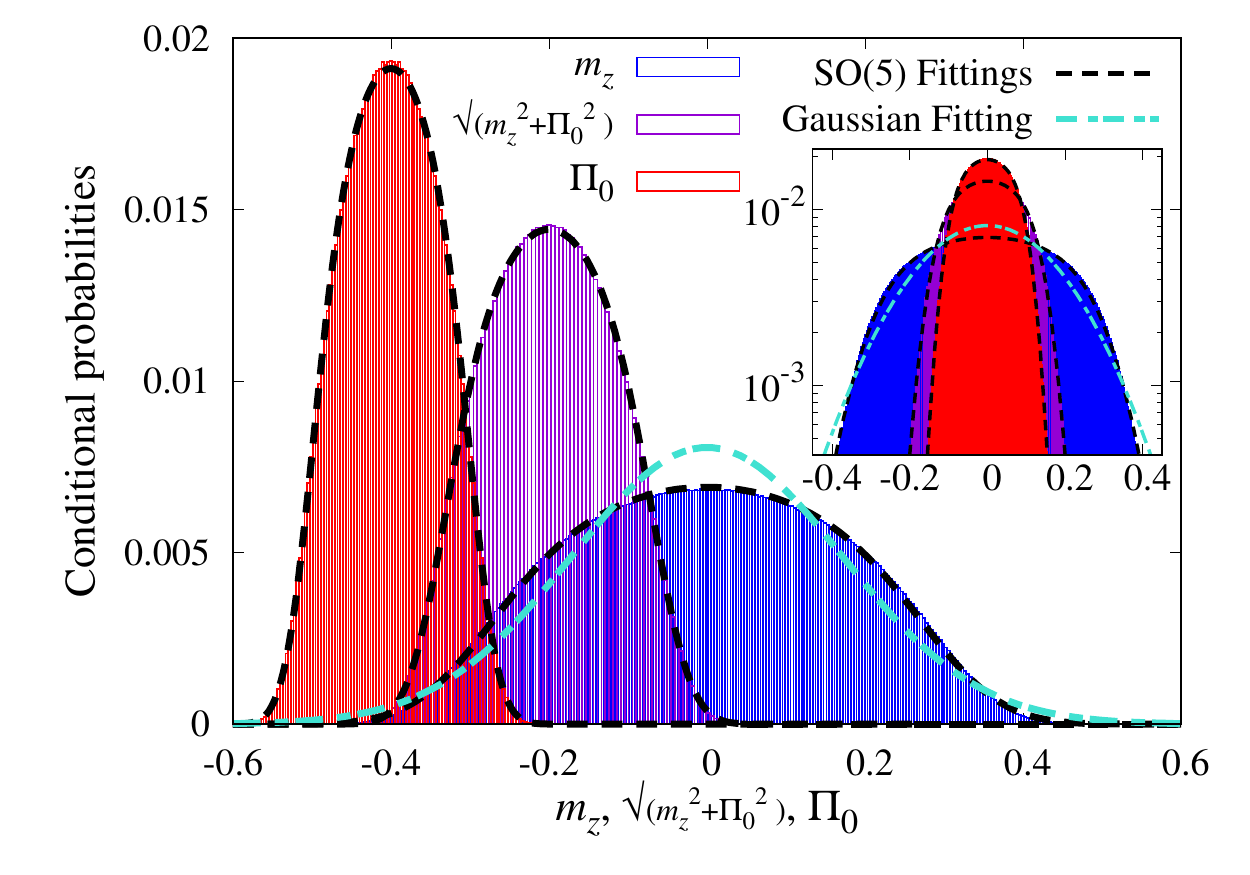}
\caption{\label{fig:SO5cuts}
  Three conditional probability distributions of the PVBS and AFM order parameters; $P(\Pi_0 | m_z=0)$ (blue),  $P(\sqrt{\Pi_0^2+m_z^2}| m_z=\Pi_0)$ (purple),
  and $P(m_z | \Pi_0=0)$ (red). The latter two histograms are shifted to the left for clarity. Fitting functions in black dotted lines are computed under the
  assumption of a uniform distribution over a 5-dimensional spherical surface with Gaussian fluctuation of its radius. By fixing the second moment of the resulting
  function to match the SSE data, we conducted a one-parameter fitting with the relative width of the Gaussian to match $P(\sqrt{\Pi_0^2+m_z^2}| m_z=\Pi_0)$.
  The resulting distribution was rescaled by the standard deviations of the other two histograms $P(\Pi_0 | m_z=0)$ and $P(m_z | \Pi_0=0)$, without any other
  adjustments. A simple Gaussian distribution with the same second moment as $P(m_z | \Pi_0=0)$ is drawn as the light blue dashed curve. The inset
  shows the same distributions on a log scale.} 
\end{figure}

Without fluctuations,
the conditional probabilities defined above would be semi-circles if there is SO(5) symmetry. To take into account the radial fluctuations, we average the
semi-circular distributions over Gaussian distributions of their radia. 
In practice, this is a one-parameter fit, because we can fix the resulting standard deviation of the distribution to that of the SSE computed histogram. 
We carry out this matching procedure
only for one out of the three computed distributions, and rescale to the other two by a factor fixed by the standard deviations of those distributions. As shown in
Fig.~\ref{fig:SO5cuts}, the distributions obtained in this way for our largest system size, $L=48$, agrees essentially perfectly with all the SSE computed
conditional probabilities. As a contrast, we also show a Gaussian distribution, which has significantly fatter tails than the broadened semi-circle.

The conclusion drawn based on the results presented in this section is that the AFM--PVBS transition in the $J$-$Q_6$ model is first-order and associated
with emergent O(5) or SO(5) symmetry, thus apparently similar to the first-order transition with emergent O(4) or SO(4) symmetry observed in the CBJQ
model \cite{Zhao19}. Numerically, if no deviations from the symmetric distribution (or if the measures of symmetry continue to improve with increasing
system size, as is expected for an emergent symmetry \cite{Zhao19}) it is not possible to prove positively that the symmetry survives asymptotically as the
system size increases, only that any perturbation that may eventually break the symmetry is weak and does not affect the system up to the largest sizes
studied so far.

\section{The alternating VBS phase}\label{sec:Altresult}

When $Q$ is further increased beyond the AFM--PVBS transition point, there is a second phase transition at $Q_{Ac}\approx 0.934$ into the eight-fold degenerate AVBS
phase illustrated in Fig.~\ref{fig:VBS}(c). This phase transition can be described as a continuous freezing of the resonating plaquette singlets into either
horizontal or vertical valence bond pairs in a checker-board pattern. The formation of the checker-board pattern involves the breaking of a $\mathbb{Z}_2$
symmetry. In the AVBS phase, the plaquette order parameter $\Pi$ does not change appreciably from its value in the PVBS phase prior to the transition, and we are
unable to detect this $\mathbb{Z}_2$ symmetry breaking in $\Pi$ (though potentially there could be a higher-order singularity).

To analyze the PVBS--AFBS transition
we define the following four-component ``alternation order parameter'' ${A} = (A_{0,0}, A_{0,1}, A_{1,1}, A_{1,0})$ with
\begin{equation}
\label{eq:AVBSop}
A_{a,b} = \frac{4}{N} \sum_{(x, y)\equiv (a, b)} e^{i\frac{\pi}{2}(x-a+y-b)} \Psi_{x, y},
\end{equation}
where the notation $(x, y)\equiv (a, b) $ stands for coordinates for which $x \equiv a ~\mathrm{mod}~2$ and $y \equiv b~\mathrm{mod}~2$ for one out of the four possible
combinations of the subscripts $a \in \{0,1\}$ and $b \in \{0,1\}$. These combinations correspond to the four different plaquette patterns in the PVBS
phase. The ``plaquette orientation'' operator $\Psi_{x,y}$ in Eq.~(\ref{eq:AVBSop}) is defined as 
\begin{equation}
\Psi_{x,y} =\frac{1}{4} (S^z_{x,y} - S^z_{x+1,y+1}) (S^z_{x+1, y} - S^z_{x, y+1}),
 \end{equation}
and, thus, it can detect the dominant direction of the valence-bond pairs in the plaquette with lower-left corner at $(x,y)$. Note that, while the
order parameter $A$ can detect the $\mathbb{Z}_2$ breaking associated with the AVBS phase (where one out of the four components $A_{a,b}$ takes a
non-zero value when the symmetry is broken), it remains zero through the AFM--PVBS phase transition. 

\begin{figure}[t]
\includegraphics[width=8.4cm]{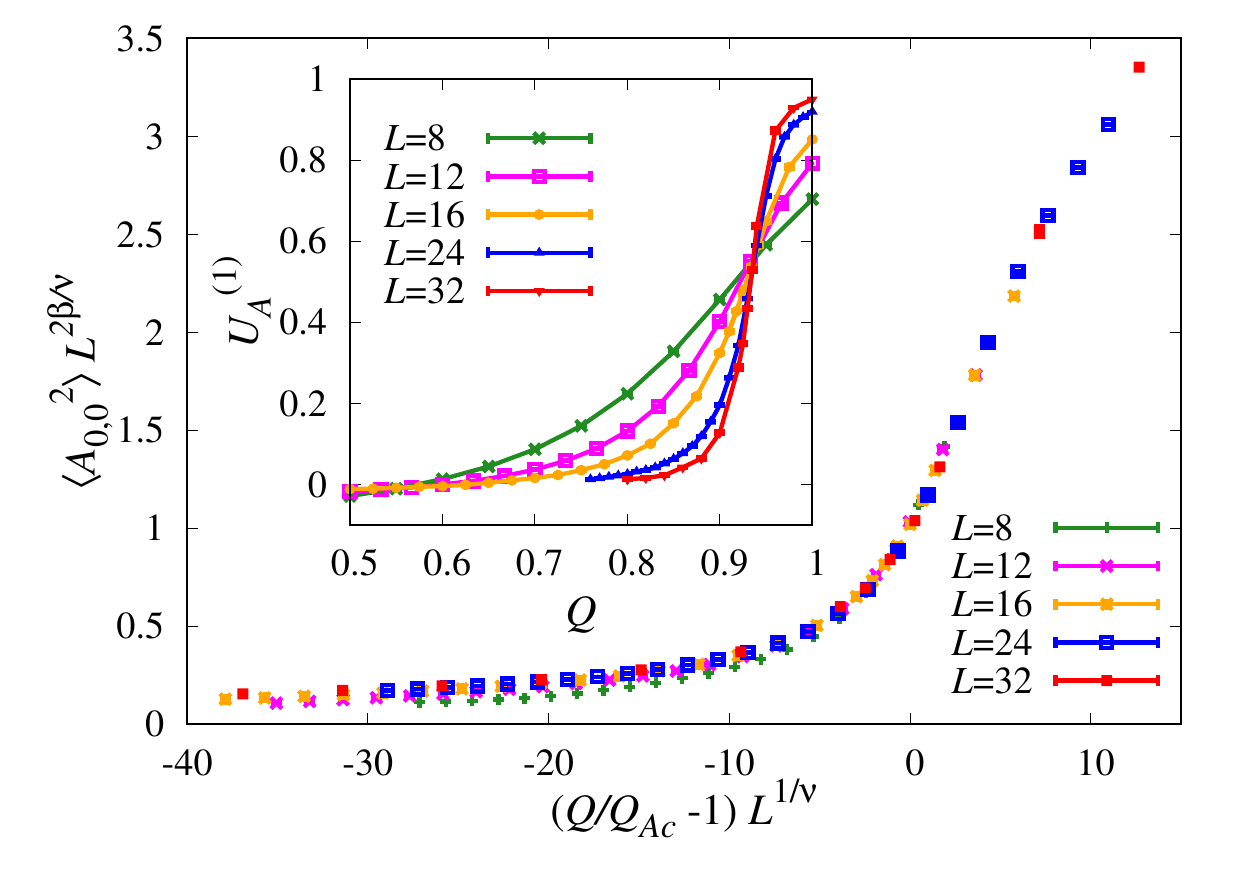}
\caption{\label{fig:AltTransition}
Squared AVBS order parameter on lattices with open boundaries close to the PVBS--AVBS transition. $A_{0,0}$ is the component in Eq.~(\ref{eq:AVBSop}) that is favored
by the boundaries. The values have been rescaled to test the expected critical scaling form $\langle A_{0,0}^2\rangle = L^{-2\beta/\nu}f(\delta L^{1/\nu})$, where
$\delta=Q/Q_{Ac}-1$ and $\beta$ and $\nu$ are 3D Ising exponents ($\beta \approx 0.607$, $\nu=0.630$ \cite{Hasenbusch99}). The scaling function $f$ is exhibited
by the data collapse for sufficiently large $L$. The inset shows the corresponding Binder cumulant $U_A^{(1)}=(3-\langle A_{0,0}^4\rangle/\langle A_{0,0}^2\rangle^2)/2$,
from which the critical point $Q_{Ac} \approx 0.934$ is obtained as the asymptotic location of the crossing points of data for different system sizes.}
\end{figure}

Since the system is already in a PVBS phase before the second phase transition takes place, in a large system, where the $\mathbb{Z}_4$ symmetry in
practice is broken in QMC simulations (i.e., the tunneling time between the different pattern is much longer than feasible simulation times \cite{vbscomment}),
it is possible to only consider the single component of $A$ corresponding to the specific plaquette pattern realized. We will later, in Sec.~\ref{sec:OrderGraph},
discuss the order parameter and symmetry breaking in more detail, but for now focus on the simplest way to detect and characterize the second transition.

The best way we found to study the PVBS--AVBS transition is to use open boundary conditions on $L\times L$ lattices with even length $L$, in which case the
PVBS plaquette pattern is non-degenerate, i.e., the boundaries themselves induce a specific plaquette pattern that is stable in the limit $L \to \infty$ if
$Q$ is above the critical value $Q_{Ac}$ for the AFM--PVBS transition (while for $Q < Q_{Ac}$ the pattern only remains at the boundaries and decays to zero in
the bulk with increasing $L$). This setup allows us to monitor only the corresponding single component of the four-component operator $A_{a,b}$ in Eq.~(\ref{eq:AVBSop}).
The choice of boundary condition does not change the universality class or any other essential physics. We can also observe the transition in systems with
periodic boundary conditions, though with considerably larger fluctuations that make it harder to study the critical behavior. We only discuss open-boundary
SSE results in this section.

\begin{figure*}[t]
\floatbox[{\capbeside\thisfloatsetup{capbesideposition={right,top},capbesidewidth=7cm}}]{figure}[\FBwidth]
{\caption{An order graph depicting the relationships between the eight degenerate AVBS states, represented as the yellow circles. The thick black bonds connect
states that can most easily fluctuate between each other, while the thinner grey bonds correspond to the second-strongest fluctuations, with all being equal due to
symmetries. States not connected by any bond have the smallest probability of fluctuating into each other. The PVBS phase forms when the tunneling barriers
between states connected by the thinner bonds vanish and the system locks into one of the four groups of two strongly-connected states. Subsequently
AVBS order forms when also the thick bonds vanish and the system locks into one of the two states in the previously chosen PVBS pair. The vectors
displayed close to the circles show an example of a faithful Euclidean embedding of the graph in six dimension (the minimal dimensionality of the combined
order parameter).
}\label{fig:AltVBSGraph}}
{\includegraphics[width=10.17cm]{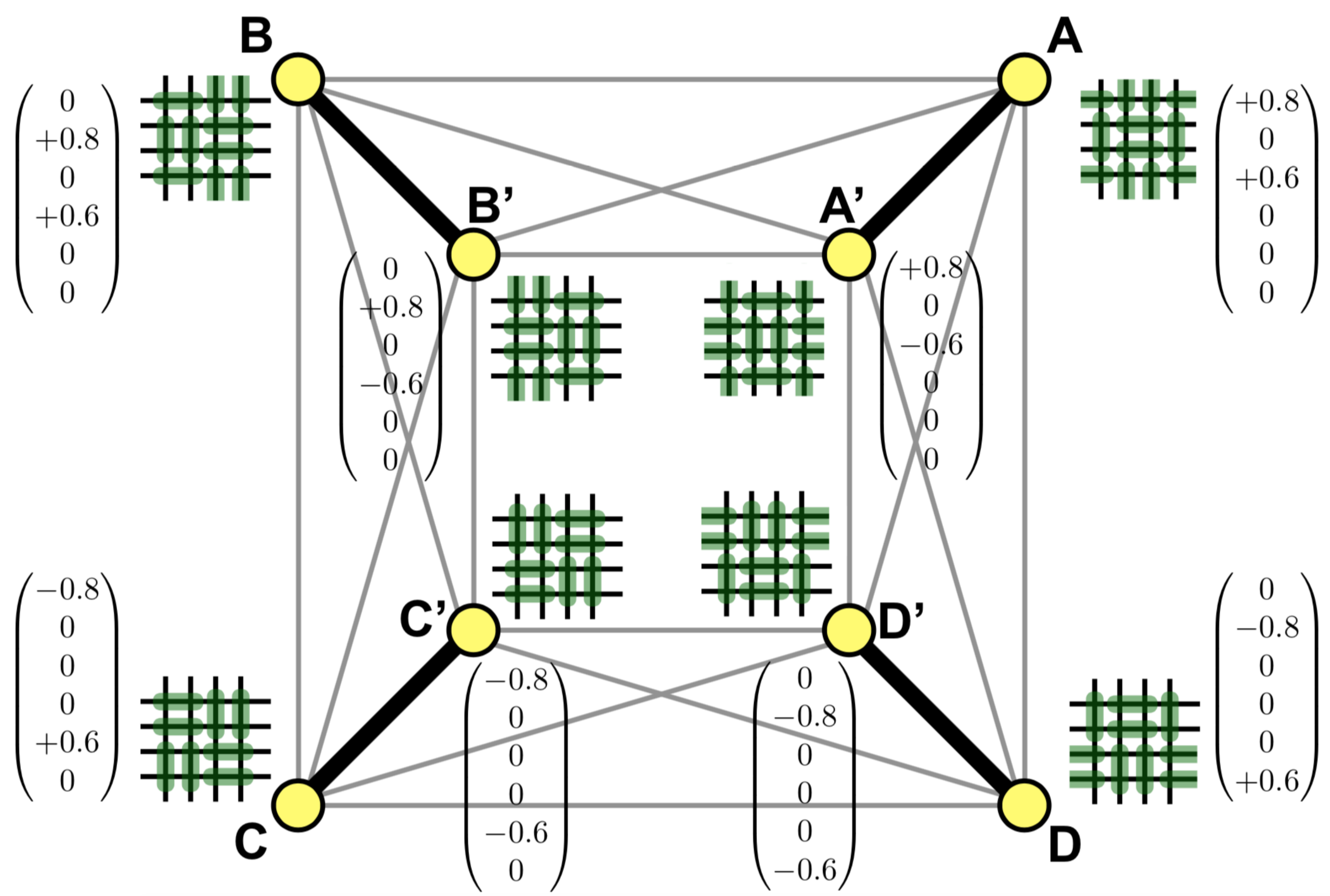}}
\end{figure*}

Since this phase transition involves an additional $\mathbb{Z}_2$ symmetry breaking in the $(2+1)$-dimensional quantum system, we expect the criticality
to be that of the 3D Ising universality class. Fig.~\ref{fig:AltTransition} shows the squared order parameter rescaled with the expected powers of the
system size, with the estimated critical point $Q_{Ac}=0.934$ obtained using the Binder crossing method (shown in the inset of Fig.~\ref{fig:AltTransition}).
Except for the smallest system size, the data points collapse well onto a single curve (the scaling function), thus supporting the expected Ising nature of
the PVBS--AVBS transition. Thus, we have demonstrated that the PVBS state, which is a well established ground state of a quantum magnet even though it has
previously not been found in sign-free models, can be unstable to a continuous freezing of the resonating plaquettes into static dimers with a checker-board
pattern on the already formed plaquette pattern. The $D_4$ symmetry of the square lattice is then fully broken in two stages, first with $\mathbb{Z}_4$ 
broken in the PVBS phase and then another $\mathbb{Z}_2$ breaking in the AVBS phase.

A state similar to our AVBS state was previously discussed in the context of a quantum dimer model with multiple potential and kinetic terms
in the Hamiltonian \cite{Nakata13}. Perhaps because the dimers lack the intrinsic singlet nature and are restricted to connecting nearest-neighbor sites (while
there are no analogous constraints in a spin model, where the dimers of a VBS are emergent objects), this dimer model only exhibited first-order transitions
of the AVBS-like state, and there was no discussion of multi-stage symmetry breaking of the kind found here. We next discuss how to formally
describe the two-stage breaking of the lattice $D_4$ symmetry.

\section{Symmetry breaking and unified VBS order parameter}
\label{sec:OrderGraph}

We here construct a unified framework for describing the two stages of discrete symmetry breaking and, within that scheme,
interpret the PVBS phase as an intermediate phase. The first step is to construct an adequate order parameter for detecting both of the phase
transitions in some way, unlike the previously discussed order parameters $D$, $\Pi$, and $A$,
which capture only one of the transitions. In order to
do that in a systematic and compact way, we introduce the concept of an {\it order graph}, which we will outline in the following and explain in more detail in
Appendix \ref{app:OrderGraph}.

\subsection{The order graph}
\label{sec:ordergraph}

The eight-fold degenerate ordered states in the AVBS phase have relative fluctuations depicted in Fig.~\ref{fig:AltVBSGraph}, which is what
we will call the order graph. In the thermodynamic limit, such fluctuations do not occur inside the AVBS phase, but they are manifested on any finite system
and become more prominent as a quantum phase transition is approached. The thick black bonds represent fluctuations among states which are separated
by the smallest tunneling barriers, and two states which have no bonds between them have the smallest probability of fluctuations among each other.

The states connected by thin bonds all have equivalent relations although they may not have the exact same symmetry transformations connecting them.
For example, starting from the state marked by $\mathsf{A}$ in Fig.~\ref{fig:AltVBSGraph}, transforming to state $\mathsf{B}$ or $\mathsf{B}^\prime$
corresponds to lattice translations in the $+x$ and $-x$ direction, respectively. These transformations are different, but since the parity symmetry
between $\pm x$ directions is not broken, these pairs of states should have equal tunneling barriers separating them. The fluctuations between primed
and unprimed states with the same letter are the easiest, because they do not involve changing the plaquettes on which the bond pairs form, only the
orientation of the bond pairs within the plaquettes. The tunneling barriers between these state pairs vanish at the AVBS--PVBS transition.

In the PVBS phase the accessible subspace of the Hilbert space corresponding to one out of the four possible plaquette patterns is much larger than just the
combined space of the two states represented by a pair of strongly connected yellow circles in Fig.~\ref{fig:AltVBSGraph}, as the bond pairs on different
plaquettes fluctuate essentially independently of each other (with some correlations that vanish with increasing distance between plaquettes) and connecting
the two states by tunneling involves the creation of domain walls. The physical
interpretation of the thick bond thus changes from representing the largest tunneling probability between the eight states in the AVBS phase on a finite
lattice (which vanishes when the system size is taken to infinity) to an enlarged ``basin'' in the Hilbert space in the PVBS phase, roughly comprising all the
states on what was previously the tunneling path. In the AFM phase the fraction
of the Hilbert space involved in the unique ground state is further enlarged to encompass equivalently all the points in the graph---the AFM phase of course
has completely different symmetry structure and fluctuations that are not described by the order graph for the PVBS and AVBS states. Some remnants of the graph
structure may still be manifested in the short-distance fluctuations and plaquette and dimer correlation functions, at least close to the AFM--PVBS transition.

Any symmetry transformation that preserves the Hamiltonian will correspond to an automorphism of the order graph. Therefore, the order parameter is required to have 
the same transformation properties as that automorphism group. One way to construct such an order parameter is to embed the order graph into an Euclidean space in
a faithful way, i.e., where all pair of vertices have equal distance if and only if they have the same type of bond in the order graph. This is possible in six
dimension (or more) in our case, which is exemplified in Fig.~\ref{fig:AltVBSGraph} with possible coordinates shown as vectors. From the embedding,
a useful six-dimensional order parameter can be constructed naturally, which we will explain in detail in Appendix \ref{app:OrderGraph}. 
The idea is to construct an order parameter that has eight peaks at those points embedded in six-dimensional space. 
The projection of the six-dimensional order parameter into lower dimensions corresponds to the 2D PVBS order parameter $\Pi$ (and/or alternatively $D$) and
the four-dimensional AVBS order parameter $A$. 

\subsection{Classification of broken symmetries}

The order graph also provides us with a convenient way to correctly classify what symmetry is broken at a given transition. As we discussed in Sec.~\ref{sec:DQCresult}
and further in Appendix \ref{app:d4}, the CVBS and PVBS phases both can be regarded as realizing a $\mathbb{Z}_4$ symmetry breaking of the square lattice $D_4$
symmetry \cite{Levin04}. The usual prescription for such symmetry breakings is to consider the symmetry group $G$ the system
originally transforms under together with the remaining symmetry group $H$ that it still transforms with after a symmetry-breaking transition
has taken place. Then the quotient $G/H$ corresponds to relevant symmetry group of the order parameter. This can be translated into the
automorphism group of the order graph $G^\prime$, and the relevant subgroup $H^\prime$ of that graph, which is the automorphism of the order
graph when we distinguish some vertices. The distinction of vertices into two kinds corresponds to the state(s) the system gets stuck into and
the remaining ones. This point of view is equivalent to classifying symmetry transformations into equivalent classes under their
operations on the order parameters, as we discuss further in Appendix \ref{app:OrderGraph}.

The automorphism group of the order graph in Fig.~\ref{fig:AltVBSGraph} is $D_4\rtimes\mathbb{Z}_2^4$, corresponding to the rotations and reflections of
the over-all graph ($D_4$) and 4 independent swaps (e.g. $\mathsf{A}\leftrightarrow\mathsf{A}^{\prime}$) possible at each corner ($\mathbb{Z}_2^4$). The
automorphism group becomes $\mathbb{Z}_2^5$ when two pairs of vertices (connected with thick black bonds) in the order graph are chosen in the PVBS phase. More
precisely, the individual swaps in the corners remain, and $D_4$ breaks down to $\mathbb{Z}_2$ because we are only left with one reflection that leaves
the chosen pair invariant. Therefore, the broken symmetry is expressed by the coset $D_4/\mathbb{Z}_2$. This is exactly the same as the broken symmetry of
the four-state clock model, which is frequently, somewhat inaccurately (in a way explained in Appendix \ref{app:d4}), referred to as $\mathbb{Z}_4$
symmetry breaking \cite{Senthil04a,Levin04}.
When the system is in the AVBS phase, it corresponds to a single vertex being selected in the order graph, and now the remaining symmetry is $\mathbb{Z}_2^4$.
Therefore, the symmetry breaking at the PVBS--AVBS transition is $\mathbb{Z}_2^5 / \mathbb{Z}_2^4 = \mathbb{Z}_2$ (as also naively expected). 

In the case of a direct transition into the AVBS phase, the order graph symmetry reduces from $D_4\rtimes\mathbb{Z}_2^4$ to $\mathbb{Z}_2\times\mathbb{Z}_2^3$, 
and the broken symmetry becomes $D_4$. 
This could also be intuitively understood in the following way. 
In the PVBS phase, all rotations around a lattice site does not conserve the macroscopic state anymore, 
but there is always a remaining spacial reflection with an axis going through a lattice site diagonally (whether $\pm 45^{\circ}$ depends on which of the four PVBS states). 
This is the remaining $\mathbb{Z}_2$ symmetry of the original square lattice point group symmetry $D_4$, 
but also breaks when the system becomes AVBS, thus the entire $D_4$ symmetry is broken now.
While $D_4$ is not a direct product $\mathbb{Z}_4\times\mathbb{Z}_2$, it still is the semidirect product $\mathbb{Z}_4\rtimes\mathbb{Z}_2$.

From the above symmetry considerations, it is clear that, as long as the individual AFM--PVBS and PVBS--AVBS phase transitions are considered, 
a 2D clock order parameter and a scalar order parameter, respectively, are sufficient. 
Note that in the latter case, we assume that the system is already completely locked
into one of the four degenerate PVBS patterns, as is the case in an infinite or, in practice, in a very large system in QMC simulations \cite{vbscomment},
or when using open boundary conditions as we did in Sec.~\ref{sec:Altresult}. Otherwise, a four-dimensional order parameter, e.g., as defined in
Eq.~(\ref{eq:AVBSop}), would be necessary for studying a transition between the PVBS and AVBS phases. 
To study a direct transition from a disordered state into AFVBS,
a six-dimensional order parameter is preferable; we will consider that in Sec.~\ref{sec:6Dim} in the context of a finite-temperature paramagnetic--AVBS
transition.

\section{Finite-temperature transition and transition paths}
\label{sec:finT}

Since the PVBS and AVBS phases break discrete symmetries they extend to finite temperature $T>0$. We here discuss the nature of the finite-temperature
transition and also summarize all the possible ways in which the symmetries can be broken simultaneously or in multiple stages leading to the AVBS phase, 
not just in the case of the $J$-$Q_6$ model but generically based on the order graph.

\subsection{Transitions at $T>0$}

\begin{figure}[t]
\includegraphics[width=8.5cm]{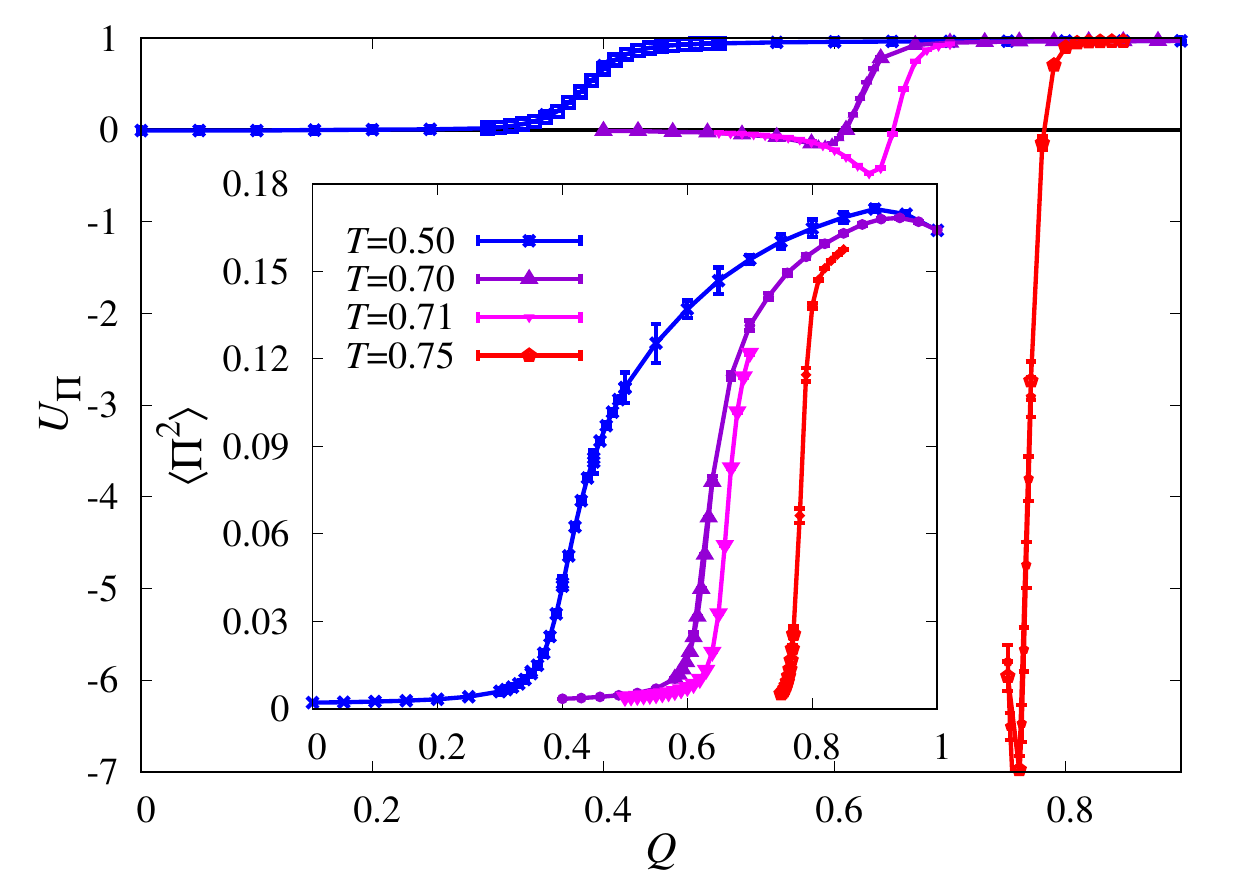}
\caption{\label{fig:FinTtrans} 
Examples of constant-$T$, $Q$-dependent PVBS order parameters (inset) and the corresponding Binder cumulants (main graph) for $L=32$ systems.
The negative peaks in the cumulants indicate first-order transitions.}
\end{figure}

Figure \ref{fig:FinTtrans} shows SSE results
for the PVBS order parameter and the corresponding Binder cumulant versus $Q$ at several different temperatures $T$. 
Here we can see clearer signs of first-order transitions as $T$ is increased. 
Not only does the order parameter grow more acutely for higher $T$, but the Binder cumulant develops an increasingly prominent negative dip in the neighborhood of the transition point at higher $T$---such a dip is a sign of conventional
phase coexistence, i.e., with no emergent higher spherical symmetry.

\begin{figure}
\includegraphics[width=8.5cm]{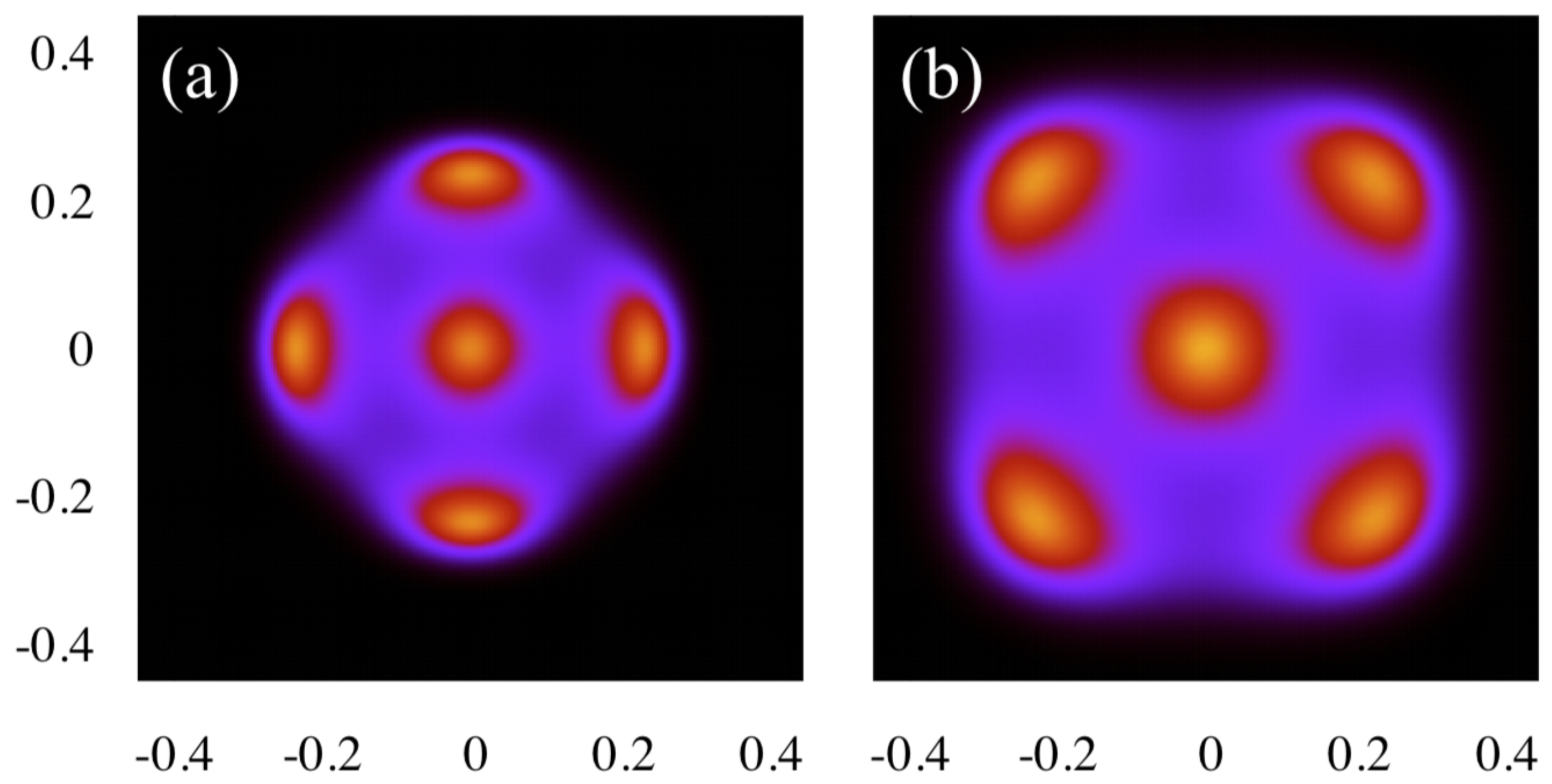}
\caption{\label{fig:FinTHist} 
 Probability distributions of the plaquette order parameter $P(\Pi_0,\Pi_1)$ in (a) and the dimer order parameter $P(D_x,D_y)$ in (b) at the
 transition value of $Q$ (defined here as the position of the negative peak in the Binder cumulant---see Fig.~\ref{fig:FinTtrans}). The system
 size is $L=32$.}
\end{figure}

In this case, we indeed observe histograms of the PVBS order parameter with five peaks (see Fig.~\ref{fig:FinTHist}); 
four of them corresponding to the ordered PVBS patterns and the one at the origin corresponding to the absence of PVBS order in the paramagnetic state. 
We can conclude that no emergent symmetry of the PVBS order parameter is present at $T>0$, and the transition is first-order
with a conventional coexistence state with free-energy barriers. This is in contrast to the CVBS case, where the $T>0$ transition stays continuous,
with exponents varying according to one of the critical branches of the Ashkin-Teller model, becoming increasingly similar to a Berizinsky-Kosterliz-Thouless
transition as $T_c \to 0$ when the quantum-critical point is approached  (e.g., the correlation-length exponent $\nu$ grows as $T_c \to 0^+$ and is
different from the DQC exponent exactly at $T=0$) \cite{Jin13}.

\begin{figure}[t]
\includegraphics[width=8.0cm]{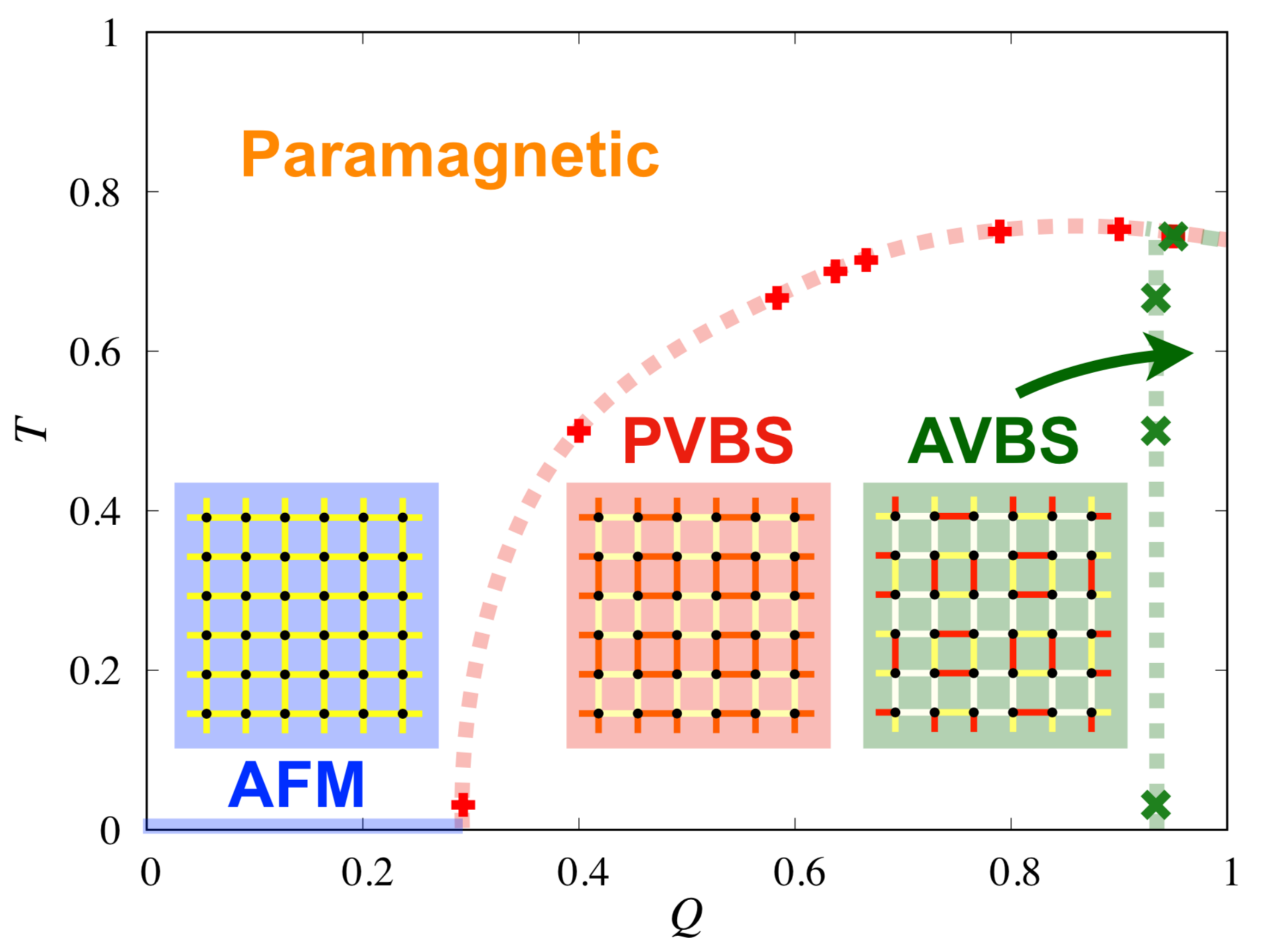}
\caption{\label{fig:PhaseDiagram} 
Estimated $T$-$Q$ phase diagram of the $J$-$Q_6$ model (with $J=1-Q$, and $T$ given in units of $J+Q=1$). The dotted lines are guide to the eye,
connecting several paramagnetic--PVBS transition points (red $\mathbf{+}$ symbols) and PVBS--AVBS points (green $\mathbf{\times}$ symbols),
all computed for $L=32$ systems (for which remaining finite-size effects on the phase boundaries are small).
The diagrams in shaded boxes illustrate the nature of the phases with bond correlation strengths $-\langle S^z_i S^z_j\rangle$
on a piece of the lattice in each phase, obtained from actual SSE simulations. White, yellow and dark red lines correspond to the weakest,
intermediate and strongest correlations, respectively.}
\end{figure}

\begin{figure*}
{\includegraphics[width=12cm]{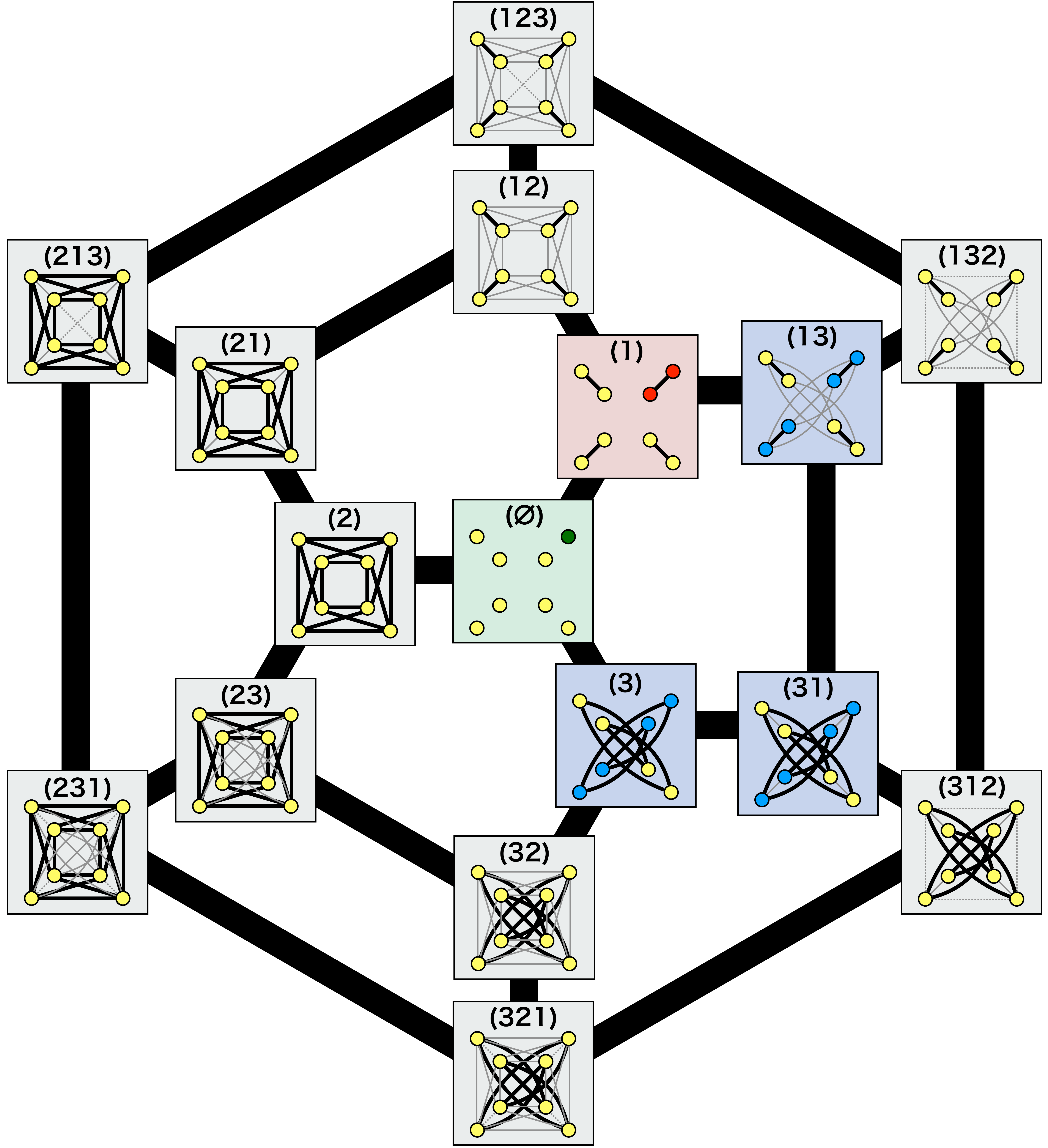}}
{\caption{All 16 possible order graphs that satisfy the symmetry of the AVBS phase. The numbers $\mathsf{1}$, $\mathsf{2}$, and $\mathsf{3}$ correspond,
respectively, to the strongest (black), second strongest (grey), and absent fluctuation paths in Fig.~\ref{fig:AltVBSGraph}. Those fluctuations that survive
in the thermodynamic limit are coded according to their relative strengths by the order of the numbers within (), with strongest to weakest fluctuations
from left to right. In the corresponding order graphs, the fluctuation paths are drawn in black (strongest), gray (second-strongest), and dotted gray
(weakest). In order to avoid cluttering, only a few of the weakest bonds are shown in the cases where they are present (in the six outermost order graphs).
Symmetry breaking corresponds to the order graph becoming disconnected because of vanishing fluctuations, i.e., the graph breaks into equivalent clusters
of connected yellow vertices (states), and the system is trapped in one of the equivalent clusters. These are shown in blue (one out of two clusters),
red (one out of four clusters), or green (one out of eight vertices). The thick bonds connect order graphs that only differ by a single swap of adjacent
fluctuation strength or by the disappearance of the weakest fluctuation.}
\label{fig:BigMap}}
\end{figure*}

We provide the full $T$-$Q$ phase diagram of the model in Fig.~\ref{fig:PhaseDiagram}, which is based on several transition points on the paramagnetic--PVBS
phase boundary, extracted using fixed-$T$ scans such as those shown in Fig.~\ref{fig:FinTtrans}. To determine points on the PVBS--AVBS boundary we used  
open-lattice calculations, as described in Sec.~\ref{sec:Altresult}. An important aspect of the phase diagram is that the AVBS phase is completely
inside the PVBS phase. A direct paramagnetic--AVBS transition takes place upon lowering the temperature for $Q$ very close to 1.

The fact that the $T>0$ transitions become more strongly first-order as the transition temperatures increase may seem counterintuitive at first sight, but
such behavior has been observed and analyzed in the context of fluctuation-induced first-order transitions of vestigial phases
\cite{Janoschek13,Fernandes12,Fernandes19}. We next provide another perspective on this phenomenon, utilizing the order graph introduced in the previous section.

\subsection{Multi-stage symmetry breaking}

We now discuss how transition paths between different states can be described systematically, which will provide generic insights into multi-stage 
discrete symmetry breaking also beyond the specific $J$-$Q_6$ model and $D_4$ symmetry considered here.

Figure \ref{fig:BigMap} shows all possible order graphs which respect the final symmetry of the
AVBS state.  This figure shows connections (thick lines) between order graphs if and only if continuously varying the fluctuation strengths in the system can
cause the order graph to change from one to another. The fluctuations can change by varying parameters in the Hamiltonian or by changing the temperature.
We require these changes to not affect the symmetries, i.e., the three sets of fluctuation bonds introduced in Fig.~\ref{fig:AltVBSGraph} are maintained.
The only changes that are allowed in general in the order graph are either (a) a swap of ranking of two adjacent fluctuation strengths, or (b) removal of
the weakest fluctuation or emergence of a new weakest fluctuation.

This kind of map, although conceptually simple, can restrict the topology of the phase diagram.
The two assumptions (a) and (b) may appear to imply that we are considering only continuous transitions, where the
evolution of the non-vanishing fluctuation strengths upon varying a parameter is smooth. Clearly, there can be first-order transitions where this assumption
is not valid, but if a transition is {\it weakly} first-order, such that the correlation length of an order parameter is rather large even before the
transition, and the transition itself is related to these fluctuations, we also expect the transition to respect the connectivity of Fig.~\ref{fig:BigMap}.
A transition that does not respect this connectivity, jumping directly between order graphs without direct connection, should be expected to be
strongly first-order.

The sequence of symmetry breakings involved in the $J$-$Q_6$ model when it transitions first into the PVBS phase (either from the AFM phase at $T=0$ or from
the paramagnet at $T>0$) and then into the AVBS corresponds to starting in Fig.~\ref{fig:BigMap} from the box marked $(\mathsf{12})$ to the one marked
$(\mathsf{1})$, and then finally to  $(\mathsf{\varnothing})$ at the center. If we add some (unknown) term in the Hamiltonian that in the notation of
Fig.~\ref{fig:AltVBSGraph} enhances the fluctuations $\mathsf{A}\leftrightarrow\mathsf{B}$, $\mathsf{A}\leftrightarrow\mathsf{B}^\prime$, etc., the order graph
should eventually change from $(\mathsf{12})$ in Fig.~\ref{fig:BigMap} to $\mathsf{(21)}$. When the order graph is altered
in this way, all eight states are still connected even when the second strongest fluctuation bonds have died out in the case ($\mathsf{2}$), and,
therefore, a phase transition will take place directly into the AVBS if the fluctuations represented by the strongest bonds are brought to zero and
the graph $\mathsf{(\varnothing)}$ is reached. This transition breaks the $D_4$ symmetry in a single step.

It is generally considered that breaking a symmetry which has a reducible representation, such as $D_4=\mathbb{Z}_4\rtimes\mathbb{Z}_2$, should only happen by way of a strongly first-order transition \cite{Landau}. 
As we have seen, our arguments based on the natural connectivity of the order graphs suggest that the direct $D_4$-breaking transition
can also be continuous in principle and not just by fine-tuning. However, it is physically not a very likely scenario, because of the requirement that all the
fluctuation paths in the graph $\mathsf{(2)}$ in Fig.~\ref{fig:BigMap} must be of equal strength, with the other ones vanishing. 
If we imagine embedding the order
graph in a higher dimensional space such that the eight points are equidistant, a minimum of six dimensions is required (as discussed further in \ref{sec:6Dim}).
In naturally occurring systems we would not expect this fluctuation map to be easily realized, and a more likely scenario (which is realized in the $J$-$Q_6$ model), 
is a first-order transition corresponding to a direct jump from the graph $\mathsf{(12)}$ to $\mathsf{(2)}$ in Fig.~\ref{fig:BigMap}.

\begin{figure}[t]
\vspace{0cm}
\includegraphics[width=8.5cm]{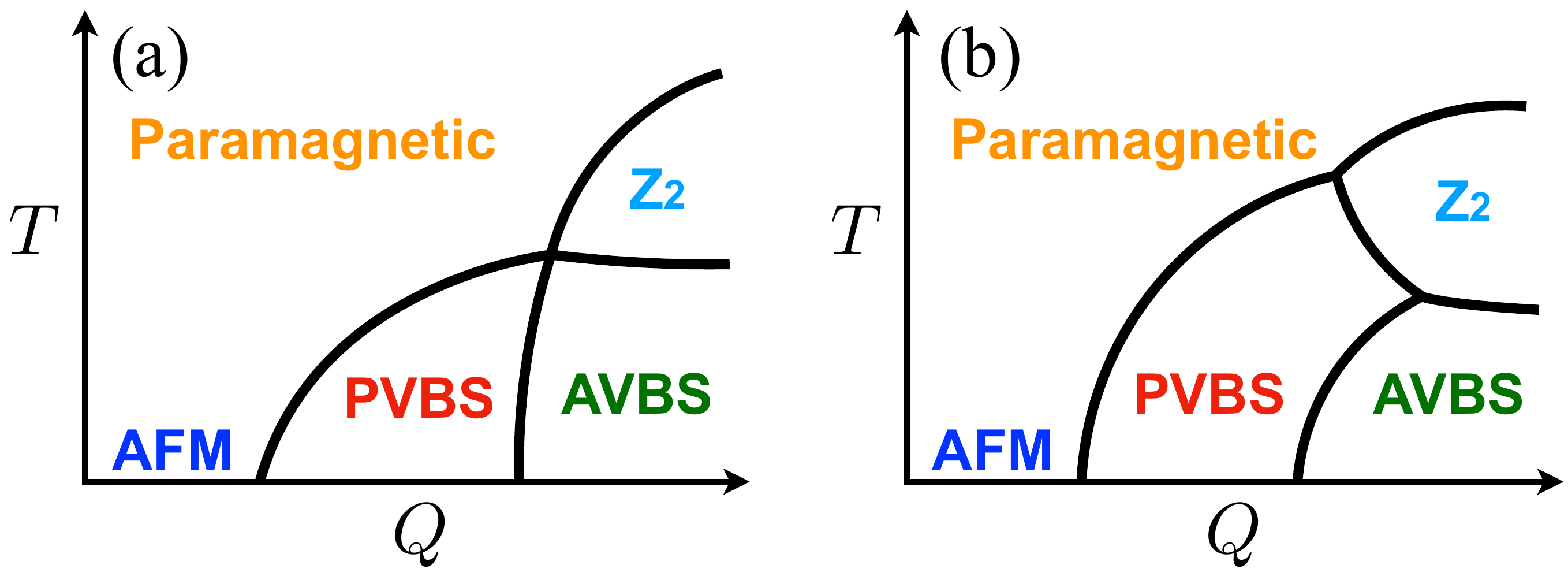}
\caption{\label{fig:2PDs} 
Two phase diagrams containing AFM, PVBS, and AVBS phases, together with a possible $\mathbb{Z}_2$ symmetry broken phase. By analyzing the order
graphs in Fig.~\ref{fig:BigMap}, we can infer that case (a) is impossible without fine tuning or altering the symmetry, while (b) is generically possible.}
\end{figure}

In contrast to the path in Fig.~\ref{fig:BigMap} where $D_4 = \mathbb{Z}_4\rtimes\mathbb{Z}_2$ is broken in a single step, the previously
considered $\mathsf{(12)\rightarrow(1)\rightarrow(\varnothing)}$ path (non-VBS to PVBS and then to AVBS) corresponds to a sequential breaking
of the $D_4$ symmetry; first $\mathbb{Z}_4$ and then $\mathbb{Z}_2$. At first sight, it may seem that the $\mathbb{Z}_4$ symmetry breaking line and
the $\mathbb{Z}_2$ symmetry breaking line could cross each other, forming a purely $\mathbb{Z}_2$ symmetry breaking phase as exemplified in
Fig.~\ref{fig:2PDs}(a), similar to a nematic phase. In fact, this is what happens in the vicinity of a tetracritical point \cite{Eichhorn13} where
four critical phase boundaries meet at one point. Also, the order graphs $(\mathsf{13})$, $(\mathsf{31})$, and $(\mathsf{3})$ in Fig.~\ref{fig:BigMap}
all correspond to a purely $\mathbb{Z}_2$ broken phase. However, this does not imply that such phase diagrams with a tetracritical point are possible. 
To the contrary, the order graph analysis tells us that, in general, it is impossible to have such a tetracritical point when the AVBS phase is set as
the final phase. This is because it is necessary that the order graph relations form a minimal loop that contains order graphs corresponding to all
of the different phases surrounding a multicritical point, in this case paramagnetic, PVBS, AVBS and the $\mathbb{Z}_2$ symmetry breaking phase. 
None of the minimal loops in Fig.~\ref{fig:BigMap}, square or pentagon, has all of those four phases. Nevertheless, we have minimal loops that contain
three phases, thus tricritical points such as those in Fig.~\ref{fig:2PDs}(b) are possible. 

\begin{figure}[t]
\vspace{0cm}
\includegraphics[width=6.17cm]{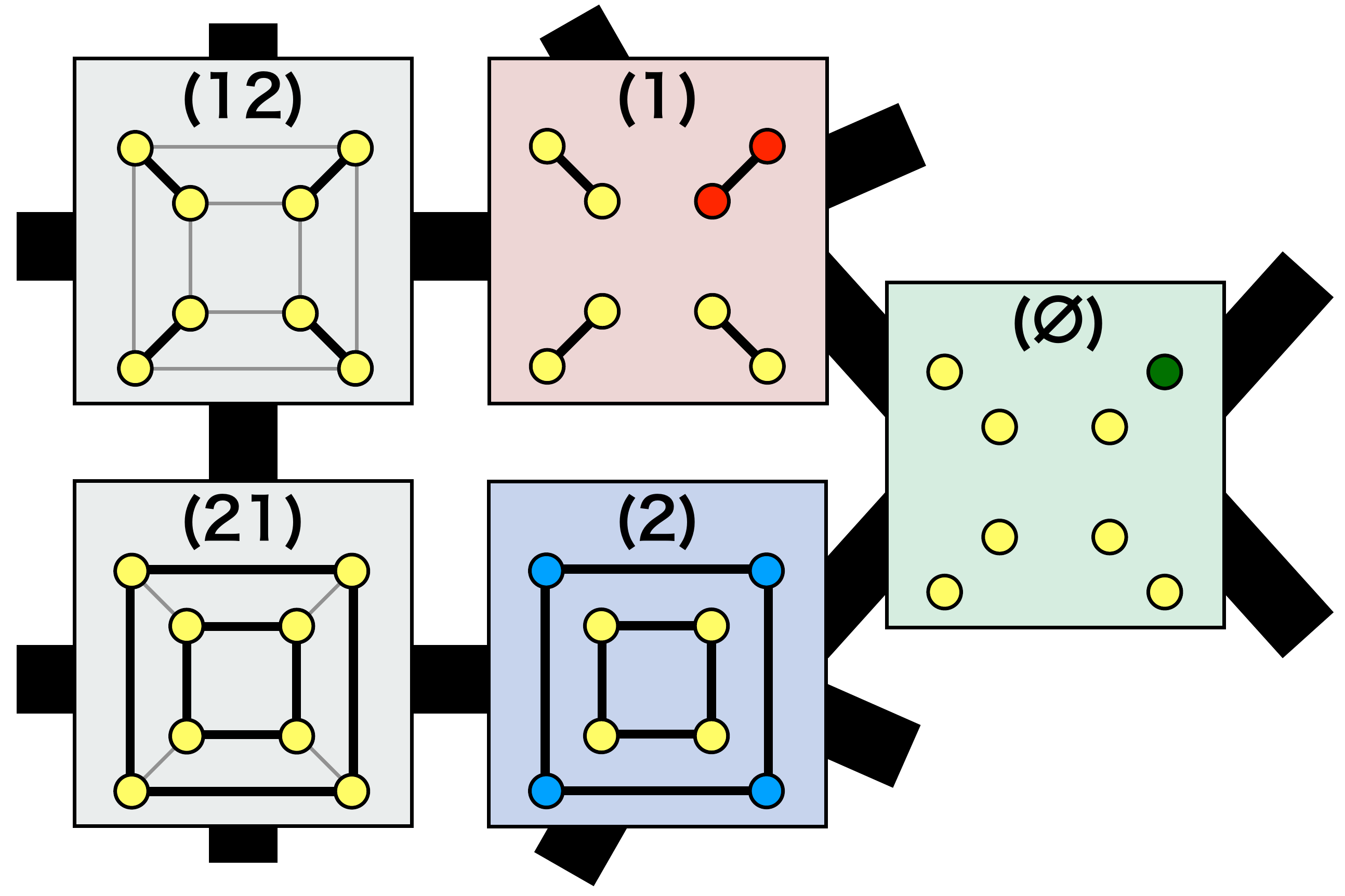}
\caption{\label{fig:Nemap}
Relationships between order graphs such as in Fig.~\ref{fig:BigMap}, but for an eight-fold degenerate final phase arising in a model with lower
symmetry than the $J$-$Q_6$ model. Only those order graphs are shown that are needed to demonstrate the possibility of a tetracritical point
such as in Fig.~\ref{fig:2PDs}(a).}
\end{figure}

It is of course possible in general to have a tetracritical point as in Fig.~\ref{fig:2PDs}(a) with $\mathbb{Z}_2$ and $\mathbb{Z}_4$ symmetry breaking
if we consider systems with less than the full $D_4$ symmetry considered so far. More specifically, if we allow order graphs such as those drawn in
Fig.~\ref{fig:Nemap}, a minimal pentagon loop contains four phases and we can have a phase diagram like Fig.~\ref{fig:2PDs}(a). However, in order to have
these order graph relations, we would need to specify eight of the type $\mathsf{2}$ bonds to be stronger than the rest, explicitly breaking the symmetry
the system had when we were considering the AVBS phase and original $D_4$ symmetry. Note that all the eight states of the less symmetric 8-fold degenerate
phase are still equivalent (i.e. vertex transitive, as we will explain in Appendix \ref{app:OrderGraph}), but they need not correspond to the AVBS patterns that we have considered explicitly in the $J$-$Q_6$ model.

To demonstrate explicitly that we cannot achieve the phase diagram in Fig.~\ref{fig:2PDs}(a) with the current symmetry of the AVBS phase,
let us look back again on Fig.~\ref{fig:AltVBSGraph}. Consider, e.g., the spatial reflection along the $x$ axis that conserves the states of
$\mathsf{A, A^{\prime},B}$, and $\mathsf{B^{\prime}}$ but inverts $\mathsf{C}$ and $\mathsf{C}^{\prime}$ (and also $\mathsf{D}$ with $\mathsf{D}^{\prime}$). 
This is an automorphism of the original order graph Fig.~\ref{fig:AltVBSGraph}, but not for the order graph shown in Fig.~\ref{fig:Nemap} $(\mathsf{12})$. 
Thus, while the order graph analysis rules out the possibility of such tetracritical points when considering the AVBS phase with current symmetry,
in models with different symmetry, they are possible in general. 
A trivial example would be a ferromagnetic model with both Ising degrees of freedom $\{\sigma_i\}$ and 4-state clock degrees of freedom $\{\theta_i\}$, 
with a Hamiltonian such as 
\begin{equation}\label{eq:trivialZ4Z2}
H=-J_{\mathrm{Ising}}\sum_{\langle i,j\rangle}\sigma_i\sigma_j - J_{\mathrm{clock}}\sum_{\langle i,j\rangle} \cos(\theta_i-\theta_j) .
\end{equation}
In this case, the $\mathbb{Z}_2$ and $\mathbb{Z}_4$ symmetry breakings are decoupled and by tuning 
$J_{\mathrm{Ising}}/J_{\mathrm{clock}}$, we can trivially obtain a phase diagram with a tetracritical point. 
Note that now the type $\mathsf{2}$ fluctuations in Fig.~\ref{fig:Nemap} connect a vertex to only two neighbors, 
namely states with the same Ising order but with $\pm \pi/2$ rotated clock order.
The $x$ axis reflection we considered above would correspond to some simultaneous transformation in the spins $\{\sigma_i\}$ and $\{\theta_i\}$ for this model,
which would not conserve the total energy. 
This shows that the naive effective Hamiltonian Eq.~(\ref{eq:trivialZ4Z2}) with both $\mathbb{Z}_4$ and $\mathbb{Z}_2$ symmetry breakings does not explain the physics of the $J$-$Q_6$ model, and we would need an effective Hamiltonian that correctly captures the symmetry of the order graphs in Fig.~\ref{fig:BigMap} instead.

Even with strongly first-order transitions that do not have to obey the above rules for the adjacency of order graphs, we can still rule out generic
phase diagrams such as Fig.~\ref{fig:2PDs}(a), since it will require fine-tuning so that all four first-order transition lines come across at one single
point. We can also predict that first-order transitions connecting order graphs that are farther away are usually stronger. This is because in general
they would correspond to farther minima of the free energy. 

Consider now again the $J$-$Q_6$ model. Ruling out a tetracritical point purely from the connectivity of order graphs is not the only aspect of
the $Q$-$T$ phase diagram that can be deduced with this tool before empirically running simulations. From Fig.~\ref{fig:BigMap}, we see that if we want a
purely $\mathbb{Z}_2$ symmetry broken phase as in Fig.~\ref{fig:2PDs}(b), then we would need to enhance the fluctuation of type $\mathsf{3}$, the weakest
one in Fig.~\ref{fig:AltVBSGraph}. However, even when we consider finite temperature, there is no reason to expect such fluctuations to become stronger. 
On the other hand, if we consider having large enough $Q$ and gradually lowering the temperature, the fluctuations of type $\mathsf{1}$ should become
relatively weaker, because they are quantum fluctuations of singlets induced by the $J$-term (remember that we set $J+Q=1$ now). Therefore, if this relative
weakening of the type $\mathsf{1}$ bonds continues enough so that the fluctuation strength ranking actually is reversed to ($\mathsf{21}$), we would have a direct
transition to the AVBS phase as explained earlier. 
As we argued before, the order graph ($\mathsf{2}$) is unlikely to occur naturally in the $J$-$Q_6$ model, 
and it is more likely that there would be a direct transition from ($\mathsf{21}$) to ($\mathsf{\varnothing}$) in such case, which should be strongly first-order. 

Even if the effect of small $J$ is not strong enough to cause such reversals in the fluctuation strength, it will definitely drag the symmetry breaking path
to the left side of Fig.~\ref{fig:BigMap} rather than the right side where the purely $\mathbb{Z}_2$ broken phase is. If there is a direct transition from
($\mathsf{12}$) to ($\mathsf{\varnothing}$), this also must be strongly first-order since they are not connected.  We will see in the following section
that this is indeed the case, by observing the six-dimensional order parameter we construct from the order graph. 

In either cases, it is expected that larger $Q$ with thermal fluctuations will result in a direct paramagnetic--AVBS transition.
Since the region where there is a direct paramagnetic--AVBS phase transition must be strongly first-order (because the transition is between nonconnected order graphs in Fig.~\ref{fig:BigMap}), the nature of the paramagnet--PVBS phase transition should also become more and more strongly first-order as we approach closer to
the paramagnetic--AVBS region, i.e., when we increase temperature. This would be the order graph perspective of understanding the so-called fluctuation
induced first-order transitions, where it has been argued previously that increasing temperature makes such transitions more strongly
first-order \cite{Janoschek13,Fernandes12,Fernandes19}. 

We should note here that we have also assumed that intermediate phases always preserve symmetries that are remaining in the final phase i.e., no
reentrant transitions are considered. We argue that this assumption, along with other requirements discussed above, are natural for analyzing the $J$-$Q_6$
model, and other models satisfying the above criteria may have an even richer set of phases and symmetry breaking paths according to Fig.~\ref{fig:BigMap}.
The order graph provides us with a compact way of enumerating all the possibilities, restricting the possible topology of the phase diagram with multicritical points. 
Furthermore, we can know that phase transitions connecting faraway order graphs will always be strongly first-order, 
and vice-versa: it provides good reasons to expect weak first-order transitions to respect the paths discussed here.

\subsection{Unified order parameter for direct $D_4$ symmetry breaking and other transitions}
\label{sec:6Dim}

Here, we further analyze the phase diagram of the $J$-$Q_6$ model by looking into the unified order parameter for PVBS and AVBS phases. 
As we mentioned in Sec.~\ref{sec:ordergraph} and further explain in Appendix \ref{app:OrderGraph}, 
the order graph can be used to construct a unified order parameter that captures both PVBS and AVBS order. 
The unified order parameter, which we name $V$, is a six-dimensional vector and takes nonzero values in both the PVBS and AVBS phases, 
but in a different way. In the two phases, the histogram of the order parameter will have 4 and 8 peaks, 
according to the degeneracy of the phases. The unified order parameter is essentially the 
direct sum of the plaquette order parameter $\Pi$ and the alternating order parameter $A$,
defined in Eq.~(\ref{Piadefs}) and (\ref{eq:AVBSop}), respectively;
\begin{equation}
V=(8\Pi_0/7, 8\Pi_1/7, 2A_{0,0}, 2A_{0,1}, 2A_{1,1}, 2A_{1,0}).
\label{vorderdef}
\end{equation}
Here the factors $8/7$ and $2$ are chosen to normalize all the components to become 1 in an ideal AVBS state that is perfectly dimerized.

\begin{figure}[t]
\includegraphics[width=8.5cm]{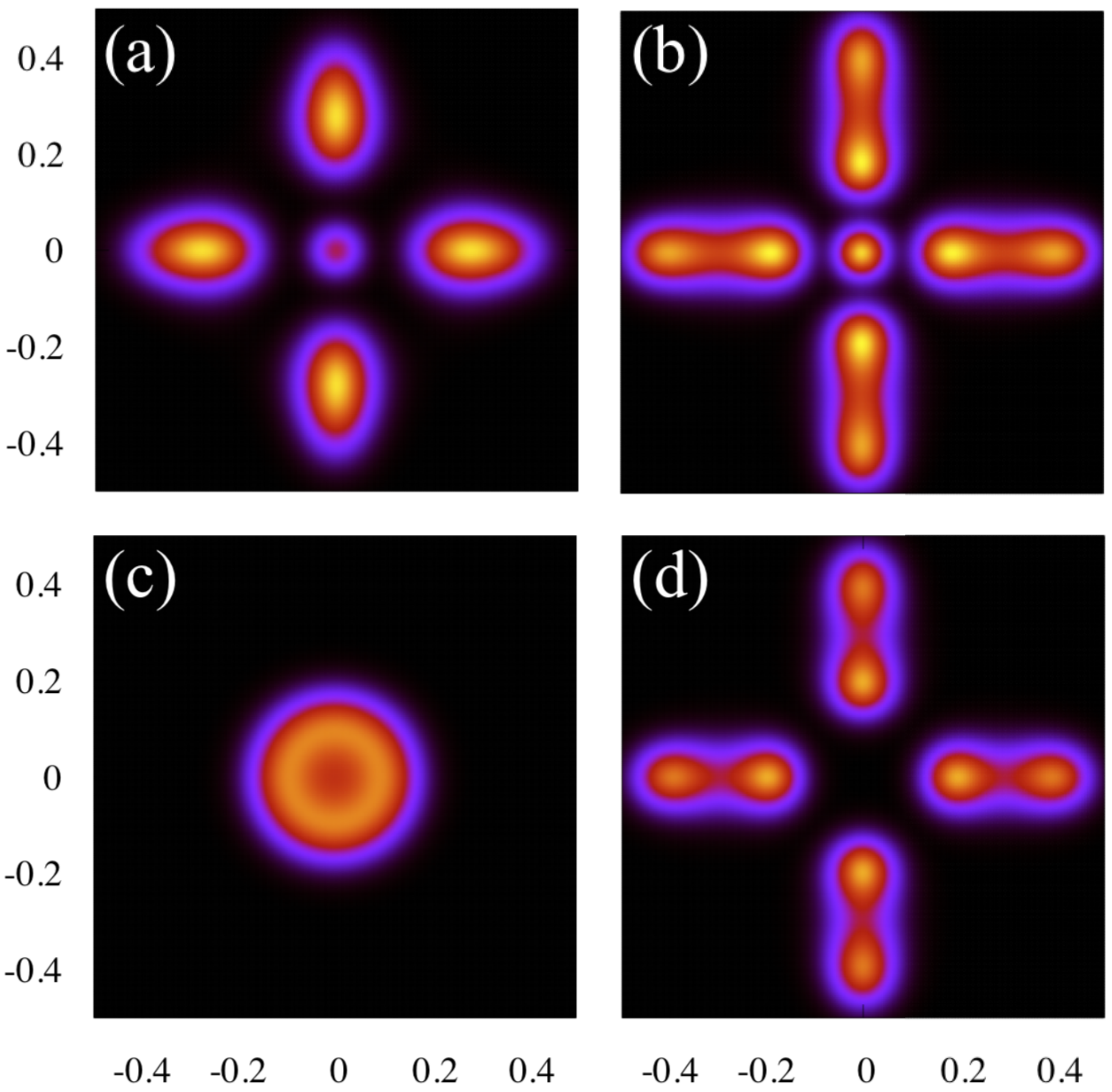}
\caption{\label{fig:6DHist} 
Probability distributions of the unified order parameter $V$ defined in Eq.~(\ref{vorderdef}), projected into $P(X,Y)$, with the components $X$
and $Y$ defined in Eq.~(\ref{xydef}). The system size is $L=32$. (a) At the first-order transition point $Q=0.9$, $T=0.75$ into the PVBS phase 
(b) At the first-order transition point $Q=0.95$, $T=0.74$ into the AVBS phase. (c) Slightly within the PVBS phase after the DQC point $Q=0.28$, $\beta=L$.
(d) Slightly within the AVBS phase $Q=0.94$, $\beta=L$.}
\end{figure}

For visualization, we project $V$ into two dimensions in a way that clearly shows the eight-fold degeneracy. 
The following projection results in a histogram similar to the depiction of the order graph in Fig.~\ref{fig:AltVBSGraph}:
\begin{equation}
\begin{pmatrix}
X\\
Y
\end{pmatrix}
=
\begin{pmatrix}
1 & 0 & 1 & 0 & 1 & 0  \\
0 & 1 & 0 & 1 & 0 & 1  
\end{pmatrix}
V^{\rm T}.
\label{xydef}
\end{equation} 
Fig.~\ref{fig:6DHist} shows the distribution $P(X,Y)$, where in panels (a) and (b)
we can observe the coexistence of thermal paramagnetic and VBS states and distinguish the type of
VBS the system transitions into from the number of peaks. The fact that Fig.~\ref{fig:6DHist}(a) has five peaks, similar to
Fig.~\ref{fig:FinTHist}(a), shows that the phase coexistence at $Q=0.9$ is indeed between the paramagnetic phase (center peak) and the PVBS phase
(four surrounding peaks). In contrast, in Fig.~\ref{fig:6DHist}(b) there are nine peaks in total at $Q=0.95$, implying a phase coexistence between the
paramagnetic phase and the eight-fold degenerate AVBS phase. We can also see that the quantum fluctuation induced by decreasing $Q$ is connecting the
domains into four pairs in the order parameter space. Because of this, the four peaks of Fig.~\ref{fig:6DHist}(a) have a rather extended shape in the
radial direction as opposed to Fig.~\ref{fig:FinTHist}(a). Fig.~\ref{fig:6DHist}(c) shows emergent U(1) symmetry at very low temperature (essentially
in the ground state; $T=1/L$) in the vicinity of the AFM--PVBS transition, just as in Fig.~\ref{fig:DQCHistogram}(c) and Fig.~\ref{fig:DQCHistogram_DxDy}(c')
but with the distribution projected in a different way. In Fig.~\ref{fig:6DHist}(d) we can see that the strongest fluctuation path inside the AVBS
phase are still within the four groups of two states.

By observing histograms of the unified order parameter
we can also confirm the corresponding order graphs at each point in the phase diagram. 
For example, the phase coexistence in Fig.~\ref{fig:6DHist}(a) suggests a transition between 
the order graphs $(\mathsf{12})$ and $(\mathsf{1})$ of Fig.~\ref{fig:BigMap}. As we argued in the previous section, 
in principle it is possible that the transition between these two order graphs is continuous, but it turns out that it
is actually first-order in this case. 
Fig.~\ref{fig:6DHist}(b) corresponds to a transition between $(\mathsf{12})$ and $(\mathsf{\varnothing})$,
and we have argued in the previous section that this must be a strong first-order transition since the two order graphs are not
directly connected in Fig.~\ref{fig:BigMap}.

\begin{figure}[t]
\includegraphics[width=8.5cm]{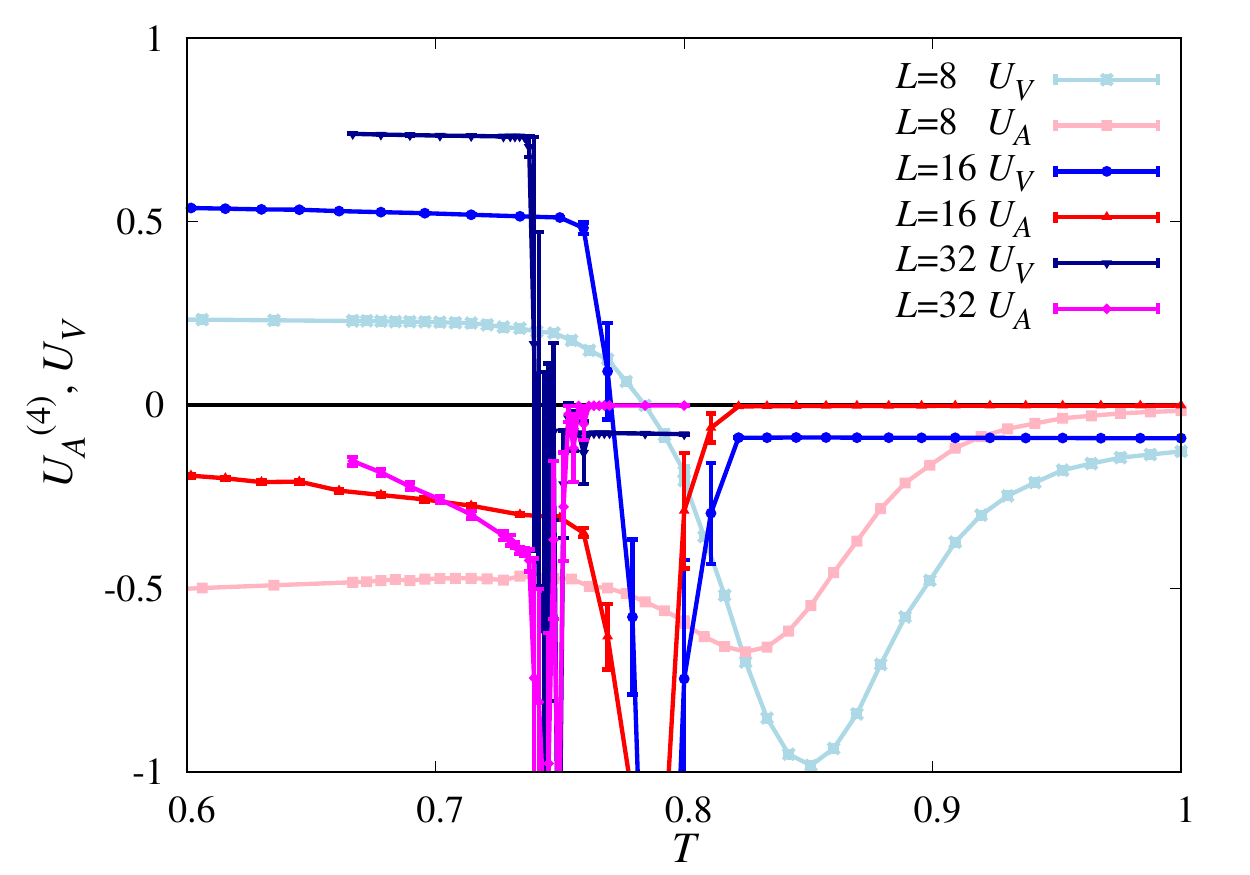}
\caption{\label{fig:verTfinT} 
Temperature dependance of the Binder cumulants of the unified order parameter, $U_V$, and the alternating order parameter, $U_A^{(4)}$,
at $Q=0.95$ for different system sizes.}
\end{figure}

Finally, in Fig.~\ref{fig:verTfinT}, we can see that the Binder cumulant of the unified order parameter, $U_V=4-3\langle V^4\rangle/\langle V^2\rangle^2$,
behaves far better than the cumulant of just the AVBS order parameter, $U_A^{(4)}=3-2\langle A^4\rangle/\langle A^2\rangle^2$, when investigating the
paramagnetic--AVBS transition with periodic boundary condition. These cumulants are again normalized
so that they must necessarily converge to $1$ as $L \to \infty$ in the ordered phase (since both order parameters have a nonzero value with a
fixed absolute value). However, even for $L=32$ we have a completely negative $U_A^{(4)}$ in the ordered phase, while we see a more rapid convergence of
the final value of $U_V$ respect to the system size (though still not reaching very close to $1$).

We close this section by noting that the construction of the unified order parameter and the analysis of the symmetry breaking paths with order graphs
should be in principle possible with standard Landau theory. However our approach with the order graph based on strengths of fluctuation paths (combined
with symmetries) provides a more intuitive and compact way of understanding multi-stage discrete symmetry breaking. We expect it to be a useful approach
in many other cases beyond the eight-fold degenerate AVBS paths analyzed here.

\section{Discussion}
\label{sec:Discussion}

In summary, the $J$-$Q_6$ quantum spin Hamiltonian introduced here exhibits several interesting physical phenomena. To begin with, it is,
to our knowledge, the first sign-problem free 2D model hosting a four-fold degenerate PVBS ground state, thus enabling detailed QMC studies of the
AFM--PVBS transition---a DQC candidate alongside the often studied AFM--CVBS transition. Instead of a continuous transition, we here found a clearly
first-order transition, the coexistence state of which hosts an enlarged symmetry of the order parameter, with the combined three-component AFM
order parameter and the two-component PVBS order parameter forming an SO(5) symmetric psudo-vector. 
We further discovered that the PVBS phase can be considered as a vestigial phase; 
an intermediate phase on the way to an eight-fold degenerate AVBS state which breaks a $\mathbb{Z}_2$ symmetry in the already
$\mathbb{Z}_4$-symmetry broken PVBS phase. The two-stage breaking of two discrete symmetries stimulated us to introduce the {\it order graph}
as an intuitive tool for analyzing such phase transitions. 

We here further discuss some of the most intriguing aspects of our findings. In Sec.~\ref{sub:spinons} we discuss the nature of spinons in the
PVBS state. At variance with the fracton scenario \cite{You19}, we argue that the spinons are mobile and similar to those in a CVBS state. In
Sec.~\ref{sub:symmfirstorder} we further discuss possible reasons for the first-order nature of the AFM--PVBS transition, using a different model
with a first-order AFM--CVBS transition for comparison; the {\it columnar} $J$-$Q_6$ model. In Sec.~\ref{sub:cuprates} we discuss the intriguing analogy
between the symmetry-enhanced AFM--PVBS transition and the superconducting transition in the SO(5) theory of the high-$T_c$ cuprates.
We discuss future prospects in Sec.~\ref{sub:future}.

\subsection{Domain-wall structure and spinons}
\label{sub:spinons}

\begin{figure}[t]
\includegraphics[width=8cm]{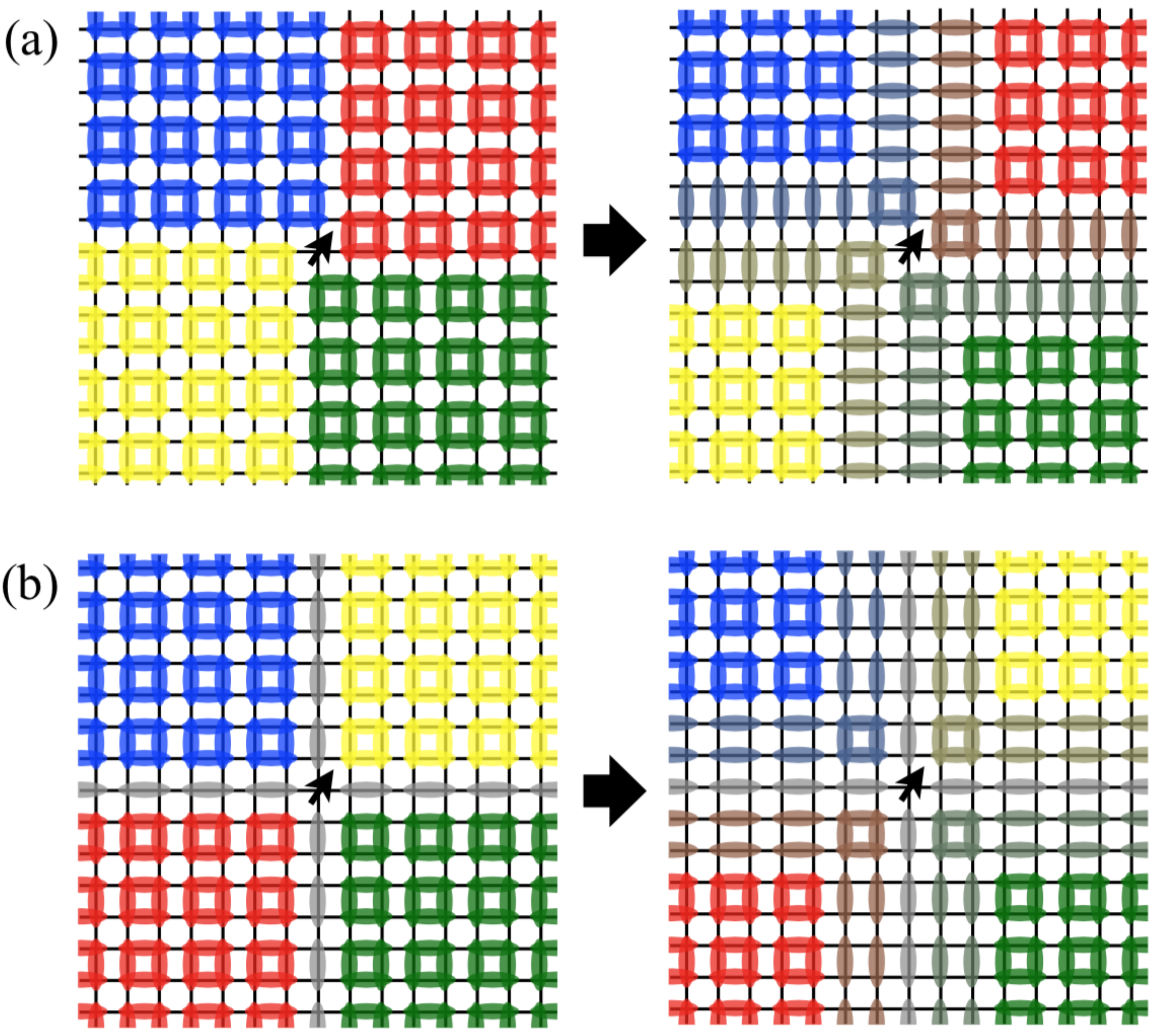}
\caption{\label{fig:TwoDomainWalls} 
  Depictions of spinons forming at the nexus of domain walls in a PVBS state in two possible ways; with dimers perpendicular and parallel to
  the walls in (a) and (b), respectively. Extreme cases of the thinnest domain walls are shown to the left,
  and to the right the walls are widened by additional dimers. Dimer pairs can dynamically convert into resonating plaquettes,
  and vice versa, thus allowing the domain wall to fluctuate. We have here drawn plaquettes closest to the unpaired spin in order to
  clearly show that all four phases meet at the spinon. Fluctuations leading to changes in the spinon location are illustrated in
  Fig.~\ref{fig:SpinonsMoving}.
  In both (a) and (b), the relative relations of the domain colors are the same, e.g., red and blue differ by a $y$ direction shift in the PVBS pattern.}
\end{figure}

\begin{figure}[t]
\includegraphics[width=8.4cm]{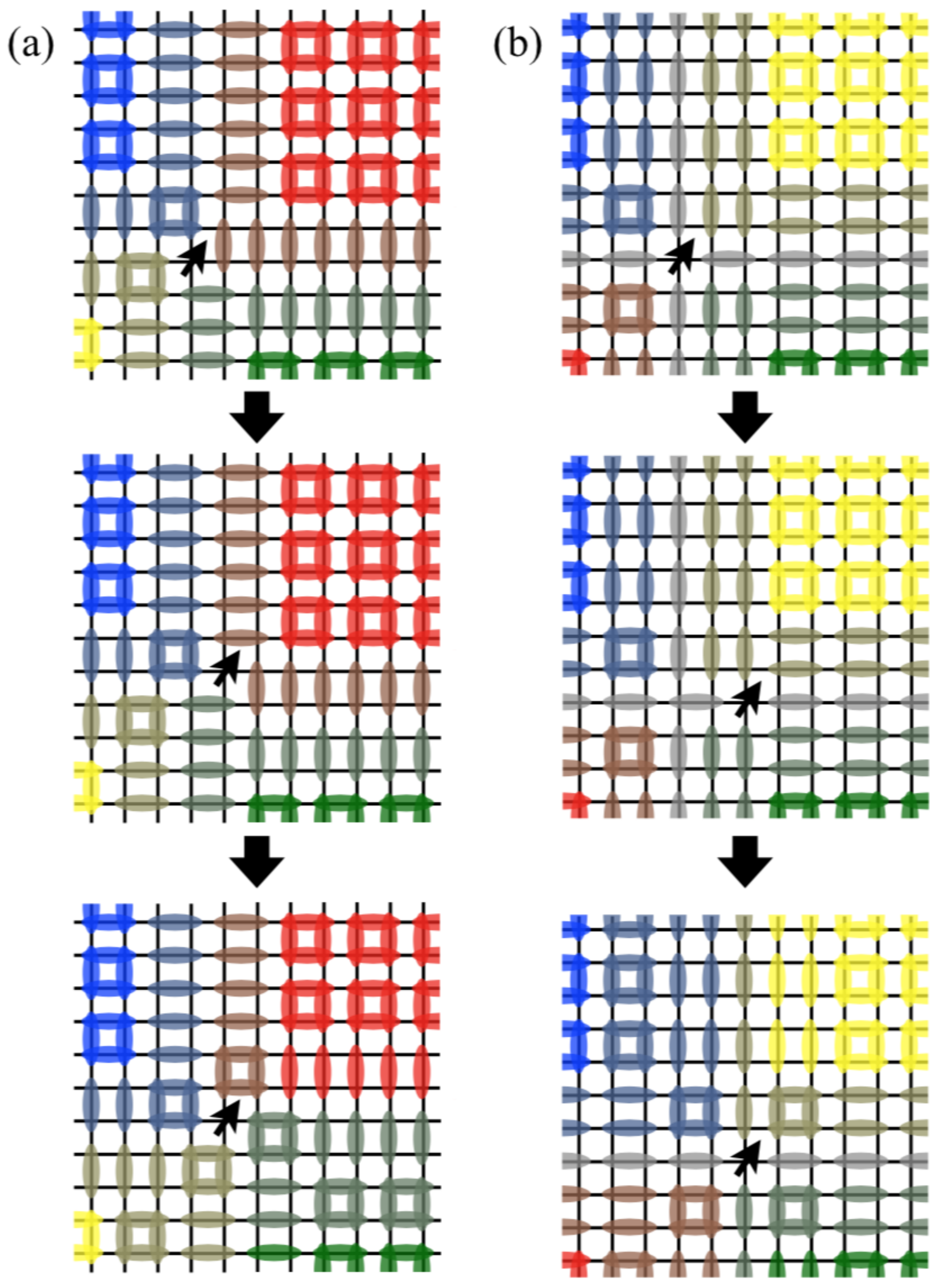}
\caption{\label{fig:SpinonsMoving}
Illustration of quantum fluctuations allowing a spinon to move in two different cases, corresponding to (a) and (b) in Fig.~\ref{fig:TwoDomainWalls}. 
Colors were changed to reflect the final domain wall configuration, while the grayed singlets reflect the initial domain wall position.
}
\end{figure}

As we discuss in detail in Appendix \ref{app:d4}, in terms of the effective symmetry being broken the CVBS and PVBS states are dual and equivalent. 
There is still a possibility that the domain wall structure and types of spinons emerging as a defects in these two states may be different,
and this could be the reason for the first-order transition found here. Recently, it was argued that the AFM--PVBS transition should in general
be first-order and not a continuous DQC transition \cite{You19}, as a consequence of the spinons in the PVBS phase being immobile fractons. At first sight,
this conjecture appears to be confirmed by our results. However, we do not believe that the spinons in the PVBS considered here are fractons,
and we will argue that the fracton scenario in general is very unlikely in quantum magnets.

In Ref.~\cite{You19} the singlet plaquettes of the PVBS state were treated as rigid objects, and it was pointed out that a spinon (a site
not belonging to any plaquette) forming at the nexus of four domain walls in such a state cannot move because the constraints prohibit local
fluctuations that gradually shift the spinon together with some of the plaquettes. Our primary objection to this scenario is that the singlet plaquettes
in an actual PVBS state in a quantum spin system are not rigid objects, except perhaps in some extreme case that is not likely to be realized in
practice with a naturally-arising spin Hamiltonian. The PVBS state hosts various quantum fluctuations, and when viewed in the valence-bond basis the singlet
plaquettes are resonating pairs of horizontal and vertical valence bonds on $2\times 2$ sites on the square lattice. 
There are also longer valence bonds, though they likely are less important for capturing the essential physics. 
In a minimal effective description, the PVBS state should be viewed as a mixture of rigid
plaquettes and dimers, and there must be processes converting a plaquette into a pair of dimers and vice versa. 
There are then two possible types of domain walls, and, therefore, two kinds of spinons. We illustrate these in Fig.~\ref{fig:TwoDomainWalls}. 
As in the DQC scenario \cite{Senthil04a,Levin04},
in the same way as a domain wall between CVBS phases can be regarded as consisting of plaquette singlets in a PVBS pattern, 
a domain wall between PVBS states should be primarily made up out of dimer singlets in CVBS pattern. 
Whichever type of domain wall is realized in a given system should depend on the details of the Hamiltonian.

In either type of domain wall, in the presence of dimers the spinon can also move locally inside the domain wall through processes where
a dimer and a spinon are shifted together as depicted in Fig.~\ref{fig:SpinonsMoving}. 
In both cases, the first step only involves the spinon itself moving as a consequence of the action of a single $J$-term (singlet projector) $P(i,j)$. 
Note that, with a bipartite Hamiltonian, the spinon is constrained to move only within a given sublattice. 
The second step exemplifies how the domain walls around the unpaired spin can adapt in order to form the configuration 
shown in Fig.~\ref{fig:TwoDomainWalls} again with the new spinon position. 
All the changes in this second step are plaquette-dimer conversions, or a pair of dimers resonating to point in the other direction. 
The dimer resonance is easily induced by a single $J$-term, and plaquette-dimer conversions happen even more easily because of the large overlap between the two states. 
Considering these processes, the domain wall itself can fluctuate, and, thus, 
the collective object (the spinon) consisting of the meeting point of the domain walls and the unpaired spin 
can move through the lattice and should not be regarded as a fracton.

\subsection{First-order transitions and emergent symmetry}
\label{sub:symmfirstorder}

The arguments in the preceding section make the fracton scenario unlikely as an explanation for the first-order AFM--PVBS transition,
and we therefore discuss other possible reasons here. First, we note that the DQC scenario itself does not guarantee that the AFM--VBS
transition is necessarily continuous (with the VBS being either a CVBS or a PVBS); the claim is that the transition is generic and not a fine-tuned multi-critical
point, but first-order transitions for some Hamiltonians can never be excluded \cite{Senthil04a}. Second, the presence of an emergent
SO(5) symmetry observed here at the transition point may suggest that the system is close to a DQC point with such a higher symmetry
\cite{Wang17}, so that expected perturbations destroying that symmetry is weak and only observable on large lattices. Another possibility
is that the DQC phenomenon is connected to first-order transitions with asymptotically exact emergent symmetries. Both of these scenarios were discussed
in the context of first-order transitions with apparent O(4) symmetry in systems with AFM--PVBS transitions where the PVBS state is
only two-fold degenerate \cite{Zhao19,Serna19}. While an exact emergent symmetry at a first-order transition would be outside previous
expectations, this scenario has some support also in a recent work claiming exact emergent supersymmetry at certain phase transitions
with fermionic and bosonic degrees of freedom \cite{Yu19}. Here we do not provide any definite conclusions on the root causes of these
unusual first-order transitions, but we address a related issue using illuminating computational results.

The first issue concerns the possibility of first-order transitions into the columnar CVBS state. So far, first-order transitions have
been observed in spin-anisotropic $J$-$Q$ models in which the AFM order parameter is O(2) symmetric, where the discontinuities weaken
as the anisotropy is decreased and there may be a change to a continuous transition at some critical value of the anisotropy \cite{Qin17,Ma19a}.
In spin-isotropic systems such as the $J$-$Q_2$ and $J$-$Q_3$ models with a columnar arrangement of the singlet projectors, no clear-cut
signs of first-order transitions have been observed, though there are unusual scaling corrections \cite{Sandvik10a,Shao16} that have also been
interpreted as a consequence of a weak first-order transition \cite{Jiang08,Chen13,Wang17,Ma19,Nahum19}. By introducing other types of $Q$ interactions in
a $J$-$Q_3$ model with columnar $Q_3$ interactions, it was recently found that the DQC point can evolve to a clearly first-order transition \cite{Zhao20}.
It is also interesting to study columnar $J$-$Q_n$ models with $Q$ terms containing more than $n=3$ singlet projectors. Here we consider the $n=6$
case, i.e., the same number of projectors as in the model studied previously in this paper but with a different spatial arrangement of the projectors.

\begin{figure}[t]
\includegraphics[width=7.5cm]{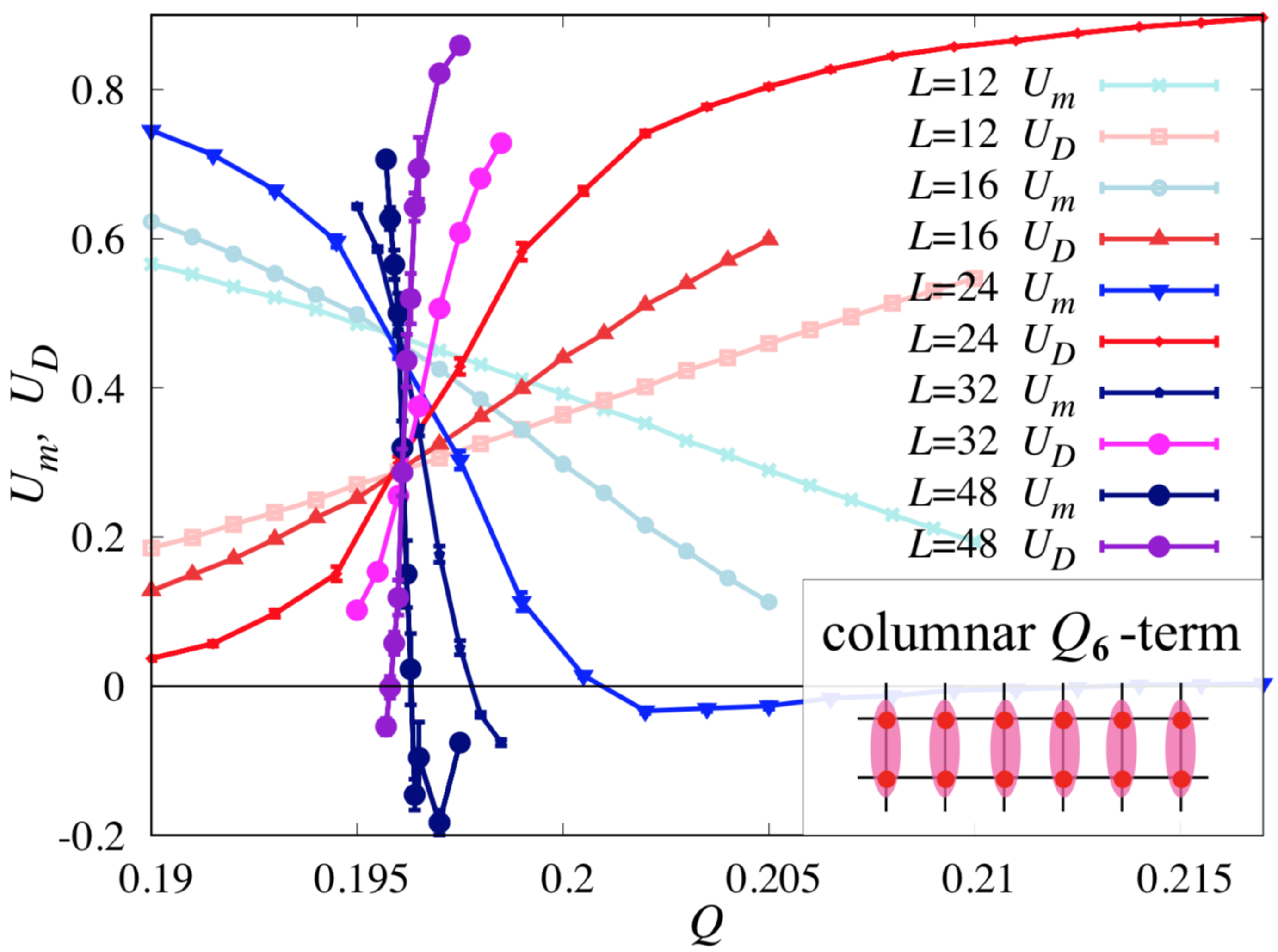}
\caption{\label{fig:ColumnarJQ6Transition} 
Binder cumulants of the AFM and and CVBS order parameters of the the columnar $J$-$Q_6$ model.
The inset shows one of the $Q_6$-terms in the model; all translations of this operator and its $\pi/2$ rotated version are included
in the Hamiltonian, defined in analogy with Eq.~(\ref{eq:Ham}).}
\end{figure}

As shown in Fig.~\ref{fig:ColumnarJQ6Transition}, the columnar $J$-$Q_6$  model exhibits CVBS order for large $Q$, but, unlike the columnar $J$-$Q_2$
and $J$-$Q_3$ models, the AFM--CVBS transition here is strongly first-order. Evidence for this type of transition is seen in the CVBS Binder cumulant,
which in Fig.~\ref{fig:ColumnarJQ6Transition} develops increasingly negative peaks as the system size increases. We also observe (not shown here) coexisting
finite AFM and VBS order parameters similar to Fig.~\ref{fig:OrderParametersAtDQC}. We believe that the first-order transition here is caused by the
large number of coupled spins in the $Q_6$ terms, which may cause the system to nucleate PVBS order locally on large enough ``droplets'' to cause an
instability of the AFM state. 

\begin{figure}[t]
\includegraphics[width=7cm]{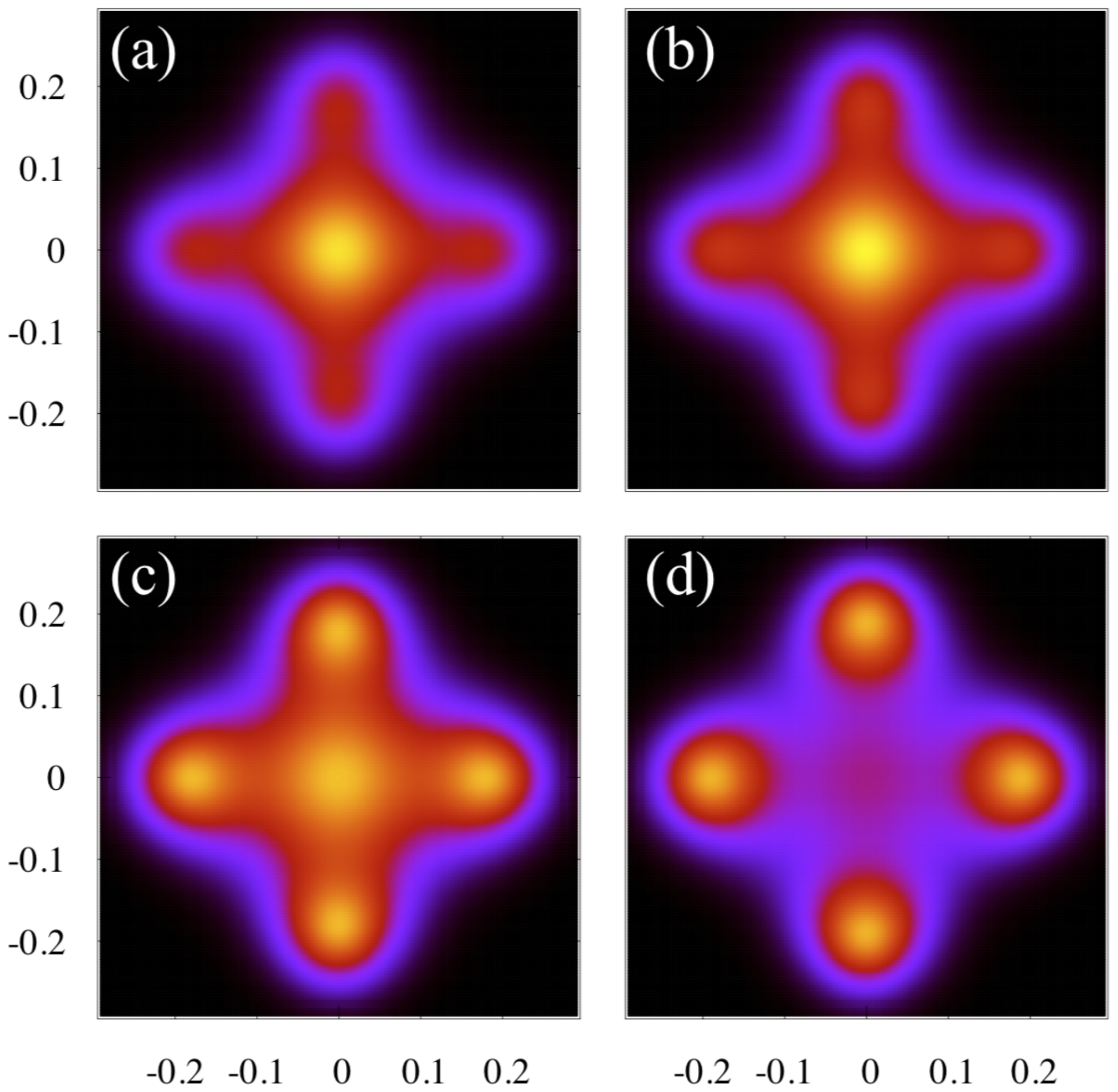}
\caption{\label{fig:ColumnarJQ6Dist} 
Probability distribution of the dimer order parameter $P(D_x,D_y)$ of the columnar $J$-$Q_6$ model at (a) $Q=0.1961$, (b) $0.1962$, (c) $0.1963$,
and (d) $0.1964$, computed in SSE simulations on lattices of size $L=48$.}
\end{figure}

Having identified a first-order AFM--CVBS transition, we can now address the issue of the generality of emergent symmetry at first-order transitions
at AFM--VBS transitions. Since the Binder cumulants in Fig.~\ref{fig:ColumnarJQ6Transition} exhibit the tell-tale signs of conventional phase coexistence,
we do not expect any emergent spherical symmetry. Indeed, there are no signs of emergent higher symmetries in order-parameter distributions.
Fig.~\ref{fig:ColumnarJQ6Dist} shows the distribution of the CVBS dimer order parameter $P(D_x,D_y)$ at four $Q$ values close to the transition
point. Here we see how the distribution evolves from a central peak in the AFM phase (where the CVBS order parameter is peaked at $0$) to a four-peak
distribution reflecting the four-fold degeneracy in the CVBS phase. In a narrow window close to the transition we observe [Fig.~\ref{fig:ColumnarJQ6Dist}(b,c)]
 five peaks reflecting coexistence of the AFM and CVBS phases.

On the one hand, these results for the columnar $J$-$Q_6$ model support a standard nucleation mechanism at play in this case, as no emergent symmetry is
present. On the other hand, in the plaquette $J$-$Q_6$ model we have the same number of interacting spins but do observe emergent symmetry, which speaks
against the standard nucleation scenario. 
The continuous nature of the PVBS--AVBS transition we observe also shows that nucleation does not necessarily follow from a large number of interacting spins. 
Nevertheless, in this case, a plaquette consisting of four spins becomes one degree of freedom in the effective Ising model, 
so the $Q$-term may be considered as an effectively three-body term.

In any case, the emergent symmetry is clearly not universal at general first-order AFM--VBS transitions (see also Ref.~\cite{Desai19}), 
and the possibility also remains that the emergent symmetry at the AFM--PVBS transition 
in the present case only holds up to some length scale larger than the systems studied here. 
If so, the transitions in the two different $J$-$Q_6$ models may be qualitatively the same but differ in the length scale at which conventional 
coexistence is apparent---in which case it still is remarkable and unexpected to have SO(5) symmetry up to such large length scales in the plaquette case. 
It is interesting to note that the two $J$-$Q$ models where
emergent symmetry of the coexistence state has  been observed  both have PVBS states; the two-fold degenerate one in Ref.~\cite{Zhao19} and the
four-fold degenerate case studied in the present paper. This may be an indication of a symmetry-breaking perturbation that exists (or is strong)
at first-order AFM--CVBS but vanishes (or is very weak) at AFM--PVBS transitions.

\subsection{SO(5) theory of high-$T_c$ superconductivity}
\label{sub:cuprates}

The possible emergence of SO(5) symmetry in condensed matter systems has received significant attention due to the fact that phases with
O(3) AFM order often exist adjacent to superconducting phases, which break U(1) symmetry. One may then speculate that the two types of orders
share a common origin in a unified degree of freedom that collectively can rotate between the two phases \cite{Demler04}. The SO(5) scenario for high-$T_c$
superconductivity \cite{Zhang97} postulates that doping away from the half-filled-band, where the cuprate materials are AFM insulators, eventually leads to
a ``flop'' on an SO(5) sphere from the a direction spanned by the three AFM components into the plane spanned by the superconducting order parameter. This
mechanism is very similar to what we have discussed here for the transition of the AFM into the PVBS state.

In the case of the cuprates, to study the SO(5) scenario with numerical simulations, the underlying Hubbard or $t$-$J$ model first has to be projected down
to an effective model (because the electronic models are too difficult to study on sufficiently large length scales), which is bosonic and can be simulated
with QMC methods. Such studies were carried out with the SSE method in Ref.~\cite{Dorneich02}. Though a first-order transition was identified at a critical
doping fraction, the coexistence state did not exhibit SO(5) symmetry, but instead conventional phase coexistence was found. It was argued that long-range
Coulomb interactions might eventually act against phase separation and lead to a quantum-critical point. This scenario thus differs from the first-order
coexistence state with SO(5) symmetry of the $J$-$Q_6$ model at the AFM--PVBS transition, where there is neither phase separation nor conventional
criticality.

Experimentally, the existence of an excitation mode at 41 meV, detected in inelastic neutron scattering experiments, was taken as support of the SO(5)
scenario \cite{Zhang97,Hu01}, but arguments to the contrary have also been voiced \cite{Baskaran98}. Since the $J$-$Q_6$ model has a different, more
exotic kind of coexistence state than what was found in the projected SO(5) model, it would be interesting to study also the dynamical spectral functions
of the $J$-$Q_6$ model (which can be done with SSE simulations in combination with numerical analytic continuation methods \cite{Shao17}) in and close to
the coexistence state. It is tempting to speculate that the SO(5) predictions for the cuprates would come out differently with the exotic coexistence
state, and further studies of the $J$-$Q_6$ model may serve as an analogy where reliable results can be obtained.

\subsection{Future prospects}
\label{sub:future}

Our work presented here illustrates the power of the $J$-$Q$ designer Hamiltonian approach in engineering sign-free Hamiltonians with exotic quantum states
and quantum phase transitions. The results and remaining open issues prompt several possible follow-up studies, some of which we summarize here.

It would clearly be interesting to design a $J$-$Q$ model with a continuous AFM--PVBS transition, which was our initial goal. This may not be so easy
however, as we have so far not even succeeded in creating a PVBS state with less than six singlet projectors (with $Q_4$ terms similar to our $Q_6$
terms in Fig.~\ref{fig:JQ6} leading to CVBS states). If nucleation due to a large number of coupled spins is indeed the root cause of the first-order
transition (which is not clear, as discussed in Sec.~\ref{sub:symmfirstorder}), an even larger number of singlet projectors will likely take us even further
away from the DQC scenario. It would still be worth trying, e.g., $Q_8$ interactions defined on $4\times 4$ lattice sites. We here also note that the
PVBS state can be regarded as a ``superposition of superposition states'' (locally resonating valence bonds), and this specific aspect of the state seems
to be what makes it more difficult to realize a PVBS than a CVBS---at least within the $J$-$Q$ approach. The broader family of models by Kaul \cite{Kaul15},
expanding on the $J$-$Q$ approach, should also be explored more extensively, though our initial attempts have not been successful in producing PVBS states
with less than twelve interacting spins.

Another interesting prospect is to mix different kinds of $Q$ terms and investigate the interplay between interactions favoring PVBS and
CVBS states (in a way similar to what was recently done with competing interactions favoring CVBS and staggered VBS states \cite{Zhao20}).
It may be possible in this way to design a Hamiltonian which can be fine-tuned to have an eight-fold degenerate VBS state, with
both CVBS and PVBS order. In histograms such as those in Fig.~\ref{fig:DQCHistogram} this kind of state would give rise to eight equidistant
peaks, and the phase would break $\mathbb{Z}_8$ symmetry. A potentially continuous transition between the AFM state and such a VBS would be very
interesting, considering that clock-like $\mathbb{Z}_q$ perturbations should be increasingly irrelevant as the number of states $q$ is increased
\cite{Oshikawa00,Shao19}. It may then be possible to observe DQC physics with less anomalies and scaling corrections than with the spin models
studied so far \cite{Shao16}. Recently, a fermionic model argued to have a DQC-type transition without any discrete perturbation was studied
\cite{Liu19}, but in this case the system sizes are very small because of the unfavorable scaling of the fermion determinant QMC algorithm.

The finding here of a first-order transition also in the columnar $J$-$Q_6$ model calls for more systematic studies of columnar $J$-$Q_n$
models. Previous studies for $n=2$ \cite{Sandvik07,Melko08,Jiang08,Sandvik10a,Harada13,Chen13,Block13,Shao16,Suwa16} and $n=3$ \cite{Lou09}
point to continuous transitions, and it would be useful to systematically increase $n$ and find the smallest $n$ for which the transition is
clearly first-order. Such a study might be helpful in determining whether the transitions for small $n$ are truly continuous or only very weakly
first-order.

The order graph approach we have introduced here to analyze multi-stage discrete symmetry breaking should have wide applicability to
both classical and quantum phase transitions. The order graph provides a natural set of possible intermediate phases, and specifies in what ways
they can be adjacent in the phase diagram. It would be interesting to explore possible phase diagrams in other challenging models, e.g., the
classical frustrated spin models discussed in Refs.~\cite{Noh02,Orth12}, which exhibit (possibly multiple-stage) discrete symmetry breaking with
large numbers of states. 

\begin{acknowledgments}
  We would like to thank Morten Christensen, Rafael Fernandes, Ying-Jer Kao, Pranay Patil, Hirokazu Tsunetsugu, and Bowen Zhao for useful discussions. 
  A.W.S. was supported by NSF Grant No.~DMR-1710170 and
  by a Simons Investigator Grant. 
  J.T. would like to thank Boston University's Condensed Matter Theory Visitors Program for support. 
  Some computations were carried out on the Shared Computing Cluster managed by Boston University's Research Computing Services.
\end{acknowledgments}

\appendix

\section{Symmetry breaking in the PVBS phase}
\label{app:d4}

The symmetry being broken in the PVBS phase has some subtlety. While it is argued in some studies that the symmetry breaking here is just the same
$\mathbb{Z}_4$ as in the  CVBS phase \cite{Senthil04a, Senthil04b}, alternatively it may appear to be breaking $\mathbb{Z}_2\times\mathbb{Z}_2$, owing
to its translational symmetry breaking in both the $x$ direction and $y$ direction of the square lattice. Here, we clarify the situation by carefully
analyzing the symmetry breakings for the CVBS and the PVBS on the square lattice. We will see that the structure of the coset group defining the 
original and remaining symmetries in both cases are exactly the same, and technically they can all be seen as either $\mathbb{Z}_4$ symmetry breaking or 
$\mathbb{Z}_2\times\mathbb{Z}_2$ symmetry breaking, leading to the potential confusion. In both cases, the spin rotational symmetry is not broken,
so we can ignore the symmetry of the spin degrees of freedom.

Let us start with the CVBS phase. The Hamiltonian originally obeys all the symmetries of the square lattice: $D_4 \rtimes \mathbb{Z}^2$.
The point group is the fourth dihedral group $D_4$ (also called $c4v$) which includes $k\pi/2$ rotations ($k=0,1,2,3$) and reflections with respect to four different axes.
$\mathbb{Z}^2$ corresponds to the translational symmetry of the infinite lattice. In the CVBS phase, the remaining symmetry becomes
$D_2 \rtimes (\mathbb{Z}\times 2\mathbb{Z})$. The usual procedure to determine the symmetry breaking is then to consider the coset
$G/H$ where $G$ is the original symmetry group and $H$ the remaining symmetry group. However, the semidirect
product denoted by $\rtimes$ of two groups is in general not uniquely determined. Therefore, it is meaningless to consider with expressions such as
$\{D_4 \rtimes \mathbb{Z}^2\}/ \{D_2 \rtimes (\mathbb{Z}\times 2\mathbb{Z})\}$ unless the way the two semidirect products are
taken is further specified. In the present case, the situation is made most clear by considering the equivalence class among symmetric transformations
according to how they act on the order parameter.

We here only consider reflections with the axis going through sites and rotations with the center located on a particular site.
Other choices of axes or centers could be realized as a combination of the restricted reflections and/or rotations combined with
translations. We construct the following equivalence classes as elements of the relevant symmetry group:

\begin{itemize}
\item[$\bm{I}$:] 	All translations which are even in both ($x$ and $y$) directions $T_{(0,0)}$, 
			translations even in $x$ direction and odd in $y$ direction followed by a reflection by the $x$ axis $M_xT_{(0,1)}$, 
			analogous in the different direction $M_yT_{(1,0)}=R_{\pi}M_xT_{(1,0)}$, 
			or a $\pi$ rotation after a translation odd in both directions $R_{\pi}T_{(1,1)}$. 
\item[$\bm{a}$:]	Rotation of $\pi/2$ after any transformation in $\bm{I}$, i.e. $R_{\pi/2}T_{(0,0)}$, $R_{\pi/2}M_xT_{(0,1)}$,
                        $R_{3\pi/2}M_xT_{(1,0)}$, and $R_{3\pi/2}T_{(1,1)}$. 
\item[$\bm{a^2}$:]      Rotation of $\pi$ after any transformation in $\bm I$, i.e. $R_{\pi}T_{(0,0)}$, $R_{\pi}M_xT_{(0,1)}$, $M_xT_{(1,0)}$, and $T_{(1,1)}$. 
\item[$\bm{a^3}$:]      Similarly, $R_{3\pi/2}T_{(0,0)}$, $R_{3\pi/2}M_xT_{(0,1)}$, $R_{\pi/2}M_xT_{(1,0)}$, and $R_{\pi/2}T_{(1,1)}$. 
\item[$\bm{b}$:]	Reflection with respect to a $\pi/4$ angle axis $M_+$ followed by any transformation in the equivalence class $\bm I$, i.e.,
                        $T_{(0,0)}M_+=R_{3\pi/2}M_xT_{(0,0)}$, $R_{\pi/2}T_{(1,0)}$, $R_{3\pi/2}T_{(0,1)}$, and $R_{\pi}T_{(1,1)}M_+=M_xT_{(1,1)}$. 
\item[$\bm{ab}$:]	Rotation of $\pi/2$ after any transformation in $\bm b$, i.e., 
			$M_xT_{(0,0)}$, $R_{\pi}T_{(1,0)}$, $T_{(0,1)}$, and $R_{\pi/2}M_xT_{(1,1)}$.
\item[$\bm{a^2b}$:]	Rotation of $\pi$ after any transformation in $\bm b$, i.e., 
			$R_{\pi/2}M_xT_{(0,0)}$, $R_{3\pi/2}T_{(1,0)}$, $R_{\pi/2}T_{(0,1)}$, and $R_{\pi}M_xT_{(1,1)}$.
\item[$\bm{a^3b}$:]	Similarly, $R_{\pi}M_xT_{(0,0)}$, $T_{(1,0)}$, $R_{\pi}T_{(0,1)}$, and $R_{3\pi/2}M_xT_{(1,1)}$.
\end{itemize}
Here, translation of $(s,t)$ in the $(x,y)$ direction is denoted as $T_{(s,t)}$ and we categorize these according to the parity of $s$ and $t$ (mod 2).
$M_x$ is reflection with respect to the $x$ axis, and $R_\theta$ is $\theta$-rotation around the site at the origin. Any symmetry transformation of the
square lattice can be written in the form $R_\theta M_x^{\{0, 1\}} T_{(s,t)}$, and thus belongs to one and only one of the above equivalence classes. 

The above eight equivalence classes form a group isomorphic to $D_4$, which exactly corresponds to the automorphism group of the order graph for
four CVBS states---the order graph is described in Sec.~\ref{sec:OrderGraph} the main text and in further detail in Appendix \ref{app:OrderGraph}.
This is also equivalent to saying that, within each equivalence class, the transformations have exactly the same effect on the order parameter. For example,
all transformations in $\bm a$ correspond to a $\pi/2$ rotation in the order parameter space of ${D}$ and $\Pi$, and $\bm b$ results in a reflection along
a diagonal axis. When the system is in the CVBS phase, the remaining symmetry is $\{\bm I, \bm{ab}\}$ or $\{\bm I, \bm{a^3b}\}$ depending on which of
the four states the system is in.  

An important point is that neither $\{\bm I, \bm{ab}\}$ nor $\{\bm I, \bm{a^3b}\}$ is a {\it normal} subgroup of $D_4$. Therefore, the symmetry being
broken here is technically $D_4/\mathbb{Z}_2$, which is just a coset and not a quotient group. In the most strict sense, the symmetry breaking is neither
a $\mathbb{Z}_4$ nor $\mathbb{Z}_2 \times \mathbb{Z}_2$ symmetry breaking. Nevertheless, we can still express the coset $D_4/\mathbb{Z}_2$ in a way that
{\it looks like} $\mathbb{Z}_4$: e.g. the left coset is
\begin{eqnarray}
  && D_4/_{\mathrm{L}}\{\bm{I}, \bm{ab}\} =  \\
  &&~~~\bigl\{ \{ \bm{I}, \bm{ab} \}, \{ \bm{a}, \bm{a^2b} \}, \{ \bm{a^2}, \bm{a^3b} \}, \{ \bm{a^3}, \bm{b} \} \bigr\}, \nonumber
\end{eqnarray}
and we can choose the representatives as $\{\bm{I, a, a^2, a^3}\}$. This corresponds to the fact that we can express the four degenerate CVBS states as 
\begin{eqnarray*}
&&\Bigl(|D_x>0\rangle,~~ R_{\pi/2}|D_x>0\rangle = |D_y<0\rangle,\\
&&R_{\pi}|D_x>0\rangle = | D_x<0\rangle,~ R_{3\pi/2}|D_x>0\rangle = |D_y>0\rangle\Bigr),
\end{eqnarray*}
just like a $\mathbb{Z}_4$ group, because $R_{\pi/2}\in\bm a$, $R_{\pi}\in\bm{a^2}$, and $R_{3\pi/2}\in\bm{a^3}$. 
However, this is just because we chose an expression that is based on the $\mathbb{Z}_4$ subgroup of $D_4$ for the left coset
$D_4/_{\mathrm{L}}\mathbb{Z}_2$. 
We can also choose the representatives to be $\{ \bm{I, a^2b, a^2, b}\}$ 
which now corresponds to expressing the four degenerate states as 
\begin{eqnarray*}
&&\Bigl(|D_x>0\rangle,~~ R_{\pi}M_+|D_x>0\rangle = | D_y<0\rangle,\\
&& R_{\pi}|D_x>0\rangle = |D_x<0\rangle,~ M_+|D_x>0\rangle = |D_y>0\rangle\Bigr),
\end{eqnarray*}
which has a $\mathbb{Z}_2\times\mathbb{Z}_2$ structure. This ambiguity directly originates from the fact that $D_4/\mathbb{Z}_2$ is a coset, 
and either representations are equally valid. Thus, the terms ``$\mathbb{Z}_4$ symmetry breaking" and ``$\mathbb{Z}_2\times\mathbb{Z}_2$ symmetry breaking" are
both valid in the sense that subgroups of the original symmetry group $G$ of the lattice that are isomorphic to $\mathbb{Z}_4$ or
$\mathbb{Z}_2\times\mathbb{Z}_2$ are completely broken in the ordered phase.

The situation is almost exactly the same for the PVBS phase, including the way the symmetry transformations are divided into equivalence classes.
The only difference is that the remaining symmetry in the PVBS phase is either $\{\bm I, \bm b\}$ or $\{\bm I, \bm{a^2b}\}$, depending on the symmetry
broken state. Similarly to the case of CVBS phase, neither of the two are normal subgroups of $D_4$ and allow multiple representations
of $D_4/_{\mathrm{L}}\mathbb{Z}_2$. In this case, the left coset is
\begin{equation}
D_4/_\mathrm{L}\{\bm{I}, \bm{b}\} =  \bigl\{ \{ \bm{I}, \bm{b} \}, \{ \bm{a}, \bm{ab} \}, \{ \bm{a^2}, \bm{a^2b} \}, \{ \bm{a^3}, \bm{a^3b} \} \bigr\},
\end{equation}
and the $\mathbb{Z}_4$ and $\mathbb{Z}_2\times\mathbb{Z}_2$ representations are 
$\{ \bm{I, a, a^2, a^3}\}$ and $\{\bm{I, ab, a^2, a^3b}\}$, respectively.
We can represent the four different PVBS states as either $\mathbb{Z}_4$-like using the representation $\{ \bm{I, a, a^2, a^3}\}$:
\begin{eqnarray*}
&&\Bigl(|\Pi_0>0\rangle,~~ R_{\pi/2}|\Pi_0>0\rangle = | \Pi_1>0\rangle,\\
&&R_{\pi}|\Pi_0>0\rangle = |\Pi_0<0\rangle,~ R_{3\pi/2}|\Pi_0>0\rangle = |\Pi_1<0\rangle\Bigr),~~
\end{eqnarray*}
or $\mathbb{Z}_2\times\mathbb{Z}_2$-like using the representation $\{\bm{I, ab, a^2, a^3b}\}$:
\begin{eqnarray*}
&&\Bigl(|\Pi_0>0\rangle,~~ M_x|\Pi_0>0\rangle = | \Pi_1>0\rangle,\\
&&R_{\pi}M_x|\Pi_0>0\rangle = |\Pi_1<0\rangle,~ R_{\pi}|\Pi_0>0\rangle = |\Pi_0<0\rangle\Bigr),
\end{eqnarray*}
analogously to the CVBS case. 

We conclude that the symmetry breaking of the CVBS and PVBS phases have exactly the same structure: $D_4/\mathbb{Z}_2$.
We have shown that this becomes apparent when we group all the symmetry transformations into equivalence classes according to how they act
on the order parameters. This approach is practically very simple if one starts from the automorphism group of the order graph. 
The difference between the two phases amounts to whether the remaining symmetry is 
$\{\bm I, \bm b\} \&\{\bm{I, a^2b}\}$ or $\{\bm I, \bm {ab}\} \&\{\bm{I, a^3b}\}$.
The ``duality" between CVBS and PVBS is most clearly understood when we observe that the remaining symmetries of $\mathbb \mathbb{Z}_2$ for either
of the phases are actually isomorphic via automorphisms, e.g., $\bm{a^{\prime}=a, b^{\prime}=ab}$ of the group $D_4$ itself. 
To be more precise, this renaming will not affect the relations of the elements in the group (e.g. $ \bm {b^\prime}\bm{a^\prime} = \bm{a^{\prime 3}b^\prime}$ just as $\bm{ba}=\bm{a^3b}$), 
and simply corresponds to changing which axis to choose for the reflection represented by $\bm b$. In this sense, we can say that the effective symmetry
breaking in the CVBS and PVBS phases are exactly the same, which can be expressed with $\mathbb{Z}_4$ or $\mathbb{Z}_2\times\mathbb{Z}_2$.

Note that the ferromagnetic clock model with four states exactly follows the above classification of symmetries, and can actually be seen as both
$\mathbb{Z}_4$ or $\mathbb{Z}_2\times\mathbb{Z}_2$ symmetry breaking. This is consistent with the fact that when we consider a ``hard'' clock model, where
the spins can only point in discrete direction (instead of using a cosine potential), the model reduces to two decoupled Ising models, which obviously
should be able to be seen as a $\mathbb{Z}_2\times\mathbb{Z}_2$ symmetry breaking. 
This ambiguity essentially comes from the peculiarity of the group $D_4$ that does not appear in $D_q$ with $q>4$. 
Thus, for clock models with the number of states $q$ larger than $4$, the
broken symmetry can only be classified as $\mathbb{Z}_q$.

\section{Rigorous construction of the order graph and its embedding in Euclidean space}
\label{app:OrderGraph}

Here, we precisely define the construction of the order graph introduced in Sec.~\ref{sec:OrderGraph} and also explain exactly what we mean by a
faithful embedding of it in Euclidean space. The topic of embedding graphs into Euclidean spaces of broad interest in fields ranging from machine
learning \cite{ML-GraphEmbedding} to pure graph theory \cite{Erdoes}, but the embedding we consider here has a constraint which has not been considered
in depth before.

Let us consider a discrete symmetry breaking with possibly multiple steps. 
By assuming that it is always a discrete symmetry that is being broken, 
we will have only finitely many ordered (or disordered) states at any point. 
We consider a situation where we lower the (classical or quantum) fluctuation, 
starting from a disordered phase with no spontaneous symmetry breaking. 
Let us assume that, after one or several symmetry breakings there are $M$ degenerate ordered states in the final phase, 
and that we already {\it know} those $M$ states.
\vspace{0.2cm}\\
{\bf Definition: Order Graph}\\
%\begin{mydefinition}
An order graph $G=(V, E)$ for a multiple discrete symmetry breaking consists of the 
vertex set $V$ where each of the vertices corresponds to one of the $M$ ordered states in the final phase, 
and an edge {\it function} $E$. 
The edge function $E : V^2\rightarrow \{1, 2, \ldots, K\}$ is a symmetric function that tells what type of 
bond a pair of vertices have between them. 
The edge function will be defined according to the strength of fluctuation between the two ordered states. 
\vspace{0.2cm}

%\end{mydefinition}
Whereas the usual definition of a graph involves the edge {\it set} $E \subset V^2$, 
our definition is a slight extension of it since now $E$ is a function that specifies 
what type of relation two ordered states have out of $K$ possibilities. 
Usual graphs can be considered as the case where $K=2$, corresponding to 
the two possible relations either having or not having an edge. 
The strength of fluctuation between two states $v_1, v_2$ in $V$ 
can be quantified in several different ways. 
One is to compute the domain wall (free) energy between the two states. 
Another way is to evaluate the transition probability from one state to another 
under some local dynamics, e.g. local Monte Carlo updates. In a quantum state,
tunneling amplitudes can be calculated in principle and are reflected in the probability distribution
between peaks in histograms such as those in Figs.~\ref{fig:DQCHistogram} and \ref{fig:DQCHistogram_DxDy}.

All of these methods should classify all $M(M-1)/2$ pairs of states into groups  with the same ``distances". 
Then $E(v_1, v_2)$ will be set to 1 if they have the strongest possible fluctuation, and 2 if it is the next strongest, and so on.
Exactly how the strength of the fluctuations are quantified may in some cases be further refined, but the essential point here
is that we can in principle construct a well-defined function $E$, which orders and classifies the pairs of states according to their
physical fluctuation strengths. For the model we analyze in this paper, if we consider the two-stage discrete symmetry breaking in the
sequence AFM--PVBS--AVBS (or paramagnetic--PVBS--AVBS), then $M=8$ and $K=3$. We only draw edges corresponding to 1 and 2 in
Fig.~\ref{fig:AltVBSGraph} for visualization, and all pairs of vertices without a drawn edge corresponds to the third type. 

An automorphism on the order graph $G$ is a map from $V$ to itself $f: V\rightarrow V$ which satisfies 
\begin{equation}\label{eq:automorph}
\forall v_1, v_2 \in V, ~~
E(v_1, v_2) = E(f(v_1), f(v_2)) . 
\end{equation}
Since this graph is representing the relations between the ordered states, 
which should have exactly the same (free) energy, 
this graph must be {\it vertex transitive}, which means that all vertices are in a way equivalent, i.e., 
\begin{equation}
\forall  v_1, v_2 \in V, ~~ \exists f : \mathrm{automorphism} ~~ s.t.  ~~ f(v_1)=v_2 . 
\end{equation}
The graph also must be {\it edge transitive} as well, since all pairs with the same value of $E$ should be equivalent in the same way, i.e.,
\begin{eqnarray}
\forall  v_1, v_2, v_3, v_4 \in V, ~~ E(v_1, v_2)=E(v_3, v_4)~~~~~~~~~~~~~~~~~~\nonumber\\
\Rightarrow \exists f : \mathrm{automorphism}   ~~ f(v_1)=v_3, ~ f(v_2)=v_4. ~~~~~~~
\end{eqnarray}
Any symmetric transformation to a state that preserves energy corresponds to 
an automorphism of the order graph. 
To demonstrate this, let us take the model we have studied in the main text as an example. 
The order graph is as shown in Fig.~\ref{fig:AltVBSGraph}, and we have $M=8$ states 
in the final AVBS phase labeled from $\mathsf{A}$ to $\mathsf{D}^{\prime}$. 
If the system locks into the pattern labeled $\mathsf{A}$,  a shift in the $+x$ direction in unit distance
transforms the state into pattern $\mathsf{B}$. The state will become $\mathsf{B}^{\prime}$ if the shift is in the $-x$ direction instead. 
Since the system is not breaking any symmetry between the $+x$ direction and $-x$ direction,
the physical fluctuation between ($\mathsf{A}\leftrightarrow\mathsf{B}$) and ($\mathsf{A}\leftrightarrow\mathsf{B^{\prime}}$) should be equivalent. 
This explains $E(\mathsf{A}, \mathsf{B}) = E(\mathsf{A}, \mathsf{B}^{\prime})$. 
The shift $+x$ also maps other states as well, resulting in the automorphism of
$(\mathsf{ A, B, A^{\prime}, B^{\prime}})(\mathsf{C, D, C^{\prime}, D^{\prime}})$. 
Note that this permutation satisfies Eq.~(\ref{eq:automorph}). 
As another example, a $\pi/2$ clock-wise spatial rotation in the depiction will correspond to the automorphism
$(\mathsf{A, A^{\prime}})(\mathsf{B, D, B^{\prime}, D^{\prime}})$.
An ideal order parameter will have the same set of symmetries as the Hamiltonian does, 
which means it should respect the symmetry of the automorphism group of the order graph,
as we explain next.

When we think of an order parameter which is an $d$ dimensional vector, we can regard it as a 
point in the $d$ dimensional Euclidean space $\mathbb{E}^d$. 
If there are $M$ different ordered states, they should correspond to $M$ different points ${\bm x}_1, {\bm x}_2, \ldots , {\bm x}_M$  
in $\mathbb{E}^d$, 
and these points should have equal distance from the origin. 
We set $\forall i, ~ | {\bm x}_i| =1 $ for simplicity. 
Furthermore, let us think of two ordered states corresponding to ${\bm x}_i$ and $ {\bm x}_j$, where they are ``adjacent", 
meaning that the fluctuation between those two ordered states are the strongest. 
This means that the corresponding vertices in the order graph $v_i$ and $v_j$ have a connecting edge $E(v_i, v_j)=1$.
Ideally, all states ($ {\bm x}_k$ for example) which have the same relation to ${\bm x}_i$ as ${\bm x}_j$ does, 
should have the same relation, which requires $ |{\bm x}_i - {\bm x}_j | = |{\bm x}_i - {\bm x}_k |$. 
Now, if we think of a symmetric transformation of the Hamiltonian, 
it transforms one ordered state to another one in general. 
Note that a symmetric transformation which does not change any ordered state to another 
corresponds to a symmetry that does {\it not} become broken even after a (partial) symmetry breaking has taken place.
As we have exemplified in the previous paragraph,
we can express the transformation by a permutation $\sigma$ among $M$ elements. 
Such transformations translate to isometric transformations of the $M$ points in $\mathbb{E}^d$, 
corresponding to some rotation and/or reflection $F$. 
This is because the relative relation between states should not change under these transformations, and thus 
$|{\bm x}_i - {\bm x}_j | = | {\bm x}_{\sigma^F(i)} - {\bm x}_{\sigma^F(j)} |$ holds for any $F$ and $i, j$. 
If we want to construct an order parameter that reflects the broken symmetry, 
all symmetries of the Hamiltonian (note that they are always expressible by automorphisms of the order graph) 
should have a corresponding isometric transformation in the order parameter space $\mathbb{E}^d$. 

From the above argument, we arrive to a general way for constructing order parameters for 
discrete symmetry breaking, by a faithful embedding of the order graph defined as the following.
\vspace{0.2cm}\\
{\bf Definition: Faithful Embedding}\\
%\begin{mydefinition}
A faithful embedding of a graph $G=(V, E)$ to a Euclidean space $\mathbb{E}^d$ is a 
mapping $\Gamma : V \rightarrow \mathbb{E}^d$ of vertices to points, such that for all automorphism $f$ of $G$, 
there is an isometry $F$ in $\mathbb{E}^d$ that satisfies 
$\forall v \in V, ~~\Gamma(f(v)) = F (\Gamma (v))$.
\vspace{0.2cm}\\
%\end{mydefinition}
With this embedding, order parameters could be defined to be the sum of the vectors $\Gamma (v_i)={\bm x}_i$ corresponding to each microscopic
degrees of freedom in a given configuration. 
This is possible because global symmetry breaking always have long range order, meaning that
looking at local degrees of freedom is enough for determining which state the system is locally in.

Let's consider the AVBS phase as an example. 
In this phase, it is sufficient to look at four spins in a plaquette to determine which of the eight-fold degenerate states the system is in. 
More precisely, if we observe two adjacent and parallel dimer singlets, depending on which direction (vertical or horizontal) they are pointing and 
which of the four plaquette patterns they are in, one of the eight states is uniquely determined to be associated with them. 
This corresponds to the local degrees of freedom determining the state, as explained above. 
Once we obtain the ordered state $i$ that corresponds to a particular local configuration, 
then the vector $\Gamma (v_i)={\bm x}_i$ is the order parameter value that corresponds. 
Thus, a sum over all the sites of such vectors will be the most natural order parameter. 
In practice, since we use the $z$ basis representation in our SSE simulation, 
it is more practical to define the correspondence 
between a local $z$ basis configuration and the eight AVBS states. 
This essentially makes the resulting six dimensional order parameter to be equivalent to the direct sum of the plaquette order parameter $\Pi$ and 
the alternating order parameter $A$ defined in Eq.~(\ref{Piadefs}) and (\ref{eq:AVBSop}), respectively.

It should be noted that we only deal with discrete symmetry breaking here. 
Continuous symmetries are naturally associated with a corresponding perturbation in the physical system, 
and has a more direct physical meaning. 
For example, two ordered states which are connected with an infinitesimally small continuous symmetric transformation (Lie group)
also are related with a perturbation (Lie algebra). 
Two spatially separated regions with different ordered states can be smoothly connected with 
a continuously varying order parameter, which is exactly the manifestation of the continuous symmetry. 
This corresponds to some shear in the order parameter space, such as spin stiffness. 
Discrete symmetry breaking on the other hand, are less connected to specific physical properties, 
since two ordered states which are connected with a symmetry transformation do not have a smooth 
connection between them characterized by a parameterized transformation of the symmetry in general. 
Discrete symmetry groups can be regarded as a subgroup of a continuous symmetry group, 
and we can expect that they become a very good approximations to the continuous symmetry when the subgroup
is ``large enough". This is indeed the case of $q$-state clock models with $q$ large enough, 
which the discrete $\mathbb{Z}_q$ symmetry becomes a good approximation of an O(2) symmetry 
which corresponds to an XY model. 
This results in those clock models with large $q$ to have emergent XY criticality. 
Our method provides a systematic and intuitive way to obtain the possible emergent symmetry by analyzing the relation of the ordered states with a graph. 

The embedding method discussed here is very close to that in Ref.~\cite{IsometricEmbedding}, which considered only the case of
$M=2$ and furthermore imposed an additional constraint that edges are not allowed to cross each other in the embedding. 
The faithful embedding defined for our purpose is also interesting in it's own right, revealing interesting properties of vertex transitive graphs. 
For example, the minimum embedding dimension of a Petersen graph in the way we defined here is 5, which is nontrivial and also coincides with the 
embedding dimension defined in a more algebraically way \cite{PetersenGraph5D}, suggesting possible connection.

\end{document}